\documentclass[12pt]{article}

\usepackage{amsfonts}
\usepackage{amsmath}
\usepackage{authblk}
\usepackage{epstopdf}
\usepackage[T1]{fontenc}
\usepackage{geometry}
\usepackage{graphicx}
\usepackage{hyperref}
\usepackage[utf8]{inputenc}
\usepackage{setspace}

\DeclareMathOperator{\rank}{rank}
\DeclareMathOperator{\coker}{coker}
\DeclareMathOperator{\Pers}{Pers}
\DeclareMathOperator{\im}{im}
\DeclareMathOperator{\spn}{span}

\begin{document}

\title{Adaptive tracking of representative cycles in regular and zigzag persistent homology}
\author[*]{Jennifer Gamble}
\author[**]{Harish Chintakunta}
\author[*]{Hamid Krim
\thanks{\texttt{jpgamble@ncsu.edu} (Jennifer Gamble, corresponding author), \texttt{hkchinta@ncsu.edu} (Harish Chintakunta), \texttt{ahk@ncsu.edu} (Hamid Krim).}}
\affil[*]{Electrical and Computer Engineering, North Carolina State University}
\affil[**]{Coordinated Science Laboratory, University of Illinois Urbana Champaign}
\date{}

\maketitle

\abstract{Persistent homology and zigzag persistent homology are techniques which track the homology over a sequence of spaces, outputting a set of intervals corresponding to birth and death times of homological features in the sequence. This paper presents a method for choosing a homology class to correspond to each of the intervals at each time point. For each homology class a specific representative cycle is stored, with the choice of homology class and representative cycle being both geometrically relevant and compatible with the birth-death interval decomposition. After describing the method in detail and proving its correctness, we illustrate the utility of the method by applying it to the study of coverage holes in time-varying sensor networks.}

%\tableofcontents

\section{Introduction}

The field of topological data analysis \cite{Carlsson2009} has been blossoming in recent years, and many more statisticians, computer scientists and engineers are beginning to use topological tools to study their data. The most popular and successful of these tools is persistent homology, a method which characterizes a space (or object) by using a multi-scale description of its topological features. These include things like the number of connected components, holes, or voids. A variant of regular persistent homology is zigzag persistent homology, which describes the topological features as they change over a sequence of spaces.

In this paper, we propose an algorithm for obtaining specific representative cycles to track homological features over a sequence of simplicial complexes in a geometrically meaningful way. We include some explorations where this method is applied to track coverage holes in time-varying sensor networks, with fruitful results.

In Section \ref{Notation}, we build up the foundational terminology and notations required for our discussions. This includes brief descriptions of simplicial homology and zigzag persistence. In Section \ref{RepCycles} we describe our method, and give an algorithm for implementing it on a sequence of simplicial complexes. Section \ref{Correctness} proves the correctness of our algorithm, and Section \ref{Examples} illustrates the utility of the method for analysis of coverage properties of time-varying sensor networks.

\section{Terminology and notation}\label{Notation}

\subsection{Simplicial complexes and homology}

We will use ideas from simplicial homology theory throughout, so some basic notations are introduced here. For a general reference on algebraic topology (including simplicial homology), see \cite{hatcher2001}. We define the notation for simplicial complexes, homology groups and classes.

A \textit{$k$-simplex} $\sigma = [v_0, v_1, \ldots, v_k]$ is a set of $k+1$ vertices. Any subset of the $k+1$ vertices forming a simplex is called a \textit{face} of the simplex, where each face is also a simplex itself. A \textit{simplicial complex}, $K$, is a set of simplices such that any simplex in $K$ also has all of its faces in $K$, and the intersection of any two simplices $\sigma_1$ and $\sigma_2$ in $K$ is a face of both $\sigma_1$ and $\sigma_2$.

Given a simplicial complex $K$, the \textit{chain spaces} $\mathcal{C}_0$, $\mathcal{C}_1$, $\mathcal{C}_2$, $\ldots$, are vector spaces, where $\mathcal{C}_k$ has the $k$-simplices as basis elements. The specific structure of the simplicial complex is encoded in the \textit{boundary maps} $\partial_1$, $\partial_2$, $\ldots$. For $k$-simplex $\sigma = [v_0, v_1, \ldots, v_k]$, the boundary map $\partial_k: \mathcal{C}_k \rightarrow \mathcal{C}_{k-1}$ maps $\sigma$ onto the alternating sum of its faces:
\[ \partial_k \sigma = \sum_{i=0}^k (-1)^i[v_0, \ldots, \hat{v}_i, \ldots, v_k] \]
where $\hat{v}_i$ indicates vertex $v_i$ is removed. Then the \textit{$k$-th homology group} is
\[ H_k(K) = \ker(\partial_k)/\mbox{im}(\partial_{k+1}) \]
and the \textit{$k$-th Betti number} (denoted $\beta_k$) of the simplicial complex $K$ is the rank of $H_k(K)$.

An element $c$ in the chain space $\mathcal{C}_k$ is a linear combination of $k$-simplices $\sigma_1, \ldots, \sigma_{n_k}$,
\[ c = \sum_{i=1}^{n_k} a_i \sigma_i \]
The coefficients $a_i$ come from a field $\mathbb{F}$ (such as the real numbers). We choose to perform our computations over the field $\mathbb{Z}_2 = \{0,1\}$. Any chain with boundary zero (i.e. any $c$ such that $\partial_k c = 0$) is called a \textit{cycle}, and so $\ker(\partial_k)$ is the set of all $k$-cycles. All boundaries of simplices are themselves cycles (i.e. $\mbox{im}(\partial_{k+1}) \subseteq \ker(\partial_k)$), so $\partial_k \partial_{k+1} = 0$, and homology corresponds to ``cycles which are not boundaries''. Two cycles $c_1$ and $c_2$ are \textit{homologous} (written $c_1 \sim c_2$) if their difference can be written as a linear combination of boundaries. The set of all cycles that are homologous to a given cycle (say $c$) is called a \textit{homology class} (denoted $[c]$). When a specific cycle is chosen to represent an entire homology class, it is called a \textit{representative cycle}.

When writing an equation about the homology of a space, we use a general notation of $\mathsf{H}(K)$, which can be taken to mean that the total homology $H_*(K)$, or a specific $H_p(K)$, could be inserted into the equation in the place of $\mathsf{H}(K)$. Similarly, we use $\mathsf{\beta}(K) = \rank(\mathsf{H}(K))$ as a general notation for the associated Betti number. For the later applications to sensor networks, and for visualization purposes, it is convenient to think of $\mathsf{H}(K)$ to mean $H_1(K)$.

\subsection{Zigzag persistence}
The theory of zigzag persistent homology is concerned with how the homology changes over a sequence of spaces. There are mathematical results \cite{carlsson2010} showing that the changing homology of such a sequence can be expressed uniquely in terms of birth and death times of homological features in the sequence. There is also an algorithm \cite{carlsson2009b} for computing this birth-death decomposition for a given sequence of simplicial complexes.

Consider a sequence of simplicial complexes $K_1, K_2, \ldots, K_n$, connected by either forward inclusion maps $K_i \rightarrow K_{i+1}$ or backward inclusion maps $K_i \leftarrow K_{i+1}$. We write this sequence as

\begin{equation*}\label{zzSequence}
K_1 \longleftrightarrow K_2 \longleftrightarrow \ldots \longleftrightarrow K_n
\end{equation*}

The inclusion maps induce linear maps between the associated homology spaces $V_i = \mathsf{H}(K_i)$, which we write as the \emph{zigzag persistence module}

\begin{equation}\label{zzModule}
\mathbb{V}  = V_1 \overset{p_1}{\longleftrightarrow} V_2 \overset{p_2}{\longleftrightarrow} \ldots \overset{p_{n-1}}{\longleftrightarrow} V_n \overset{p_n}{\longrightarrow} V_{n+1}
\end{equation}
where $K_i \longrightarrow K_{i+1}$ induces the forward map $V_i \overset{f_i}{\longrightarrow} V_{i+1}$, and $K_i \longleftarrow K_{i+1}$ induces the backward map $V_i \overset{g_i}{\longleftarrow} V_{i+1}$. Regardless of the direction, we use $i(\cdot)$ to denote the inclusion map between consecutive simplicial complexes. We further assume that consecutive simplicial complexes $K_i$ and $K_{i+1}$ differ by exactly one simplex, so $K_{i+1} = K_i \cup \{\sigma\}$ (in the forward case), or $K_{i+1} = K_i - \{\sigma\}$ (in the backward case).

When $\sigma$ is a $d$-simplex, its addition results in either an increase in the dimension of the $d$-dimensional homology space, or a decrease in the dimension of the $(d-1)$-dimensional homology space. Similarly, the removal of a $d$-simplex $\sigma$ results in either an increase in the $(d-1)$-dimensional homology, or a decrease in the $d$-dimensional homology. When the dimension of the homology space increases, we refer to this as a \emph{birth}, and when the dimension decreases, we refer to this as a \emph{death}.
\begin{eqnarray}
Birth: \mbox{ } \dim(V_{i+1}) & = & \dim(V_i) + 1  \nonumber \\
Death: \mbox{ } \dim(V_{i+1}) & = & \dim(V_i) - 1 \nonumber
\end{eqnarray}

Each inclusion between simplicial complexes will induce maps between the homology spaces of all dimensions, but these maps will be simple identity maps in all dimensions except for one. This will depend on whether the addition or removal of $d$-simplex $\sigma$ results in a birth or a death. For the addition or removal of a $d$-simplex $\sigma$, the map $V_i \overset{p_i}{\longleftrightarrow} V_{i+1}$ on the corresponding homology zigzag module, will be interpreted as a forward or backward linear map between the appropriate-dimensional homology spaces, as summarized in Table \ref{DimTable}.

\vspace{2mm}
\begin{table}
\begin{center}
\begin{tabular}{|l|c|}
\hline
\multicolumn{1}{|c|}{Case} & $V_i \overset{p_i}{\longleftrightarrow} V_{i+1}$ \\
\hline
1. Birth by addition & $H_d(K_i) \overset{f_i}{\longrightarrow} H_d(K_{i+1})$ \\
2. Birth by removal & $H_{d-1}(K_i) \overset{g_i}{\longleftarrow} H_{d-1}(K_{i+1})$ \\
3. Death by addition & $H_{d}(K_i) \overset{f_i}{\longrightarrow} H_{d}(K_{i+1})$ \\
4. Death by removal & $H_{d-1}(K_i) \overset{g_i}{\longleftarrow} H_{d-1}(K_{i+1})$ \\
\hline
\end{tabular}
\caption{Indicates the relationship between the dimension $d$ of simplex $\sigma$ being added or removed, and the dimension of homology space which is affected. \label{DimTable}}
\end{center}
\end{table}
\vspace{2mm}

A main result from the theory of zigzag persistence, is that a zigzag module such as in Equation \ref{zzModule} has an interval decomposition,

\begin{equation}\label{IntervalDecomp}
\mathbb{V} \cong \mathbb{I}(b_1,d_1) \otimes \mathbb{I}(b_2,d_2) \otimes \ldots \otimes \mathbb{I}(b_m,d_m)
\end{equation}
which is unique up to isomorphism, and is equivalently expressed as the multiset of pairings of births and deaths in the sequence, and represented as integer intervals, called the \emph{zigzag persistence} of $\mathbb{V}$ \cite{carlsson2010}

\begin{equation}\label{bdpairs}
\Pers(\mathbb{V}) = \{[b_j,d_j] \mbox{ } | \mbox{ } j = 1,\ldots,m \}.
\end{equation}
These are interpreted as birth and death times of homological features in the sequence.

\subsection{Right filtration}

Given a zigzag module $\mathbb{V}$ as in Equation \ref{zzModule}, the zigzag persistence algorithm \cite{carlsson2009b} computes the interval decomposition in Equation \ref{IntervalDecomp} by keeping track of a right filtration $R(\mathbb{V})$ on the spaces. The right filtration $R(\mathbb{V})$ is computed incrementally, and results in a filtration (a nested sequence of subspaces) on $V_n$, along with a birth time associated to each quotient space. A right filtration on $V_i$ is denoted

\begin{equation}\label{RightFiltration}
\mathcal{R}_i = (R_i^0, R_i^1, \ldots, R_i^i)
\end{equation}
where $R_i^0 \leq R_i^1 \leq \ldots \leq R_i^i$ and $R_i^i = V_i$. The quotients $R_i^1/R_i^0$, $R_i^2/R_i^1$, $\ldots$, $R_i^i/R_i^{i-1}$ are each associated with a birth time $b_i^j$ (for $j=0, \ldots, i$), which are recorded in the vector

\begin{equation}\label{BirthVector}
\mathbf{b}_i = (b_i^1, b_i^2, \ldots, b_i^i)
\end{equation}

We may write the quotients as
\[ \mathcal{R}_i' = (R_i^1/R_i^0, R_i^2/R_i^1, \ldots, R_i^i/R_i^{i-1}) \]
The computation of a right filtration is defined inductively, depending on whether the map from $V_i$ to $V_{i+1}$ is a forward map $\overset{f_i}{\longrightarrow}$ or a backward map $\overset{g_i}{\longleftarrow}$. For a single vector space $V_1$, we have the base case of $i=1$, and we define
\[ \mathcal{R}_1 = (\textbf{0}, V_1) \mbox{ and  } \mathbf{b}_1 = (0) \]
In the inductive step, if we are given $\mathcal{R}_i$ and $\mathbf{b}_i$ as in Equations \ref{RightFiltration} and \ref{BirthVector} above, then
\begin{itemize}
\item If $V_i \overset{f_i}{\longrightarrow} V_{i+1}$, then
\begin{eqnarray}\label{fRight}
\mathcal{R}_{i+1} & = & (f_i(R_i^0), f_i(R_i^1), \ldots, f_i(R_i^i), V_{i+1}) \\
\mathbf{b}_{i+1} & = & (b_i^1, b_i^2, \ldots, b_i^i, i+1) \nonumber
\end{eqnarray}
\item If $V_i \overset{g_i}{\longleftarrow} V_{i+1}$, then
\begin{eqnarray}\label{gRight}
\mathcal{R}_{i+1} & = & (\textbf{0}, g_i^{-1}(R_i^0), g_i^{-1}(R_i^1), \ldots, g_i^{-1}(R_i^i)) \\
\mathbf{b}_{i+1} & = & (i+1, b_i^1, b_i^2, \ldots, b_i^i) \nonumber
\end{eqnarray}
\end{itemize}

Since we assume that consecutive simplicial complexes differ by at most one simplex, the change in dimension between $V_i$ and $V_{i+1}$ is at most 1. Similarly, the dimension of the quotient space $R_i/R_{i+1}$ is either 0 or 1, for $i = 1,\ldots, n$, with their total dimension equaling that of $V_i$. The dimension of $V_i$ is the rank of the homology group for $K_i$ (the Betti number, $\beta(K_i)$), which is at most $i$:
\[ \dim(V_i) = \rank(\mathsf{H}(K_i)) = \beta(K_i) \leq i \]
For example, the dimension of the quotient spaces will be a sequence of 0's and 1's
\[ \dim(R_i^1/R_i^0, R_i^2/R_i^1, \ldots, R_i^i/R_i^{i-1}) = (0, 0, 1, 1, 0, \ldots, 1, 0) \]

Note that choosing one homology class from each of the nonzero quotient spaces results in a basis for $V_i$. The right filtration on $V_i$ can then be described using the unique subspaces in the right filtration (which have corresponding quotient spaces of dimension 1). Indexing the nonzero quotient spaces by $j_1,\ldots, j_{\beta(K_i)}$, define $W_i^k = R_i^{j_k}$ for the spaces $R_i^{j_1},\ldots,R_i^{j_{\beta(K_i)}}$ to obtain a more compact representation of the right filtration $\mathcal{R}$:

\begin{eqnarray}\label{Wfiltration}
\mathcal{W}_i & = & (W_i^1, \ldots, W_i^{\beta(K_i)}) \\
& = & (R_i^{j_1}, \ldots, R_i^{j_{\beta(K_i)}}) \nonumber
\end{eqnarray}
where the $R_i^{j_k}$ are those with $\dim(R_i^{j_k}/R_i^{j_k - 1}) = 1$, therefore the quotient spaces $W_i^j/W_i^{j-1}$ are all one-dimensional. We note that a basis $\{[w_i^j]\}_{j=1}^{\beta(K_i)}$ for $V_i$ is \emph{compatible} with the right filtration $\mathcal{W}_i$ if there is one basis element in each quotient space:
\[ [w_i^j] \in W_i^j/W_i^{j-1} \]
for $j = 1,\ldots,\beta(K_i)$. We return to this concept in Section \ref{Algorithm}.

Additionally, let
\[ \mathbf{b}_i^W = (b_i^{j_1}, b_i^{j_2}, \ldots, b_i^{j_{\beta(K_i)}}) \]
contain the birth times of the non-zero quotient spaces, which is the birth vector for $\mathcal{W}$. So $\mathbf{b}_i^W$ is a subset of the birth vector $\mathbf{b}_i$ for the full right filtration $\mathcal{R}$.

The zigzag persistence algorithm is implemented by determining whether a birth or a death is occurring with each simplex addition or deletion. The right filtration and birth vector are then updated accordingly, and when a death occurs, the quotient space $R_i^j/R_i^{j-1}$ corresponding to it is determined, and the associated birth time $b_i^j$ used to output the interval $[b_i^j,i]$.

While the output of intervals $\{[b_j,d_j] \mbox{ } | \mbox{ } j=1,\ldots,m \}$ is unique, there may be more than one way to choose homology classes corresponding each interval. In Section \ref{RepCycles} we will propose a method for choosing a homology class (by choosing a specific representative cycle for it) for each interval at each time point in a way that is geometrically motivated, but still compatible with the right filtration.

\section{Tracking representative cycles}\label{RepCycles}

\subsection{Motivation}
Our interest in choosing and tracking representative cycles over a sequence of spaces stems from analysis of coverage holes in time-varying sensor networks. The idea of using homological methods to study coverage in sensor networks was proposed by de Silva and Ghrist (\cite{deSilva2006}, \cite{deSilva2007}), and the use of zigzag persistent homology allows some of these ideas to be employed in the dynamic network setting. The set of intervals output from the zigzag persistence algorithm describes the birth and death times of homological features, and these features do not necessarily correspond to individual coverage holes \cite{Adams2013}. Ideally, we are interested in tracking coverage holes over time, but this is not possible in general, given the constraints on the limited geometric information available with the sensor network model. Instead, we try and obtain a `good' representative cycle for a hole as it appears in the network, and then propagate this cycle over time as best as possible. Below we describe in more detail the model for the sensor network (\ref{Networks}), the representative cycles we would ideally like to obtain (\ref{Canonical}), and those that we are able to compute (\ref{Partial}).

\subsubsection{Homology for sensor networks}\label{Networks}

\begin{figure}[htp]
\begin{center}
\begin{tabular}{ccc}
\includegraphics[scale=0.3]{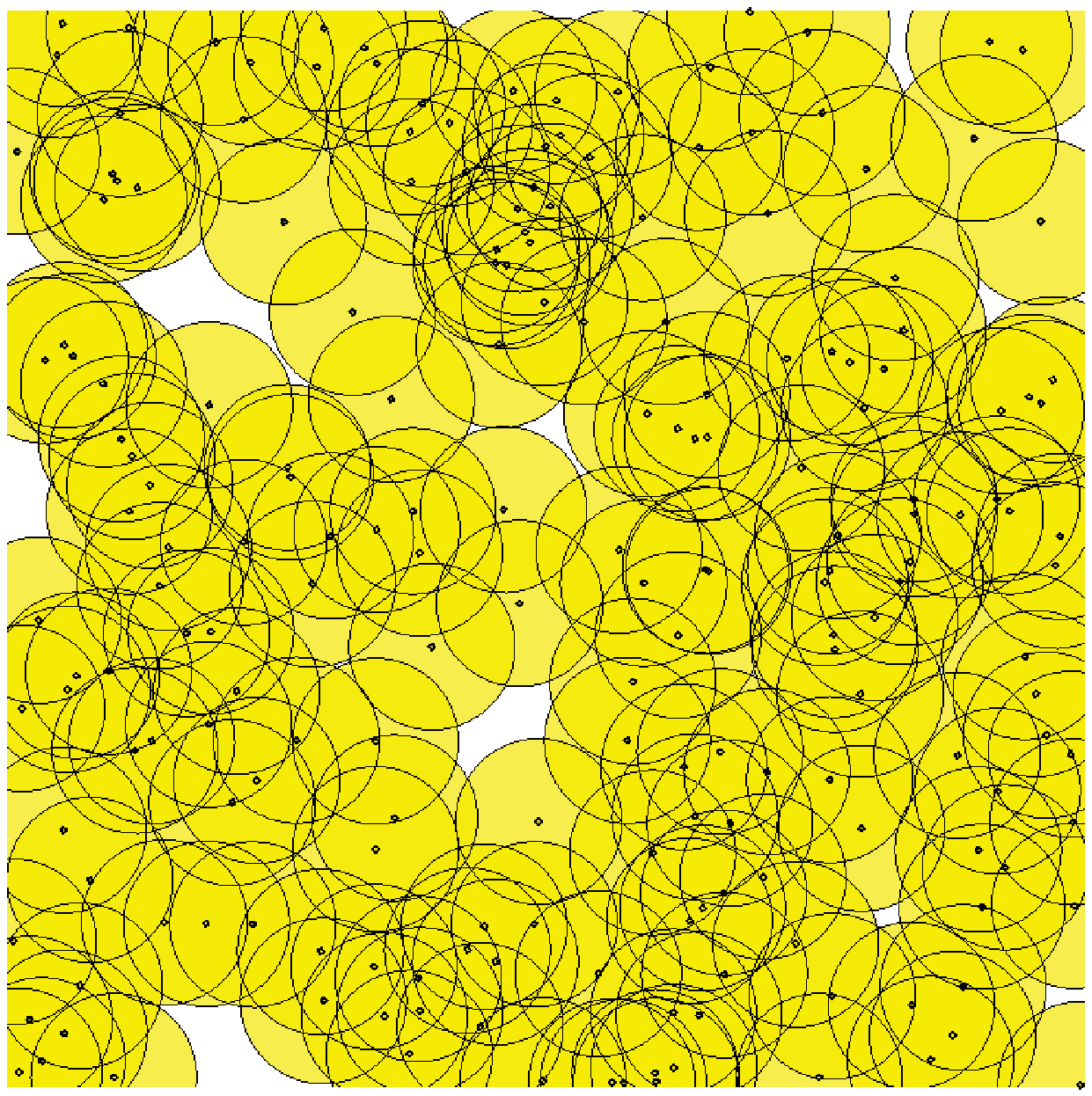} & \includegraphics[scale=0.3]{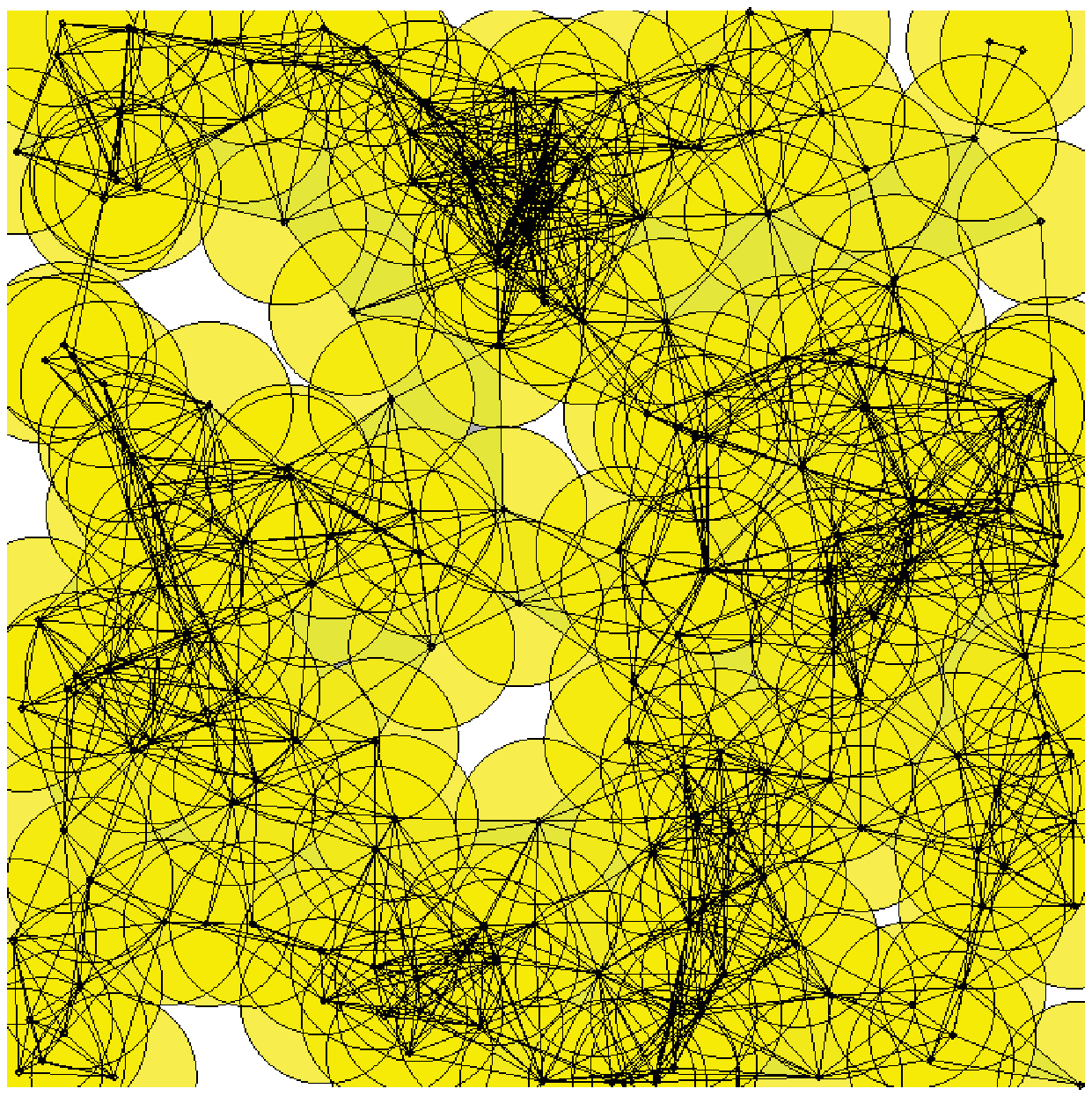}  & \includegraphics[scale=0.3]{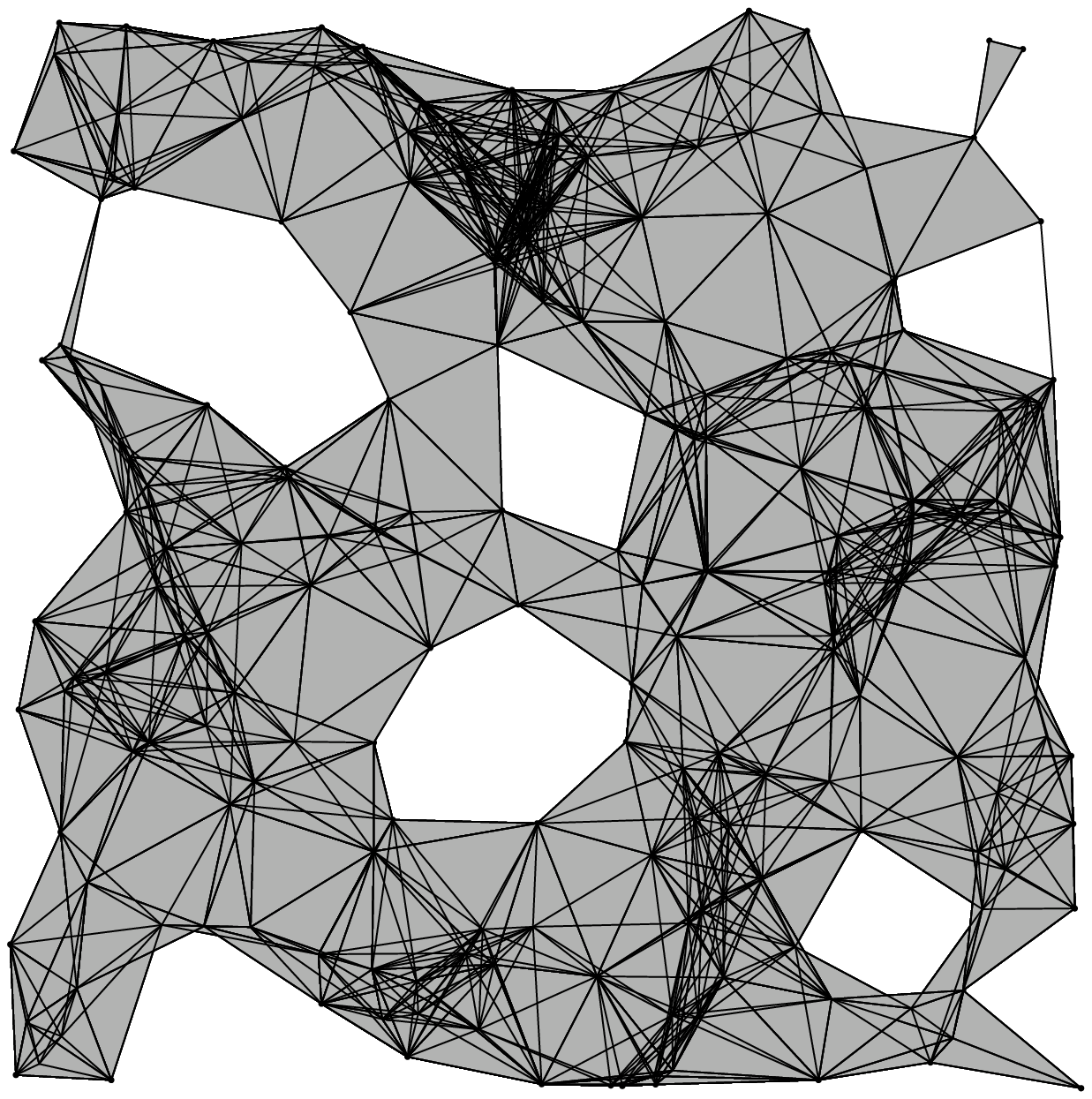} \\
\end{tabular}
\end{center}
\caption{(Left) The coverage region for a sensor network. (Center) The communication graph. (Right) The associated Rips complex. \label{Network}}
\end{figure}

\begin{figure}[htp]
\begin{center}
\begin{tabular}{cc}
\includegraphics[scale=0.3]{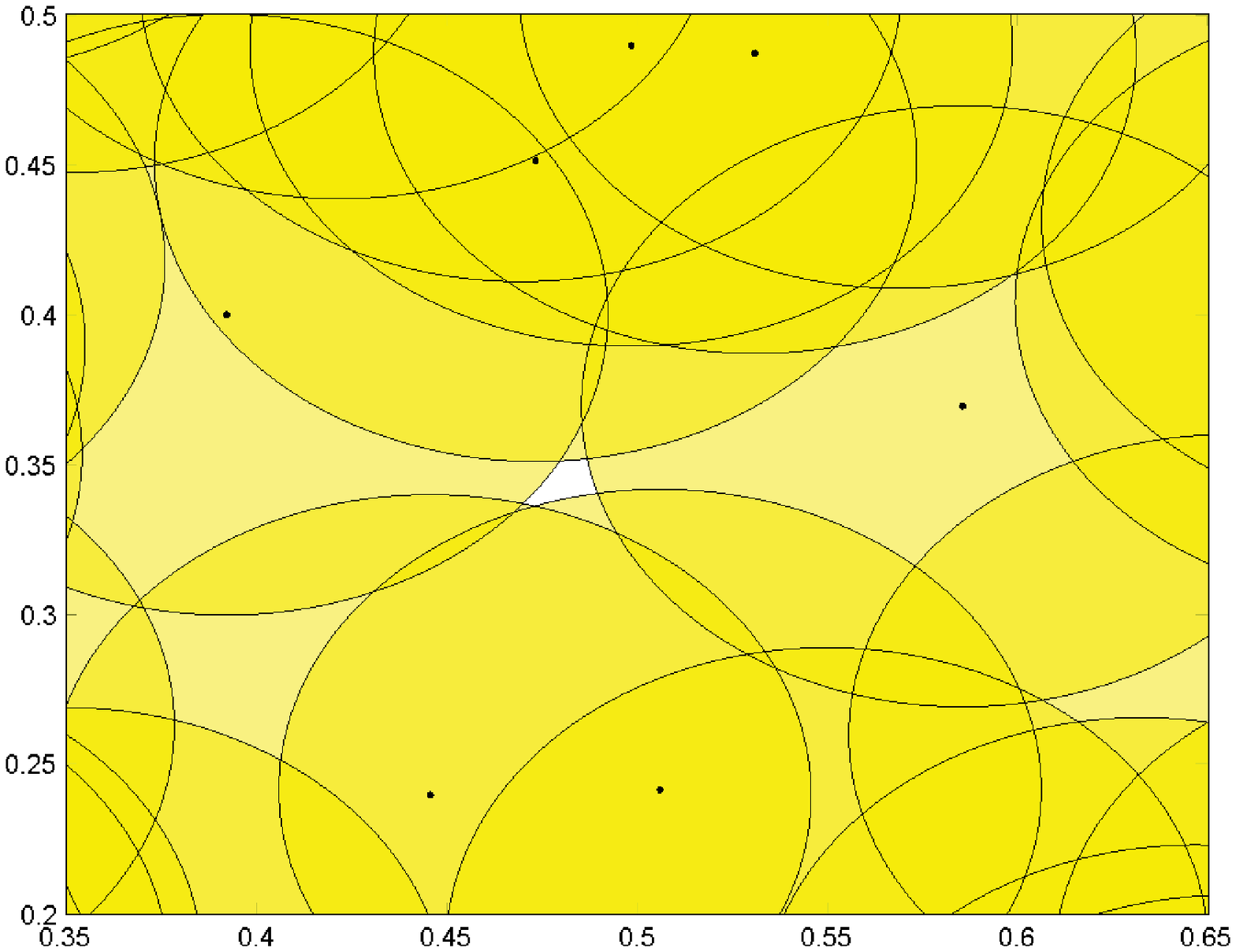} & \includegraphics[scale=0.35]{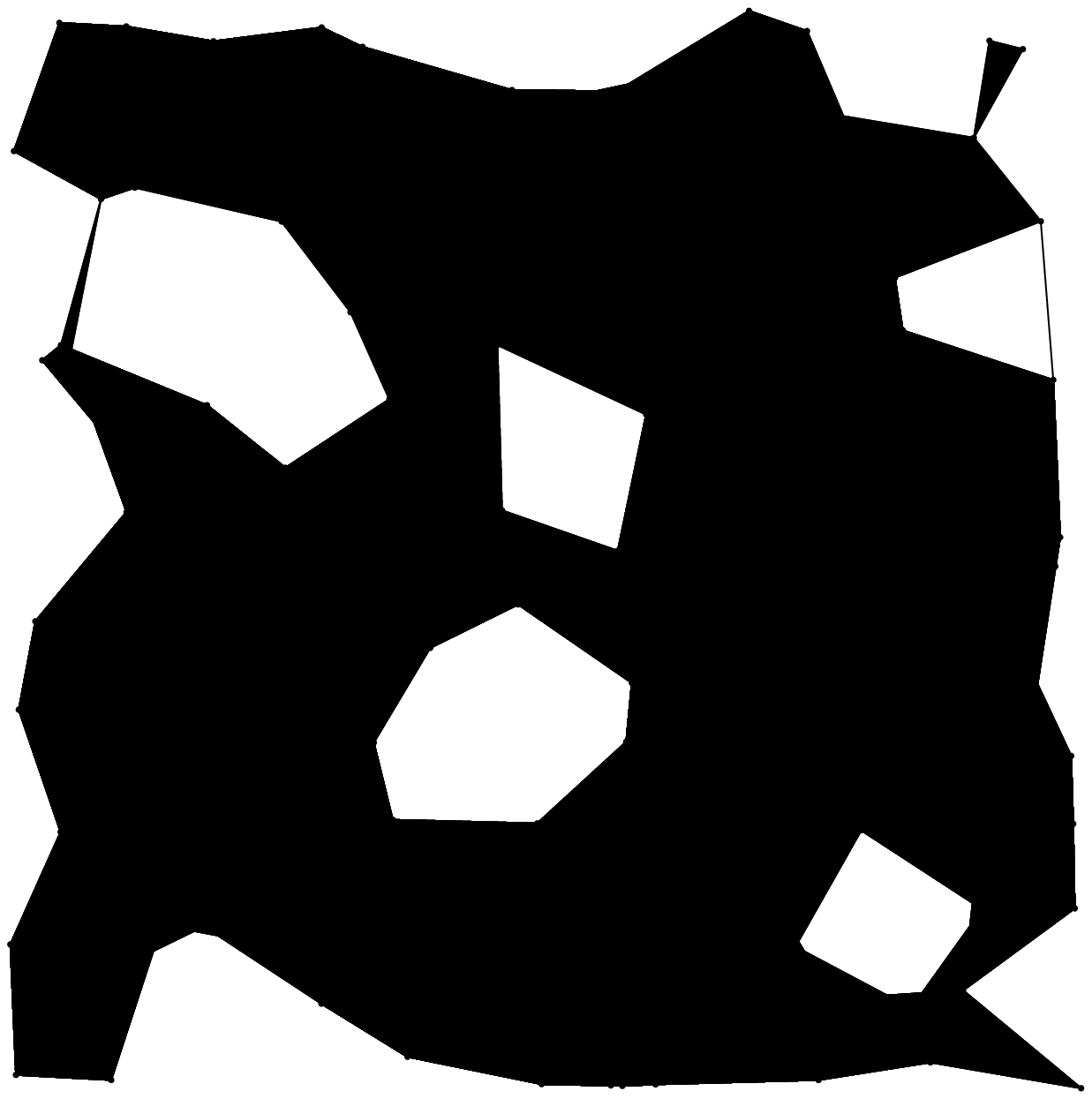}\\
\end{tabular}
\end{center}
\caption{(Left) A situation where three coverage disks each intersect pairwise, but there is no point where all three intersect. These three sensors will form a `filled-in' triangle in the Rips complex, but there is a hole in the coverage region.
(Right) The Rips shadow corresponding to the Rips complex in Figure \ref{Network}. \label{CechvsRips}}
\end{figure}

A network consists of a set of sensors, each at the center of an isotropic coverage disk of radius $r$. The union of the disks creates the coverage region for the entire network, and we are interested in making statements about coverage properties of this network, as the sensors are allowed to move over time. A communication graph is created by connecting any two sensors by an edge when they are less than distance $2r$ from one another, and the homology of the Rips complex of this graph is used to approximate the homology of the coverage region of the network. Figure \ref{Network} shows the coverage region (left), communication graph (center) and associated Rips complex (right) for a given sensor network. Note that a Rips complex is the maximal simplicial complex that can be built from a given graph, but since we are only interested in computing the first homology we only need to consider the 2-skeleton of the Rips complex. The Rips complex includes a 2-simplex defined by three sensors whenever their coverage disks have nonempty pairwise intersections, so if the disks have no triplet-wise intersection then a small hole may be present in the coverage region which is not detected by the Rips complex. See Figure \ref{CechvsRips} (left) for an example of this. For our purposes we designate such holes as too small to be of importance, and work with the homology as it is defined by the Rips complex. See \cite{deSilva2006} for an alternative approach, which allows false alarms (holes in the complex which do not exist in the coverage region), but is able to give coverage guarantees.

A final key result that we mention is by Chambers et al \cite{Chambers2010}, who show that the first homology of the Rips complex (a combinatorial object) is the same as the first homology of the projection of the Rips complex onto the plane (this projection is referred to as the Rips shadow). The Rips shadow corresponding to the sensor network from Figure \ref{Network} is shown in Figure \ref{CechvsRips} (right). For a Rips complex $K$, we denote its shadow as $K^S$.

\subsubsection{Zigzag persistence for dynamic networks}

In a time-varying network (represented by a sequence of simplicial complexes $K_{t_1},\ldots,K_{t_T}$), homology classes  may be tracked over time using zigzag persistent homology by mapping through the union complexes. So the sequence of simplicial complexes

\begin{equation}\label{zzComplexes}
\begin{array}{rlrclrccclrl}
K_{t_1} & & & K_{t_2} & & & \mbox{ } \cdot \mbox{ } & \cdot & \mbox{ } \cdot \mbox{ } & & & K_{t_T} \\
& \searrow & \swarrow & & \searrow & \swarrow & & & & \searrow & \swarrow & \\
& \multicolumn{2}{c}{(K_{t_1} \cup K_{t_2})} & & \multicolumn{2}{c}{(K_{t_2} \cup K_{t_3})} & & & & \multicolumn{2}{c}{(K_{t_{T-1}} \cup K_{t_T})} & \\
\end{array}
\end{equation}
gives rise to the associated zigzag persistence module:

\begin{footnotesize}
\[ \begin{array}{rlrclrccclrl}
H_1(K_{t_1}) & & & H_1(K_{t_2}) & & & \mbox{ } \cdot \mbox{ } & \cdot & \mbox{ } \cdot \mbox{ } & & & H_1(K_{t_T}) \\
& \searrow & \swarrow & & \searrow & \swarrow & & & & \searrow & \swarrow & \\
& \multicolumn{2}{c}{H_1(K_{t_1} \cup K_{t_2})} & & \multicolumn{2}{c}{H_1(K_{t_2} \cup K_{t_3})} & & & & \multicolumn{2}{c}{H_1(K_{t_{T-1}} \cup K_{t_T})} & \\
\end{array} \]
\end{footnotesize}

For implementational and theoretical purposes the sequence in Equation \ref{zzComplexes} is broken down, with each forward map re-written as a series of single simplex additions, and each backward map as a series of single simplex deletions. This refinement induces the analogous refinement on the zigzag module.

\subsubsection{Canonical basis}\label{Canonical}

Given a compact region in the plane such as the Rips shadow $K^S$, there exists a `canonical basis' for its first homology space, where each basis homology class surrounds a single hole. Consider $\overline{K^S}$, the complement of $K^S$ in $\mathbb{R}^2$, then the number of separate components in $\overline{K^S}$ (ignoring the infinite component) is equal to the number of holes in $K^S$ (i.e. the rank of $H_1(K^S)$). This result is a specific case of the more general principle of Alexander Duality (see, for example Ch. 5 of \cite{Miller2005}), which for a certain class of spaces, relates the $k$-th reduced homology of a space to the $n-q-1$-th cohomology of the complement of the space (where $n$ is the embedding dimension). We do not go into details here, but the salient point is that a canonical basis exists for the first homology of a space in the plane, with one homology class surrounding each hole.

Since the Rips complex $K$ and its Rips shadow $K^S$ have the same homology, a desirable goal would be to have a homology basis for the Rips complex, where projection of this basis onto the Rips shadow gives the canonical basis. In particular, we would like a representative cycle for each homology class in the basis, where the projection of the representative cycle onto the Rips shadow is homologous to the boundary of one of the holes. \emph{In general, this desirable goal is not possible.} The Rips complex itself is not embeddable in two dimensions, so Alexander Duality cannot be applied to obtain a canonical basis for its first homology. Moreover, although $K$ has the same homology as $K^S$, it is impossible to know whether a given homology basis for $K$ corresponds to the canonical basis or not (without knowing coordinates for the vertices, or the projection map from $K$ onto $K^S$).

We will see in Section \ref{Partial} that taking the dynamic nature of the network into account, there are some cases where it is possible to make a canonical choice for a homology class (with corresponding representative cycle) at its birth or death time. In Section \ref{Algorithm} we present a method for obtaining these cycles and updating them as the network changes over time, along with an explicit algorithm for doing so.

\subsubsection{Partial canonical information}\label{Partial}

As described in Table \ref{DimTable}, a homology class can be born by either the addition or removal of a simplex, and similarly a death is caused by either the addition or removal of a simplex, resulting in four distinct cases for how the homology can change. In this section we illustrate the two cases corresponding to `births', and how one of them allows a canonical choice of homology class. Our discussions here are with respect to the first homology, but the same principles hold for $d$-dimensional homology.

In the sequence of simplicial complexes, the birth of a homology class occurs at time $i$ when either the forward map $V_i \overset{f_i}{\longrightarrow} V_{i+1}$ has nonzero cokernel, or the backward map $V_i \overset{g_i}{\longleftarrow} V_{i+1}$ has nonzero kernel. Of these two cases, $\ker(g_i) \neq \textbf{0}$ is the only one which indicates the specific homology class that is being born.

\begin{figure}[htp]
\begin{center}
\begin{tabular}{cccc}
$K_i$ & $K_{i+1}$ & Updated cycle \\
\multicolumn{1}{l}{\textbf{Birth by Addition}} & & \\
\includegraphics[scale=0.25]{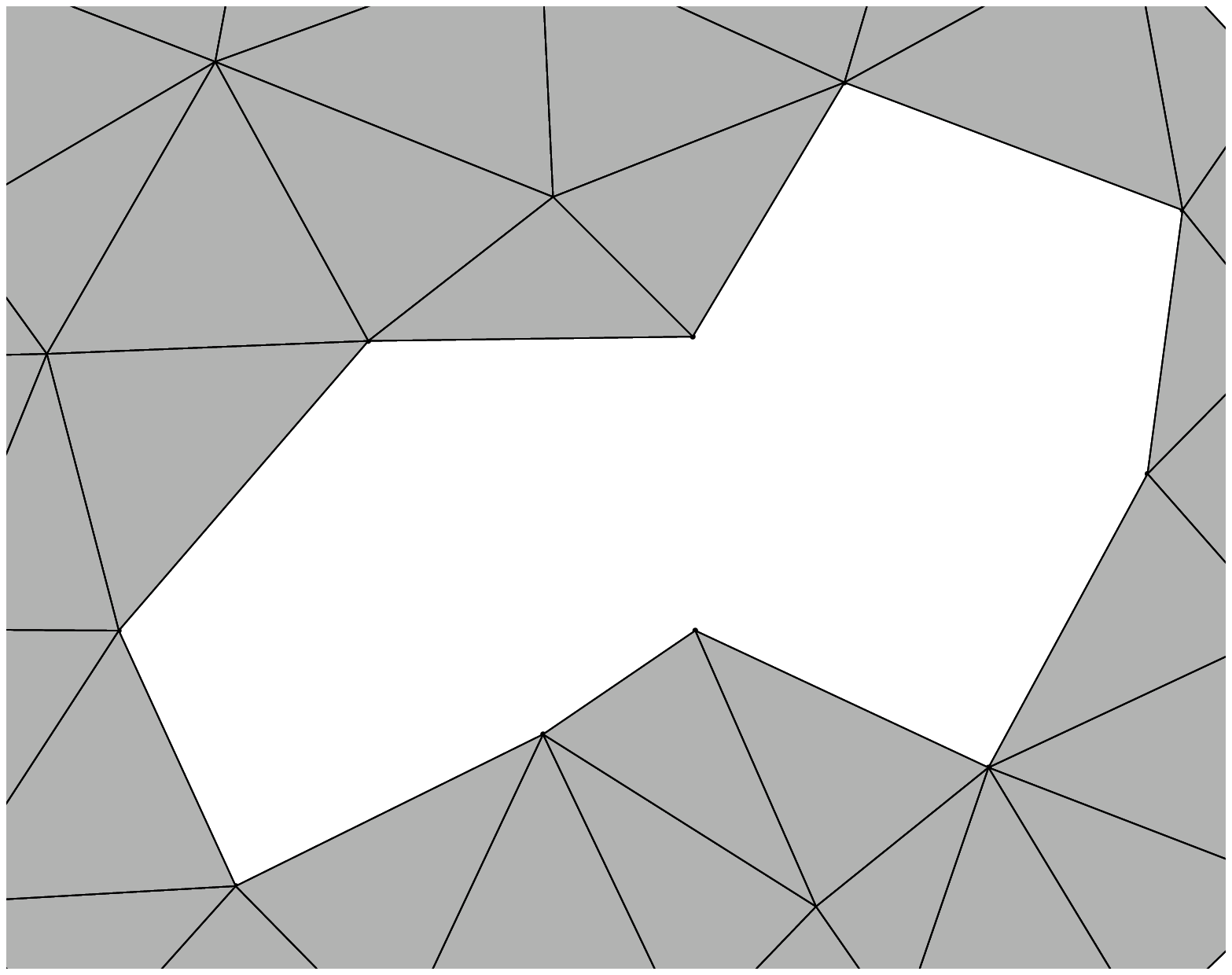} & \includegraphics[scale=0.25]{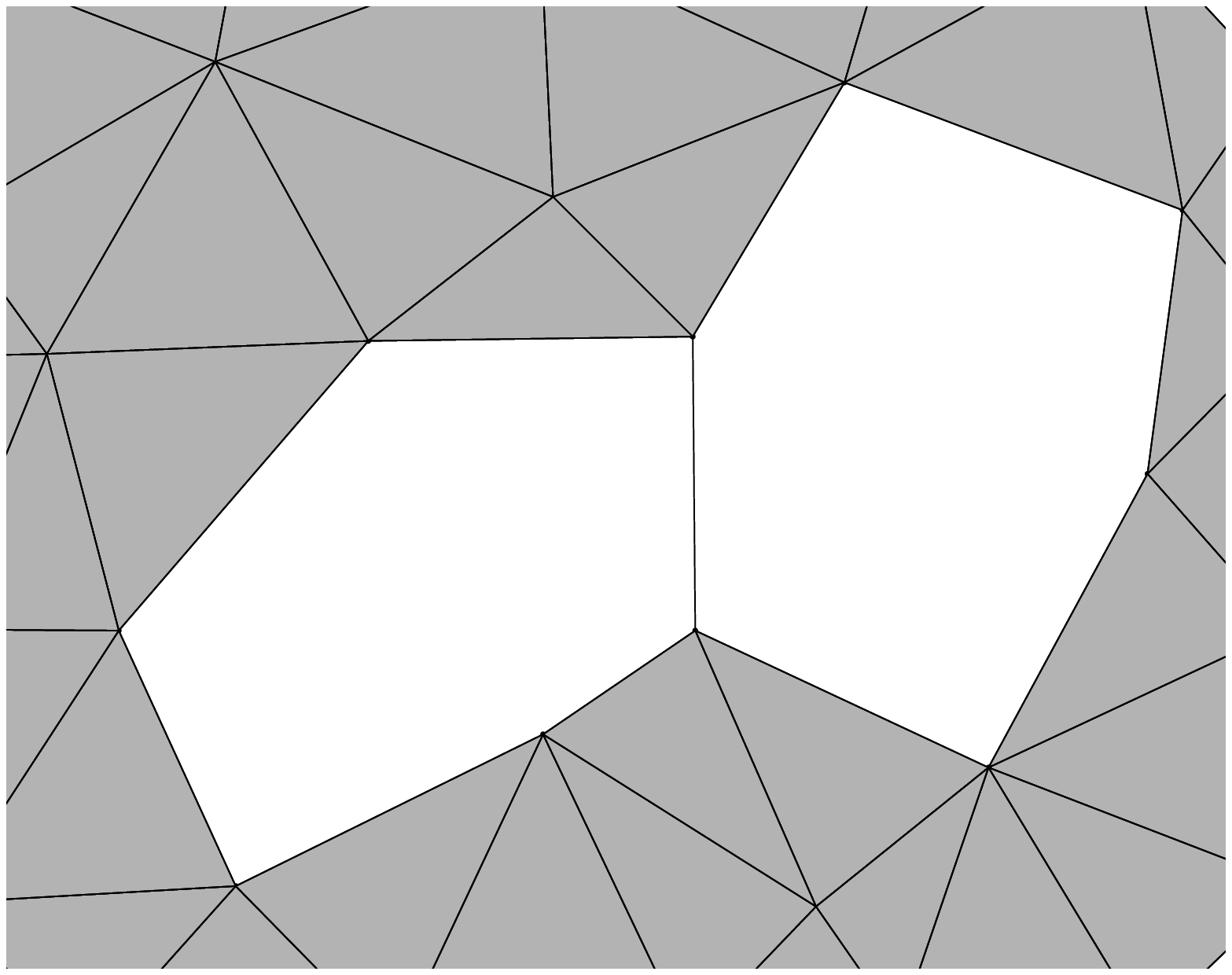} & \includegraphics[scale=0.25]{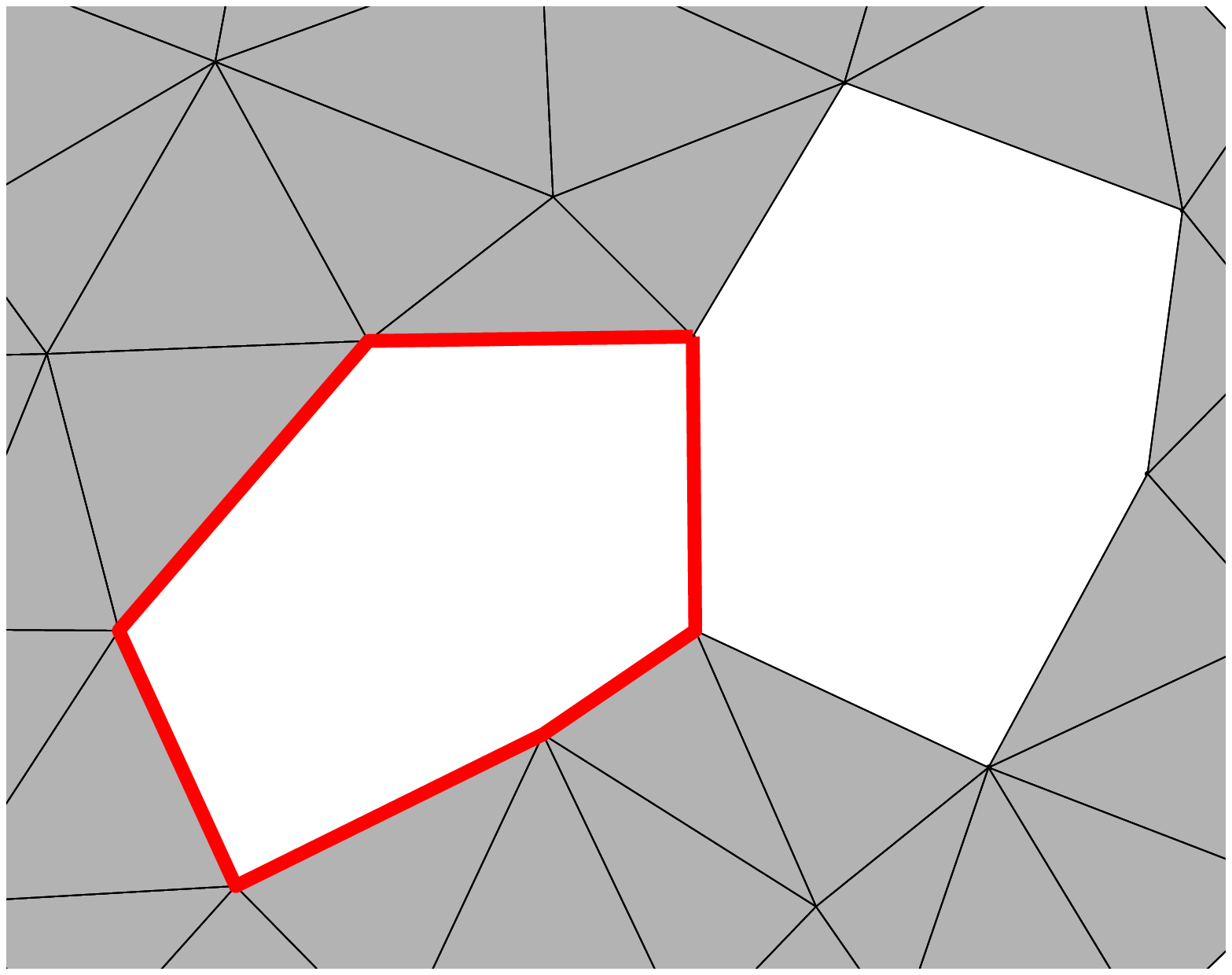} \\
\multicolumn{1}{l}{\textbf{Birth by Removal}} & & \\
\includegraphics[scale=0.25]{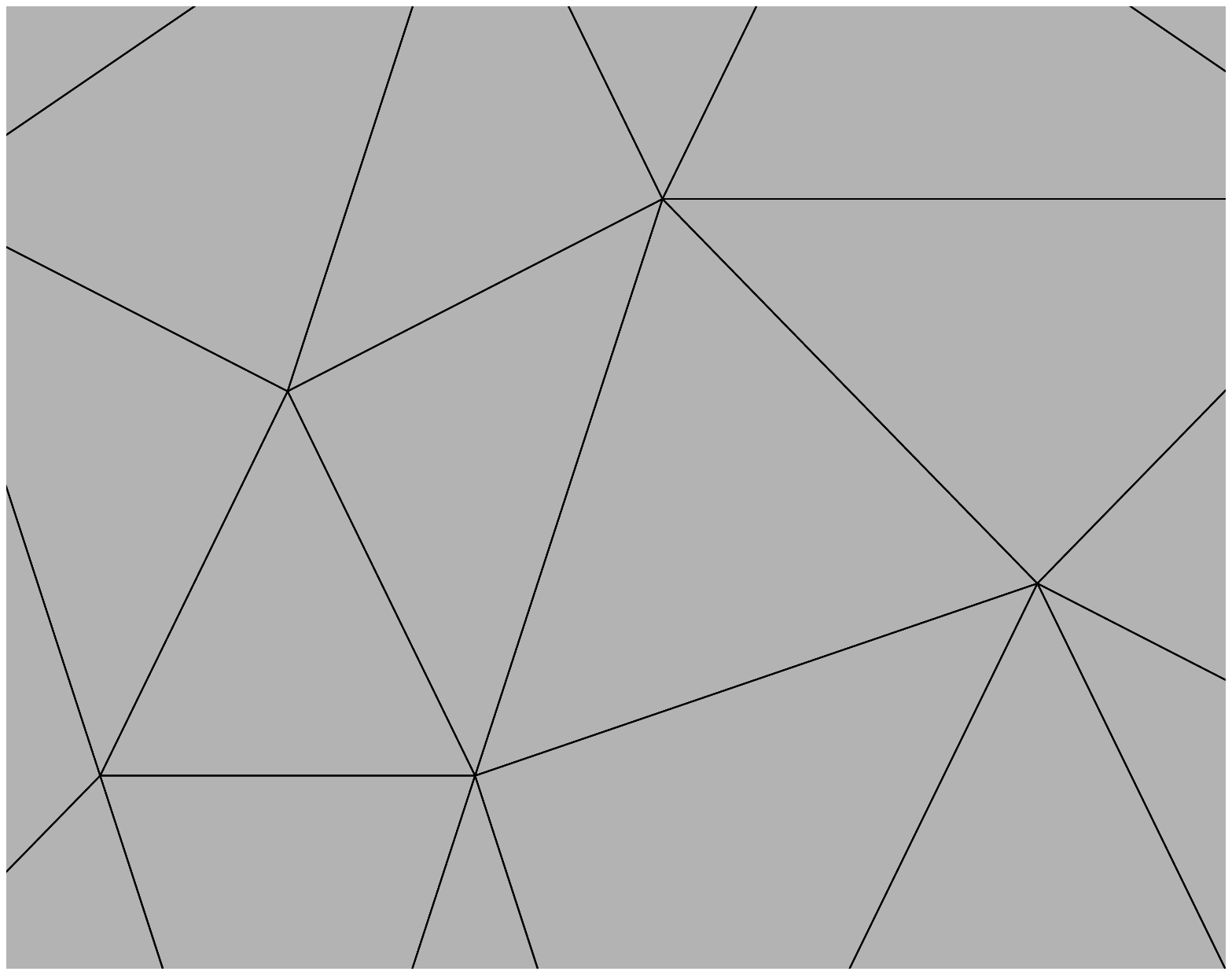} & \includegraphics[scale=0.25]{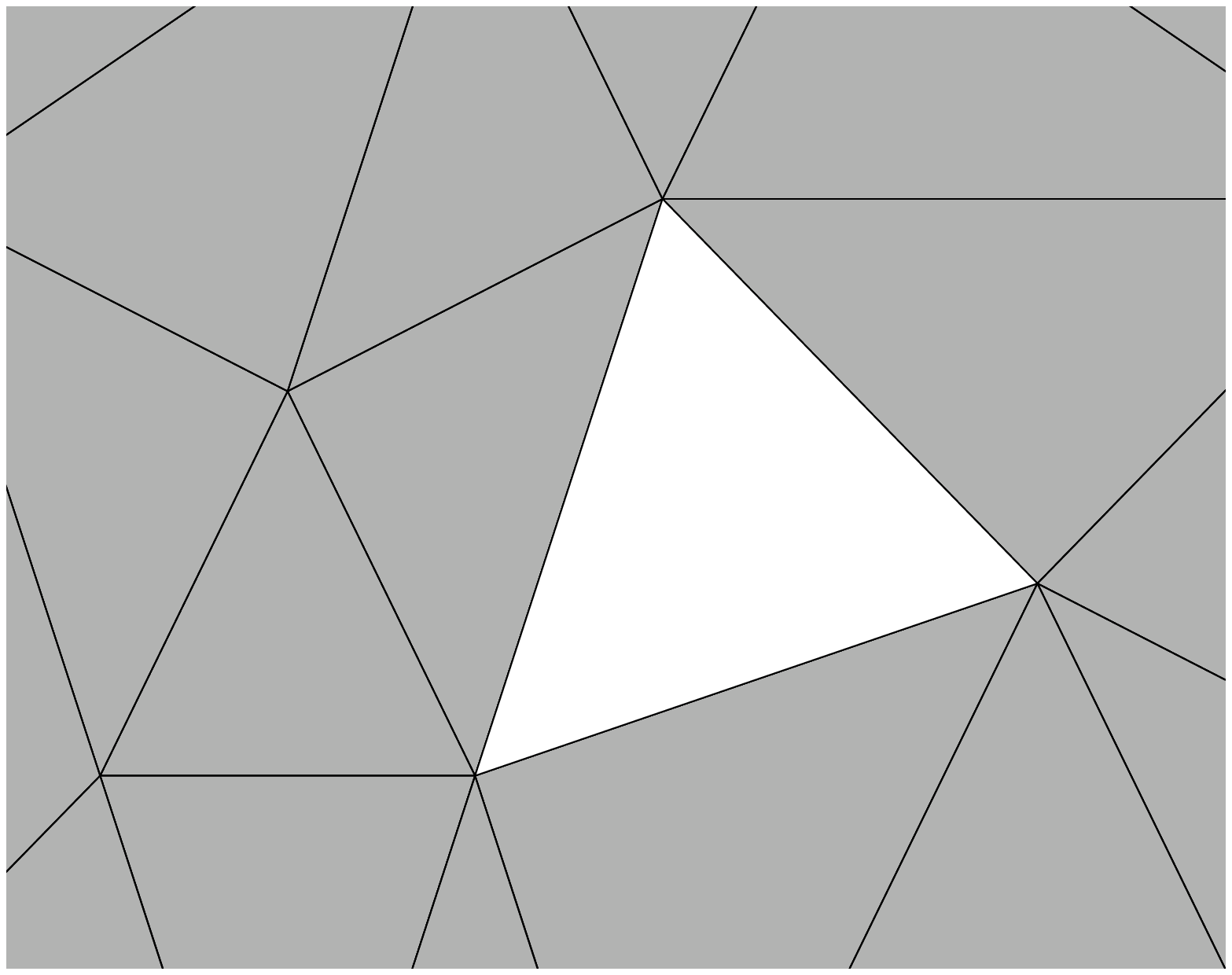} & \includegraphics[scale=0.25]{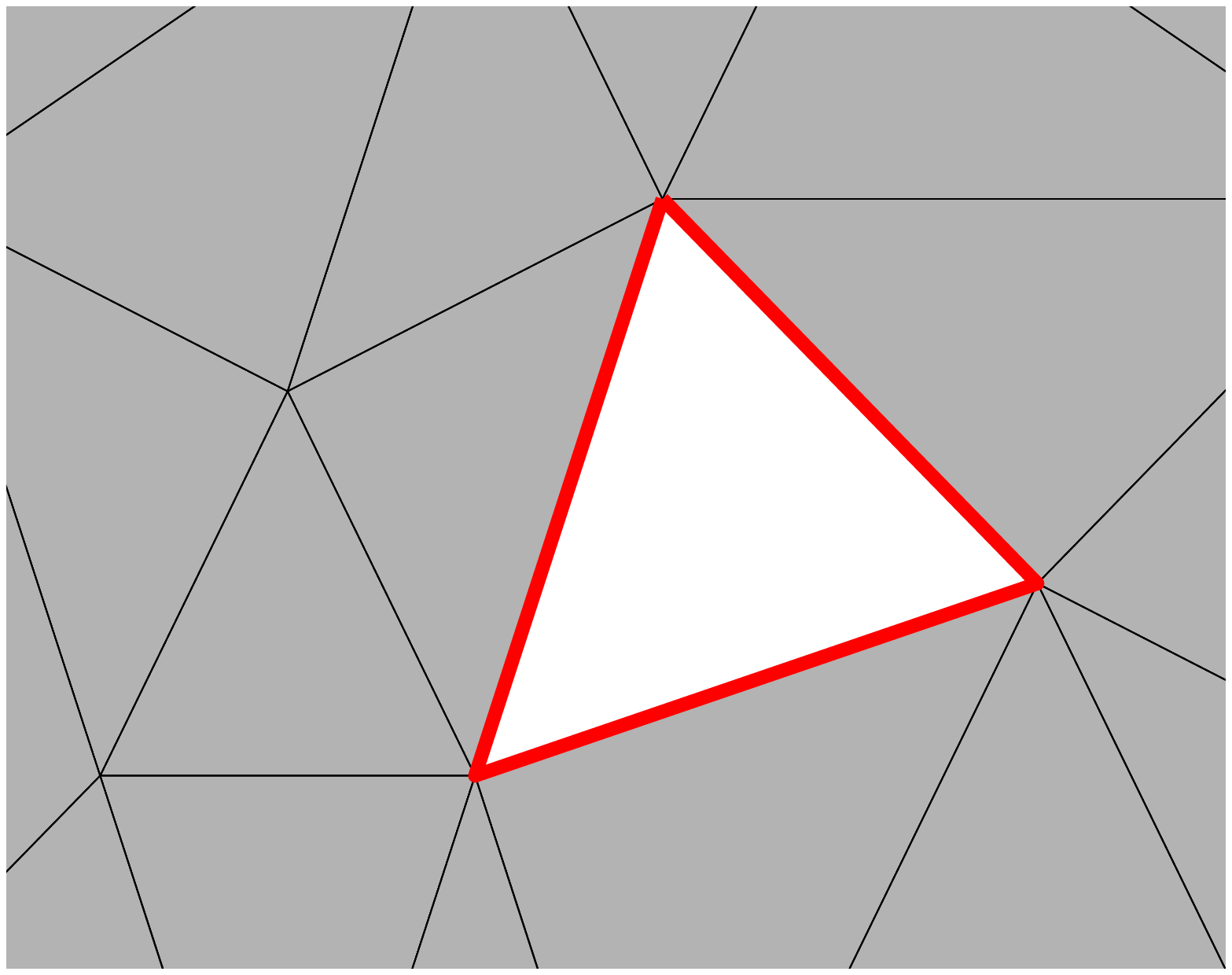} \\
\multicolumn{1}{l}{\textbf{Death by Addition}} & & \\
\includegraphics[scale=0.25]{TriangleRemovedBoundary} & \includegraphics[scale=0.25]{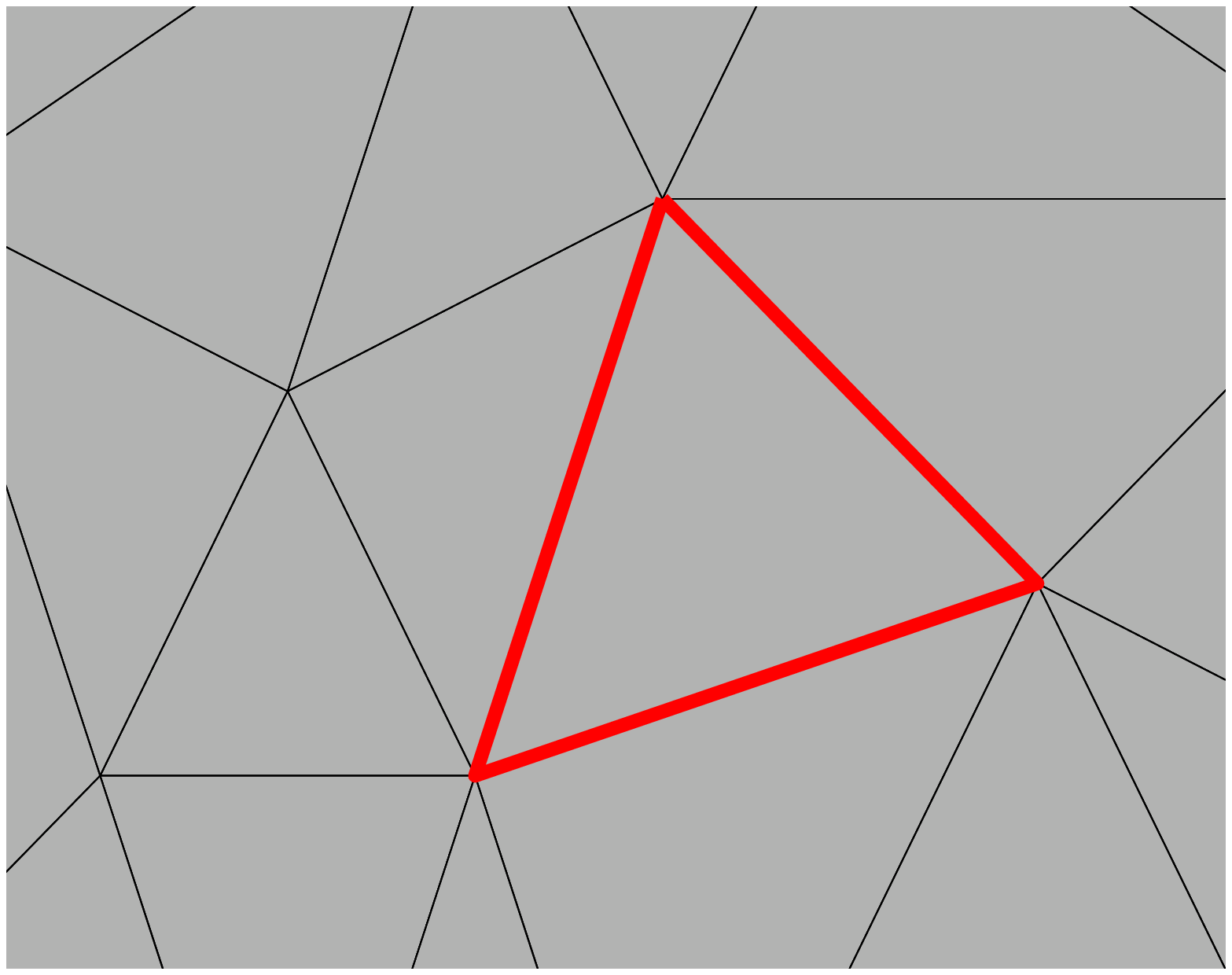} & \includegraphics[scale=0.25]{TrianglePresent} \\
\multicolumn{1}{l}{\textbf{Death by Removal}} & & \\
\includegraphics[scale=0.25]{EdgePresentCycle1} & \includegraphics[scale=0.25]{EdgeRemoved} & \includegraphics[scale=0.25]{EdgeRemoved} \\
\end{tabular}
\end{center}
\caption{Simplicial complexes which illustrate the four cases where the first homology changes, and the corresponding changes to the representative cycles. \label{CasesFigure}}
\end{figure}

Consider the case where the birth is in first homology. If a hole is formed by the removal of a 2-simplex, then there is a unique homology class (the one surrounding the hole) which is born. This homology class also the unique homology class in $\ker(g_i)$ (i.e.: the only homology class that is nontrivial in $K_{i+1}$ but trivial in $K_i$). On the other hand, if a hole is formed by the addition of an edge, there are many choices for which homology class is being born, with no choice being canonical. For example, if a hole is split into two by the addition of an edge, then which of them is the `new' hole? See the first two rows of Figure \ref{CasesFigure} for an illustration of these cases.

Our approach then, is to maintain a basis for the homology at each time point, making the canonical choice of homology class whenever available, and tracking that choice through the sequence of complexes as best as possible. Our method for implementing this, along with the specific basis we maintain and its relation to the zigzag algorithm, is detailed in Section \ref{Algorithm}.

\subsection{Algorithm}\label{Algorithm}

As mentioned in Section \ref{RightFiltration}, a zigzag module $\mathbb{V}$ (Equation \ref{zzModule}) has unique interval decomposition $\Pers(\mathbb{V}) = \{[b_j,d_j] \mbox{ } | \mbox{ } j=1,\ldots,m \}$, which describes the birth and death times of homological features in the sequence. This decomposition is determined through the maintenance of a right filtration $\mathcal{W}_i$ (Equation \ref{Wfiltration}) on the space $V_i$, and a birth vector $\mathbf{b}_i$ for $i = 1,\ldots,n$. The zigzag persistence algorithm performs this task by determining whether a birth or a death is occurring for each simplex $\sigma$ being added or removed, and updating the right filtration and birth vector accordingly (and outputting the appropriate birth-death interval whenever a death occurs).

At each stage in our algorithm, we maintain a basis for the homology that attempts to approximate the canonical basis as best as possible. Further, the basis homology classes are compatible with the right filtration $\mathcal{W}_i$, in the sense that the $j^{th}$  basis homology class is an element of the $j^{th}$ quotient space $W_i^j/W_i^{j-1}$. This means that the span of the first $j$ homology classes in the basis is equal to the $j^{th}$ subspace $W_i^j$ in the right filtration $\mathcal{W}_i$. This property is necessary if we wish to interpret our basis homology classes as corresponding to particular intervals in the birth-death decomposition. The intervals are really describing specific quotient spaces in the right filtration that have persisted over the sequence, so our homology classes need be assigned one-to-one to the quotient spaces. The proof that this property is maintained during the algorithm is presented in Section \ref{Algorithm}.

The method we present here is is computed using the regular zigzag persistent homology algorithm, but keeps an explicit record of the homology basis chosen for the right filtration at each time point. The homology basis is stored by choosing a specific representative cycle for each basis homology class. The choice of basis homology classes is not unique, so it is made in a geometrically meaningful way, attempting to approximate the canonical basis. The zigzag persistence algorithm supplies information about whether the addition or removal of simplex $\sigma$ is causing a birth or a death. If it is a birth, $\sigma$ is called a positive simplex, denoted $\sigma^+$, and if it is a death, $\sigma$ is called a negative simplex, denoted $\sigma^-$.

When a birth occurs, we must add a new representative cycle to our list. As mentioned in Section \ref{Partial}, when a birth occurs due to the removal of a simplex $\sigma$, there is a canonical choice available for the new homology class. We choose the boundary $\partial\sigma$ of the removed simplex as the representative cycle for this homology class, since it is the shortest cycle surrounding the new hole. When the birth occurs due to the addition of a simplex $\sigma$, there is no canonical choice for which is the `new' homology class, but any cycle containing $\sigma$ will have its homology class in $\coker(f_i)$. As a heuristic, we choose the shortest cycle containing $\sigma$ as the new representative cycle.

When a death occurs, we must remove a representative cycle from our list. Analogous to the two ways in which a birth can occur, a death occurs when either the forward map $V_i \overset{f_i}{\longrightarrow} V_{i+1}$ has nonzero kernel, or the backward map $V_i \overset{g_i}{\longleftarrow} V_{i+1}$ has nonzero cokernel. Of these two cases, $\ker(f_i) \neq \textbf{0}$ is the only one which indicates the specific homology class $[c] = \ker(f_i)$ that is being killed (becoming trivial). In this case, the death occurs due to the addition of a simplex, and we reduce the matrix storing the representative cycles with respect to the boundary matrix $\partial$, and remove the cycle which becomes trivial. If the death occurs due to the removal of a simplex $\sigma$, then the first representative cycle containing $\sigma$ is removed, and a change of basis is performed to remove $\sigma$ from any remaining representative cycles. This is done in the same way as the change of basis operation in the regular zigzag persistence algorithm.

The four cases are illustrated in Figure \ref{CasesFigure}. It is worth noting the distinction between the two types of deaths. In a death by addition, the representative cycle is still present in the complex, but becomes homologous to zero. In a death by removal, the representative cycle whose homology class is getting killed is no longer present in the complex, and it is possible that other representative cycles are also no longer present in the complex, so another representative cycle which is from the same quotient space must be chosen.

We store the representative cycles for time $i$ in the matrix $W_i$, which is retained for all time points. The algorithm is summarized below.

\boxed{
\begin{array}{l}
\% Notation \\
w_i^l = \textrm{ column } l \textrm{ of } W_i \\
w_i^j[\sigma] = \mbox{ coefficient of } \sigma \mbox{ in } w_i^j \\
\partial_d = \mbox{ boundary matrix } \\
\vspace{1mm} \\
\% Initialize \\
W_0 = n \times 0 \textrm{ matrix } \\
b_0 = \textrm{ empty vector } \\
\vspace{1mm} \\
\% Perform \mbox{ } updates \\
\textrm{for } i = 1 \textrm{ to } n \\
\indent \textrm{if } K_i = K_{i-1} - \{\sigma\}  \hspace{33mm}                                                          \%simplex \mbox{ } removal\\
\indent \indent \textrm{if } \sigma^+  \hspace{48mm}                                                                      \% birth \\
\indent \indent \indent W_i = [\partial\sigma_i \mbox{ } W_{i-1} ] \hspace{20mm}                                                 \% prepend \mbox{ } \partial\sigma \\
\indent \indent \indent b_i = [b_{i-1} \mbox{ } i] \\
\indent \indent \textrm{else if } \sigma^- \hspace{40mm}                                                                  \% death \\
\indent \indent \indent l = \textrm{ index of first nonzero entry in } r_{\sigma} \\
\indent \indent \indent c_l = r_{\sigma}(l)\mbox{, the coefficient of } \sigma \mbox{ in } w_l \\
\indent \indent \indent \textrm{for } j=1 \mbox{ to (\# columns of } W_{i-1}) \hspace{20mm}                                     \% change \mbox{ } of \mbox{ } basis \\
\indent \indent \indent \indent c_j = r_{\sigma}(j) \mbox{, the coefficient of } \sigma \mbox{ in } w_j \\
\indent \indent \indent \indent w_j = w_j - \frac{c_j}{c}w_l \\
\indent \indent \indent \textrm{end} \\
\indent \indent \indent W_i = W_{i-1} \mbox{ with column } l \mbox{ and row } r_{\sigma} \mbox{ removed } \\
\indent \indent \indent b_i = b_{i-1} \mbox{ with entry } l \mbox{ removed } \\
\indent \indent \textrm{end} \\
\indent \textrm{end} \\
\indent \textrm{if } K_{i+1} = K_i \cup \sigma \hspace{33mm}                                                              \%simplex \mbox{ } addition\\
\indent \indent \textrm{if } \sigma^+  \hspace{48mm}                                                                      \% birth \\
\indent \indent \indent W_i = [W_{i-1} \mbox{ } (Cu - \sigma)] \hspace{10mm}                                                 \% append \mbox{ } Cu - \sigma \\
\indent \indent \indent b_i = [b_{i-1} \mbox{ } i] \\
\indent \indent \textrm{else if } \sigma^- \hspace{40mm}                                                                  \% death \\
\indent \indent \indent l = \mbox{ index for col of } W_{i-1} \mbox{ trivial when }[\partial_d \mbox{ } W_{i-1}] \mbox{ reduced } \\
\indent \indent \indent W_i = W_{i+1} \mbox{ with column } w_l \mbox{ removed } \\
\indent \indent \textrm{end} \\
\indent \textrm{end} \\
\textrm{end} \\
\end{array}}

\section{Correctness}\label{Correctness}

The adaptive representative cycles that we track using the algorithm described in Section \ref{Algorithm} are stored as column vectors $w_i^k$ in a matrix $\mathbf{W}_i$
\[ \mathbf{W}_i = [ w_i^1 \mbox{ } w_i^2 \mbox{ } \ldots \mbox{ } w_i^{\beta(K_i)}] \]
In this section we show that the homology classes represented by the cycles $w_i^k$ form a basis for $V_i$, and further that their order in $\mathbf{W}_i$ corresponds to the order of the right filtration $\mathcal{W}_i$ (Equation \ref{Wfiltration}). In other words, the span of the homology classes of the first $j$ representative cycles is equal to the $j^{th}$ space $W_i^k$ in the filtration $\mathcal{W}_i$ of $V_i$.

i.e.:

\begin{equation}
\spn \{[w_i^k]\}_{k=1}^j = W_i^j
\end{equation}
for $j = 1,\ldots,\beta(K_i)$.

We prove this by induction on $i$, beginning with the base case of a single vector space $\mathbb{V} = V_1$, which results from a simplicial complex of one vertex $K_1 = \sigma$. This yields

\begin{eqnarray}
\mathcal{W}_1 & = & (V_1) \nonumber \\
\mathbf{W}_1 & = & [w_1^1] \nonumber
\end{eqnarray}
where $w_1^1 = [1]$ is the column vector of length 1 representing the cycle consisting of the vertex $\sigma$. The homology class $[w_1^1]$ spans the one-dimensional homology space $W_1^1 = V_1$.

In the inductive step, we assume that for

\begin{eqnarray}
\mathcal{W}_i & = & (W_i^1, \ldots, W_i^{\beta(K_i)}) \\
\mathbf{W}_i & = & [ w_i^1 \mbox{ } w_i^2 \mbox{ } \ldots \mbox{ } w_i^{\beta(K_i)}]
\end{eqnarray}
we have (for $j = 1,\ldots,\beta(K_i)$)
\[ \spn \{[w_i^k]\}_{k=1}^j = W_i^j \]

We will show then that (for $j = 1,\ldots,\beta(K_{i+1})$)
\begin{equation}
\spn \{[w_{i+1}^k]\}_{k=1}^j = W_{i+1}^j
\end{equation}
for all four of the cases described in the algorithm (Section \ref{Algorithm}). In all cases we use $\sigma$ to denote the $d$-simplex being added or removed, and the updates are performed on the representative cycles and right filtration of the appropriate dimension (see Table \ref{DimTable}).

\textbf{1. Birth by addition.} The map $V_i \overset{f_i}{\longrightarrow} V_{i+1}$ has $\coker(f_i) \neq \textbf{0}$, and the new right filtration is
\[ \mathcal{W}_{i+1} = (f_i(W_i^0), f_i(W_i^1), \ldots, f_i(W_i^{\beta(K_i)}), V_{i+1}) \]
where $V_{i+1}/f_i(W_i^{\beta(K_i)}) = \coker(f_i)$.

The new list of representative cycles is
\[ \mathbf{W}_{i+1} = [\mathbf{W}_i^{\sigma^+} \mbox{ } w_{new}] \]
where $\mathbf{W}_i^{\sigma^+}$ is the matrix $\mathbf{W}_i$ with an additional row of zeros added, corresponding to simplex $\sigma$ (so the cycles are now written in terms of simplices of $K_{i+1}$ instead of simplices of $K_i$), and $w_{new}$ is a cycle in $K_{i+1}$ containing $\sigma$. There is no canonical choice for which cycle containing $\sigma$ should be chosen, and our proof holds regardless of the choice. As mentioned in Section \ref{Algorithm}, we make this choice based on shortest hop length.

Since $w_{i+1}^k = w_i^k$ as chains (with the appropriate row for $\sigma$ added containing a 0 coefficient), we get $[w_{i+1}^k] = f_i([w_i^k])$, for $k = 1,\ldots,\beta(K_i)$  because $f_i$ is the map induced by inclusion. Therefore
\begin{eqnarray*}
W_{i+1}^j & = & f_i(W_i^j) \\
& = & f_i\left(\spn \{[w_i^k]\}_{k=1}^j \right) \\
& = & \spn \{f_i([w_i^k])\}_{k=1}^j \\
& = & \spn \{[w_{i+1}^k]\}_{k=1}^j
\end{eqnarray*}
for $j=1,\ldots,\beta(K_i)$.

Finally, we must show that $[w_{new}]$ is nontrivial and in $\coker(f_i)$, and therefore linearly independent from $\{[w_i^j]\}_{j=1}^{\beta(K_i)}$, so they together span the $\beta(K_1)+1 = \beta(K_{i+1})$-dimensional vector space $V_{i+1} = W_{i+1}^{\beta(K_{i+1})}$. First, note that having a nonzero coefficient for $\sigma$ in $w_{new}$: that $[w_{new}] \neq 0$; and that any cycle $c$ in the same homology class $[w_{new}]$ will also have a nonzero coefficient for $\sigma$. These are due to the fact that $\sigma$ is not contained in the boundary of any other simplex, and the difference between homologous cycles must be written as a linear combination of boundaries (therefore the coefficient for $\sigma$ is zero in the difference $c - w_{new}$, but is nonzero in $w_{new}$, so must also be nonzero in $c$). Moreover, note that $[w_{new}] \not\subseteq \im(f_i)$, since any homology class in $\im(f_i)$ must have a representative cycle in the image under inclusion $i(K_i) \subset K_{i+1}$, and all cycles in $[w_{new}]$ contain $\sigma \not\in i(K_i)$. Therefore, we have
\begin{eqnarray*}
W_{i+1}^{\beta(K_{i+1})} & = & V_i \\
& = & \im(f_i) \oplus \coker(f_i) \\
& = & \left(\spn \{[w_{i+1}^k]\}_{k=1}^{\beta(K_i)} \right) \oplus [w_{new}] \\
& = & \spn \{[w_{i+1}^k]\}_{k=1}^{\beta(K_{i+1})}.
\end{eqnarray*}
as desired.

\textbf{2. Birth by removal.} The map $V_i \overset{g_i}{\longleftarrow} V_{i+1}$ has $\ker(g_i) \neq \textbf{0}$, and the new right filtration is
\[ \mathcal{W}_{i+1} = (\ker(g_i), g_i^{-1}(W_i^1), g_i^{-1}(W_i^2), \ldots, g_i^{-1}(W_i^{\beta(K_i)})). \]
This is because in the full right filtration
\begin{eqnarray}
\mathcal{R}_{i+1} & = & (\mathbf{0}, g_i^{-1}(R_i^0), g_i^{-1}(R_i^1), \ldots, g_i^{-1}(R_i^i)) \nonumber
\end{eqnarray}
if $R_i^j/ R_i^{j-1}$ was nontrivial in $\mathcal{R}_i$, then $g_i^{-1}(R_i^j)/ g_i^{-1}(R_i^{j-1})$ will be nontrivial in $\mathcal{R}_{i+1}$ for $j = 1,\ldots,i$. This means that if $W_i^j$ is a subspace in $\mathcal{W}_i$ then $g_i^{-1}(W_i^j)$ is a subspace in $\mathcal{W}_{i+1}$. Also, the new term
\[ g_i^{-1}(R_i^0)/\textbf{0} = g_i^{-1}(\textbf{0})/\textbf{0} = \ker(g_i)/\textbf{0} = \ker(g_i) \]
is nontrivial, and yields the first term $\ker(g_i)$ in $\mathcal{W}_{i+1}$.

The new list of representative cycles is
\[ \mathbf{W}_{i+1} = [\partial\sigma \mbox{ } \mathbf{W}_i] \]
where $\partial\sigma$ are the simplices that make up the boundary of $\sigma$, but considered in $K_{i+1}$, instead of $K_i$.

First we note that $\spn\{[\partial\sigma]\} = \ker(g_i)$. This is because $i(\partial\sigma)$ is the boundary of simplex $\sigma$ in $K_i$ and hence homologous to zero, thus
\[ g_i([\partial\sigma]_{i+1}) = [\partial\sigma]_i = \textbf{0} \]
which means $[\partial\sigma] \subseteq \ker(g_i)$. The cycle $\partial\sigma$ is also nontrivial in $K_{i+1}$, because if there exists a $d$-chain $c$ in $K_{i+1}$ that had $\partial\sigma$ as its boundary, then in $K_i$ the union of $\sigma$ with $i(c)$ in $K_i$ would form a $d$-cycle, and the removal of $\sigma$ would result in the death of that $d$-cycle, instead of the birth of a $(d-1)$-cycle, which is a contradiction. Therefore, $[\partial\sigma]$ spans a one-dimensional subspace of the one-dimensional space $\ker(g_i)$, so $\spn\{[\partial\sigma]\} = \ker(g_i)$.

Now we show that $W_{i+1}^j = \spn \{[w_{i+1}^k]\}_{k=1}^j$ for $j = 1,\ldots, {\beta(K_{i+1})}$. First note the index change, so
\[ w_{i+1}^{k+1} = w_i^k \]
for $k = 1,\ldots,\beta(K_i)$. Consider the representative cycle $w_i^k$, and another cycle $c$ which is homologous to $w_i^k$ in $K_i$. Since $c$ and $w_i^k$ are both $(d-1)$-cycles, they are also present in $K_{i+1}$. Then $[c]_i = [w_i^k]_i$ implies $[c]_{i+1} = [w_i^k]_{i+1} + a[\partial\sigma]_{i+1}$, where $a = $ 0 or 1. Therefore
\[ g_i^{-1}([w_i^k]) = [w_{i+1}^{k+1}]\oplus[\partial\sigma]. \]
So
\begin{eqnarray*}
W_{i+1}^j & = & g_i^{-1}(W_i^{j-1}) \\
& = & \spn \{g_i^{-1}[w_i^k]\}_{k=1}^{j-1} \\
& = & \spn \left\{[w_{i+1}^{k+1}]\oplus[\partial\sigma]\right\}_{k=1}^{j-1}
\end{eqnarray*}
for $j = 2,\ldots,\beta(K_{i+1})$. Combining this with
\[ W_{i+1}^1 = \ker(g_i) = [\partial\sigma] = [w_{i+1}^1] \]
we obtain
\[ W_{i+1}^j = \spn \{[w_{i+1}^k]\}_{k=1}^j \]
for $j = 1\ldots,\beta(K_{i+1})$, as desired.

\textbf{3. Death by addition.} For the map $f_i: V_i \to V_{i+1}$ we get $\ker(f_i) = [\partial\sigma]$ with a similar proof to that of case \textbf{2} above.

Since $\ker(f_i) \neq \textbf{0}$, we have $\coker(f_i) = \textbf{0}$, so $V_{i+1}/f_i(V_i) = \textbf{0}$. Also, there exists an index $l \in \{1,\ldots,\beta(K_i) \}$ such that $[\partial\sigma] \in W_i^l$, but $[\partial\sigma] \not\in W_i^{l-1}$ (using the convention $W_i^0 = \textbf{0}$), so
\[ f_i(W_i^l/ W_i^{l-1}) = \textbf{0} \]
This gives
\[ \mathcal{W}_{i+1} = (f_i(W_i^1), \ldots, f_i(W_i^{l-1}), f_i(W_i^{l+1}), \ldots, f_i(W_i^{\beta(K_i)})) \]
so we have
\begin{align}\label{Wcase3}
W_{i+1}^j = \left\{ \begin{array}{ll}
                f_i(W_i^j) & \mbox{ if $j < l$};\\
                f_i(W_i^{j+1}) & \mbox{ if $j \geq l$}.\end{array} \right.
\end{align}

Considering now the representative cycles, we need to determine the index $l$. Since the elements $\{[w_i^k]\}_{k=1}^{\beta(K_i)}$ form a basis for $V_i$, we can write uniquely
\begin{equation}\label{sum}
[\partial\sigma] = \sum_{k=1}^{\beta(K_i)}\alpha_k [w_i^k].
\end{equation}
Then $[\partial\sigma] \in \spn \{[w_i^k]\}_{k=1}^l = W_i^l$, but $[\partial\sigma] \not\in \spn \{[w_i^k]\}_{k=1}^{l-1} = W_i^{l-1}$ implies that $\alpha_l$ is the last nonzero coefficient in this sum. We now define
\[ w_{remove} = w_i^l \]
and obtain
\[ \mathbf{W}_{i+1} =  [w_i^1 \mbox{ } \ldots \mbox{ } w_i^{l-1} \mbox{ } w_i^{l+1} \mbox{ } \ldots \mbox{ } w_i^{\beta(K_i)}] \]
noting that all of the simplices in the $(d-1)$-cycles $w_i^k$ are present in $K_{i+1}$. Therefore the corresponding homology classes are related by
\[ [w_{i+1}^j] = \left\{ \begin{array}{ll}
                f_i([w_i^j]) & \mbox{ if $j < l$};\\
                f_i([w_i^{j+1}]) & \mbox{ if $j \geq l$}.\end{array} \right. \]
for $j = 1,\ldots,\beta(K_{i+1})$, since $f_i$ is the map induced by inclusion. This, together with Equation \ref{Wcase3} yields
\[ W_{i+1}^j = \spn \{[w_{i+1}^k]\}_{k=1}^j \]
for $j = 1,\ldots,\beta(K_{i+1})$, as desired.

Note that the index $l$ indicating the last nonzero coefficient in Equation \ref{sum} also determines the birth-death interval: $[\mathbf{b}_i^W[l],i]$.

\textbf{4. Death by removal.} The map $V_i \overset{g_i}{\longleftarrow} V_{i+1}$ has $\coker(g_i) \neq \textbf{0}$. There exists an index $l$ such that $W_i^j \subseteq \im(g_i)$, for all $j < l$, but $W_i^l \not\subseteq \im(g_i)$. Then
\[ g_i^{-1}(W_i^l/W_i^{l-1}) = \textbf{0} \]
so
\[ \mathcal{W}_{i+1} = (g_i^{-1}(W_i^1), \ldots, g_i^{-1}(W_i^{l-1}), g_i^{-1}(W_i^{l+1}), \ldots, g_i^{-1}(W_i^{\beta(K_i)})). \]
We note that the image of this in $V_i$ is
\begin{eqnarray}\label{gWi}
g_i(\mathcal{W}_{i+1}) & = & \mathcal{W}_i /_{\coker(g_i)}  \\
& = & (W_i^1,\ldots, W_i^{l-1}, W_i^{l+1}/_{\coker(g_i)}, \ldots, W_i^{\beta(K_i)}/_{\coker(g_i)} ) \nonumber
\end{eqnarray}

Considering now the representative cycles, $l$ is the index of the first representative cycle $w_i^l$ which contains $\sigma$. To see that this is the same index $l$ as described above, note that since $w_i^k$ doesn't contain $\sigma$ for $k<l$, we have $[w_i^k] \in \im(g_i)$, and $\spn \{[w_i^k]\}_{k=1}^j = W_i^j \subseteq \im(g_i)$, for all $j < l$, but $\spn \{[w_i^k]\}_{k=1}^l = W_i^l \not\subseteq \im(g_i)$.

Denoting the coefficient for $\sigma$ in representative cycle $w_i^k$ by $w_i^k[\sigma]$, we consider another set of representative cycles in $K_i$
\[ \hat{w}_i^k = w_i^k - \frac{w_i^k[\sigma]}{w_i^l[\sigma]}w_i^l. \]
By definition, $\sigma$ is not present in any $\hat{w}_i^k$, so we are able to define
\[ w_{i+1}^k = \left\{ \begin{array}{ll}
                \hat{w}_i^k & \mbox{ if $k < l$};\\
                \hat{w}_i^{k+1} & \mbox{ if $k \geq l$}.\end{array} \right. \]
to be our representative cycles in $K_{i+1}$, with the row corresponding to $\sigma$ removed. Then
\begin{eqnarray*}
\mathbf{W}_{i+1} & = & [\hat{w}_i^1 \mbox{ } \ldots \mbox{ } \hat{w}_i^{l-1} \mbox{ } \hat{w}_i^{l+1} \mbox{ } \ldots \mbox{ } \hat{w}_i^{\beta(K_i)}] \\
& = & [w_{i+1}^1 \mbox{ } \ldots \mbox{ } w_{i+1}^{l-1} \mbox{ } w_{i+1}^l \mbox{ } \ldots \mbox{ } w_{i+1}^{\beta(K_{i+1})}]
\end{eqnarray*}

We proceed by showing that the $\hat{w}_i^k$ completely determine the quotiented filtration $\mathcal{W}_i/_{\coker(g_i)}$ in Equation \ref{gWi}, in the sense that
\begin{equation}\label{case4lemma}
W_i^j /_{\coker(g_i)} = \spn \{[\hat{w}_i^k]\}_{k=1}^j
\end{equation}
for $j = 1,\ldots,\beta(K_{i+1}).$

To show that Equation \ref{case4lemma} holds, we show it separately for $j<l$, $j=l$, and $j>l$. For the first case, note that when $w_i^k$ does not contain $\sigma$, we have $\hat{w}_i^k = w_i^k$. In particular, for $k<l$ we have $\hat{w}_i^k = w_i^k$, therefore
\[ W_i^j /_{\coker(g_i)} = W_i^j  = \spn \{[w_i^k]\}_{k=1}^j = \spn \{[\hat{w}_i^k]\}_{k=1}^j  \]
for $j = 1,\ldots, l-1$.

By assumption $W_i^l \not\subseteq \im(g_i)$, but $W_i^{l-1} \subseteq \im(g_i)$, so
\[ W_i^l/_{\coker(g_i)} = W_i^{l-1} = \spn \{[\hat{w}_i^k]\}_{k=1}^{l-1} = \spn \{[\hat{w}_i^k]\}_{k=1}^l \]
since $\hat{w}_i^l = \vec{0}$.

For $j > l$, we first note that the homology elements $\{[\hat{w}_i^k]\}$ are linearly independent for $k\in \{1,\ldots,l-1,l+1,\ldots,\beta(K_i) \}$. This is because each $[\hat{w}_i^k]$ is a subset of $[w_i^k] \oplus [w_i^l]$ (but not equal to $[w_i^l]$), and the $\{[w_i^k]\}$ are linearly independent. So $\{[\hat{w}_i^k]\}_{k=1}^j$ span a $(j-1)$-dimensional space when $j>l$ (since $[w_i^l]$ is trivial). Also, because all the $\hat{w}_i^k$ have a zero coefficient for $\sigma$, they are not in the $\coker(g_i)$. So
\[ \spn \{[\hat{w}_i^k]\}_{k=1}^j \subseteq W_i^j /_{\coker(g_i)} \]
Moreover, we note that $W_i^j/_{\coker(g_i)}$ is also a $(j-1)$-dimensional space for $j>l$. So $\spn \{[\hat{w}_i^k]\}_{k=1}^j = W_i^j /_{\coker(g_i)}$.

It now follows that since
\[ g_i([w_{i+1}^k]) = \left\{ \begin{array}{ll}
                [\hat{w}_i^k] & \mbox{ if $k < l$};\\
                \mbox{[} \hat{w}_i^{k+1}] & \mbox{ if $k \geq l$} \end{array} \right. \]
then
\begin{align*}
W_{i+1}^j & = \left\{ \begin{array}{ll}
                g_i^{-1}(W_i^j) = g_i^{-1}(W_i^j/_{\coker(g_i)}) & \mbox{ if $j < l$};\\
                g_i^{-1}(W_i^{j+1}) = g_i^{-1}(W_i^{j+1} /_{\coker(g_i)})& \mbox{ if $j \geq l$}\end{array} \right. \\
          & = \spn \{[w_{i+1}^j]\}_{i+1}^j
\end{align*}
completing the induction.

\section{Applications to dynamic sensor networks}\label{Examples}

In this Section, we illustrate some of the types of results possible when using the representative cycles to `track' coverage holes in dynamic sensor networks. A more detailed description of these applications is available in \cite{Gamble2014b}.

\subsection{Tracking coverage holes}

A typical model for a dynamic sensor network has $n$ nodes distributed randomly uniformly over a region of interest (say the unit square, for simulation purposes). The nodes are then allowed to move independently and randomly according to the same stochastic process, such as a discrete-time Brownian motion. The coverage region and Rips complex for this network are described in Section \ref{Networks}, and we note that the only information necessary to build the Rips complex is the binary adjacency matrix indicating which nodes are within distance $2r$ of each other. The input to the algorithm then, is simply a sequence of adjacency matrices, representing the communication graph of the network at each time point. No information about the locations of the nodes, or distances between them is used. The output is a barcode, along with a representative cycle for each bar at each time point. These representative cycles can be used to track holes in many circumstances.

\begin{figure}[htp]
\begin{center}
\begin{tabular}{ccc}
\includegraphics[scale=0.22]{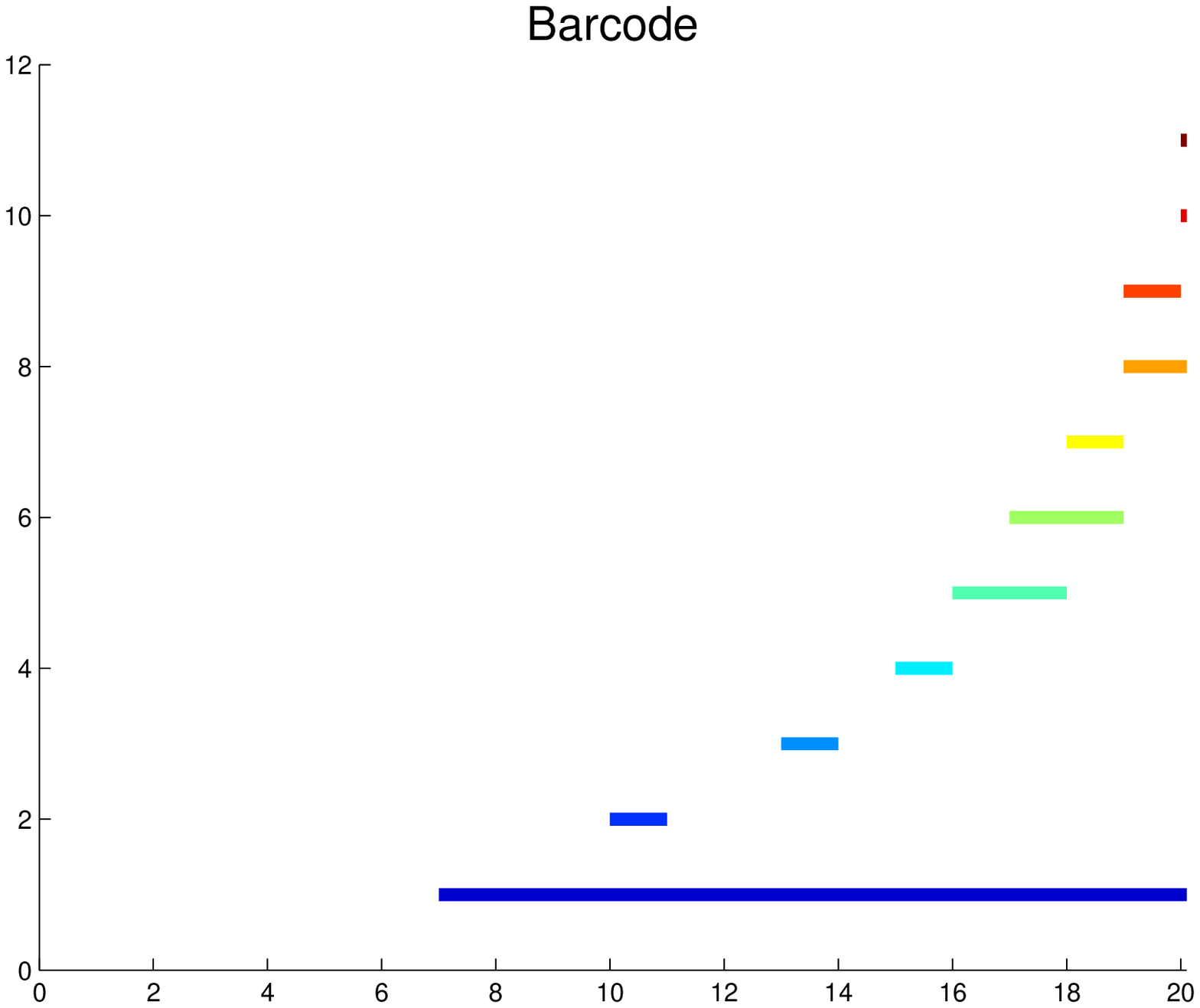} & \includegraphics[scale=0.22]{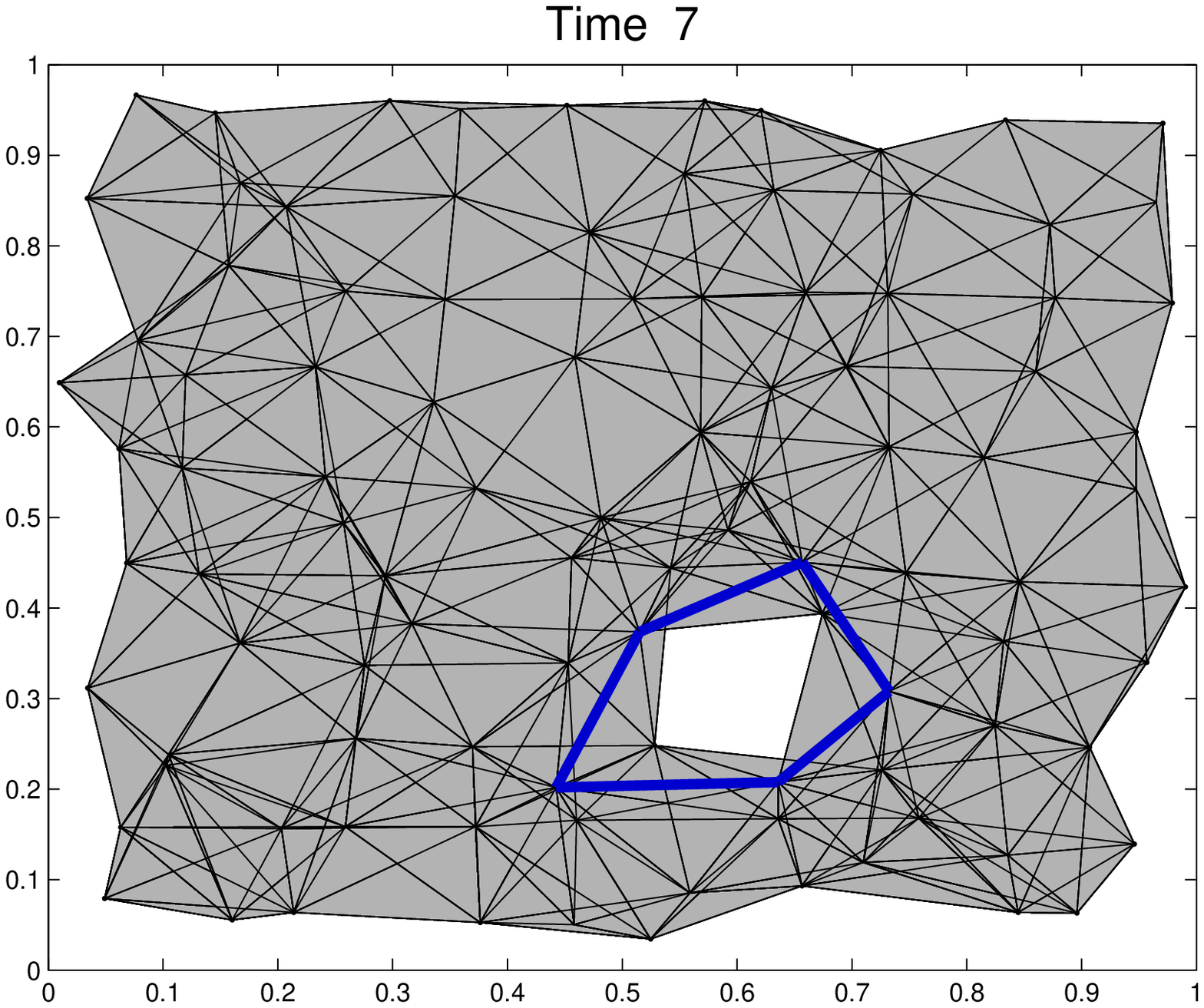} & \includegraphics[scale=0.22]{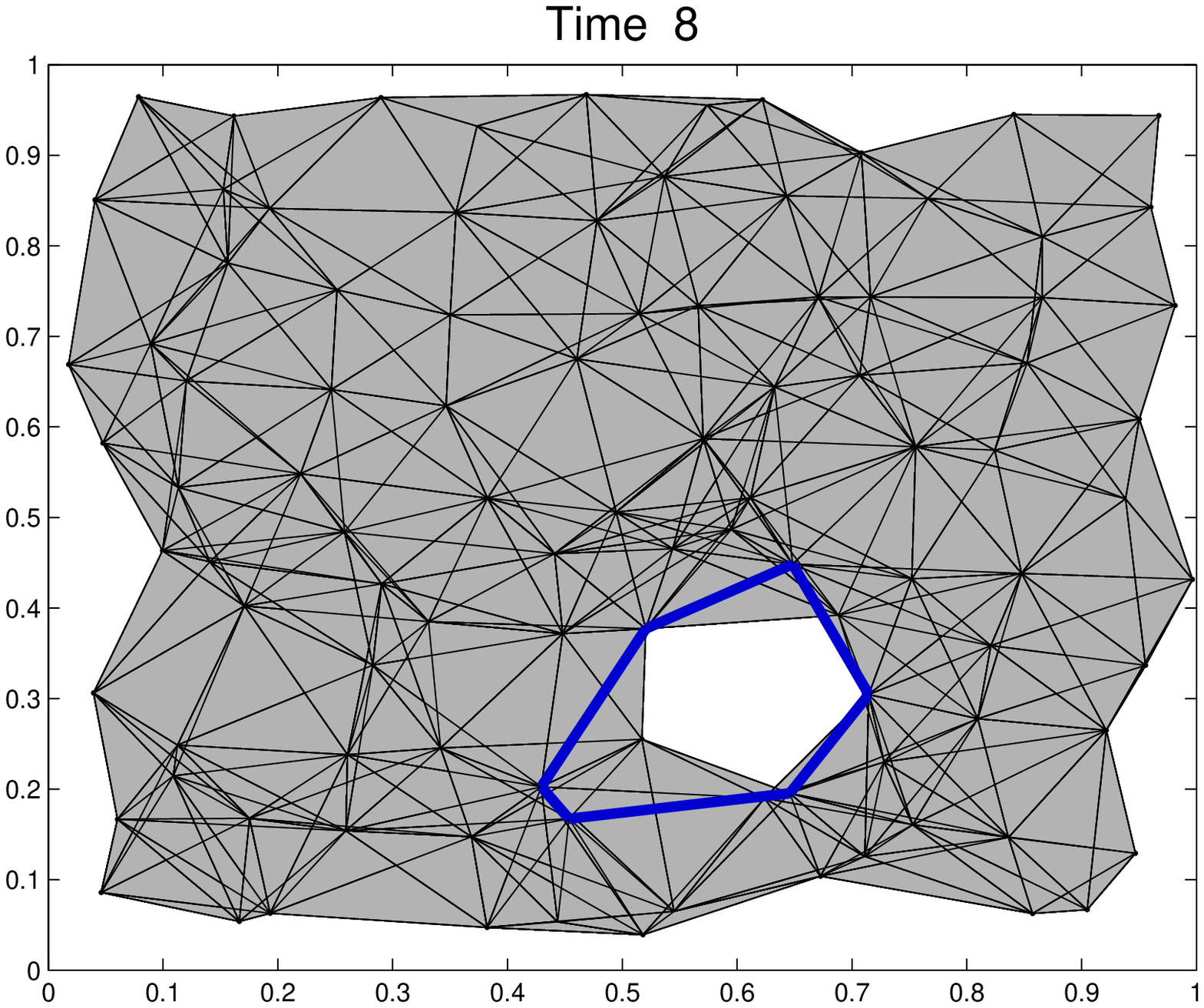} \\
\includegraphics[scale=0.22]{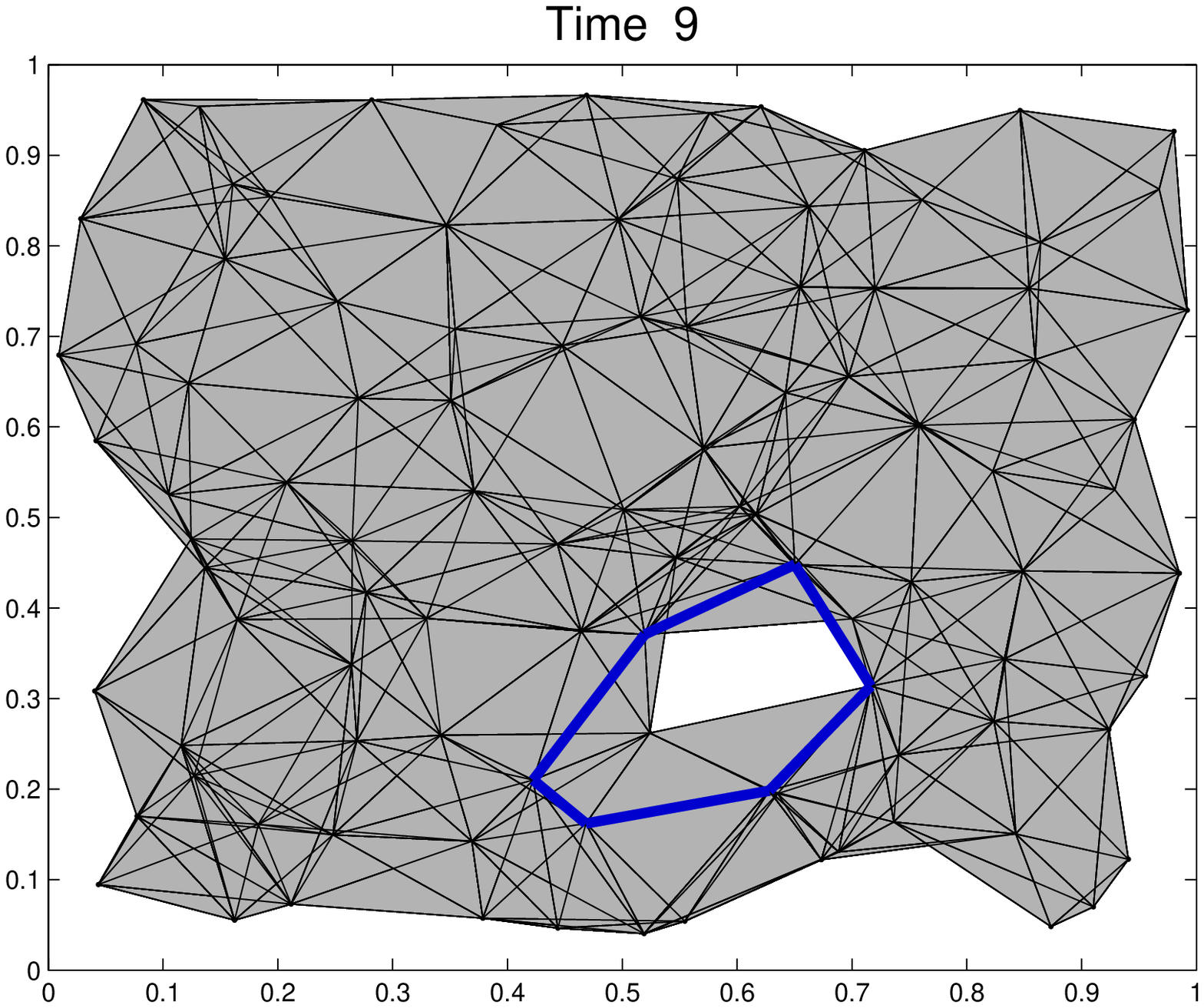} & \includegraphics[scale=0.22]{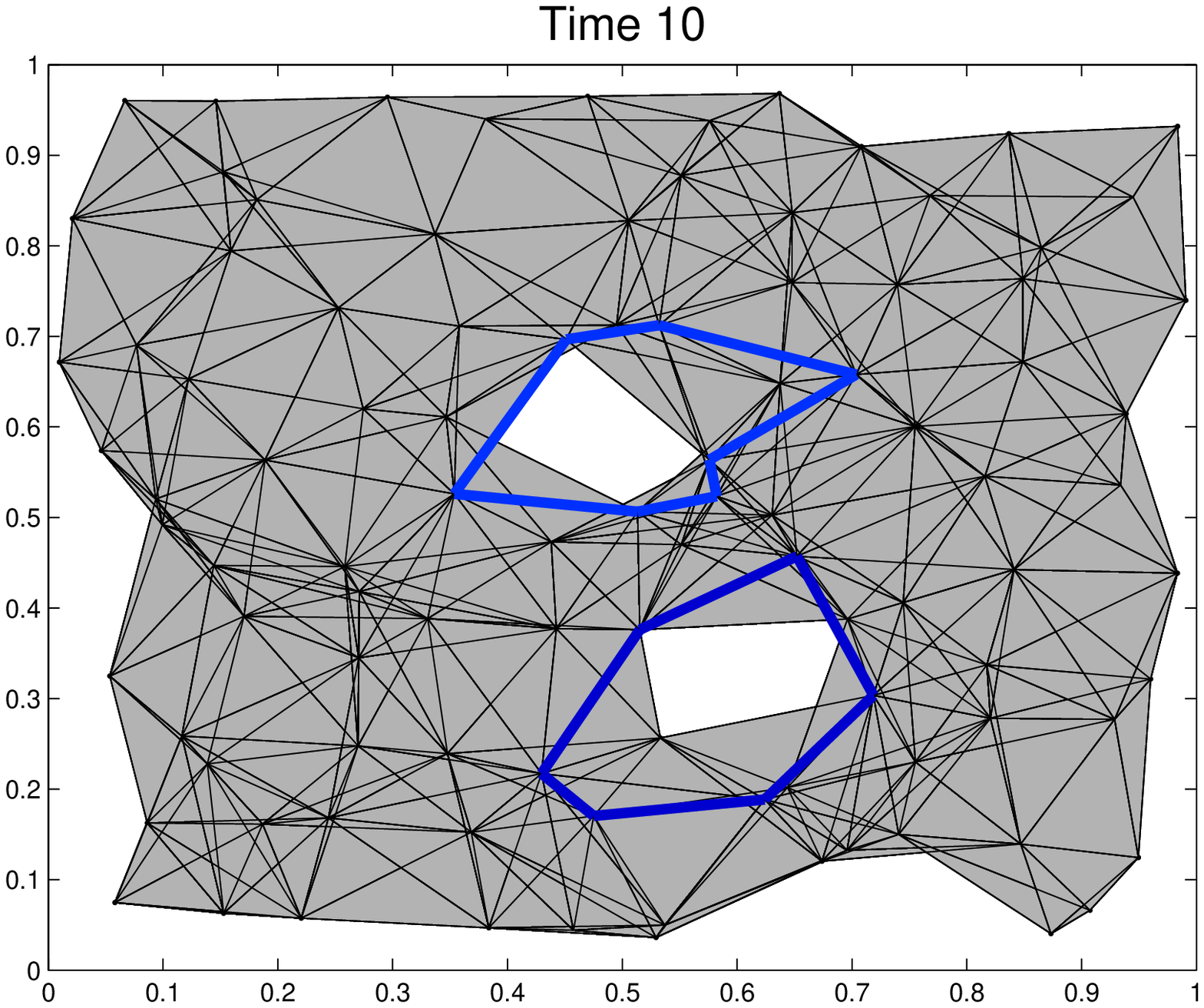} & \includegraphics[scale=0.22]{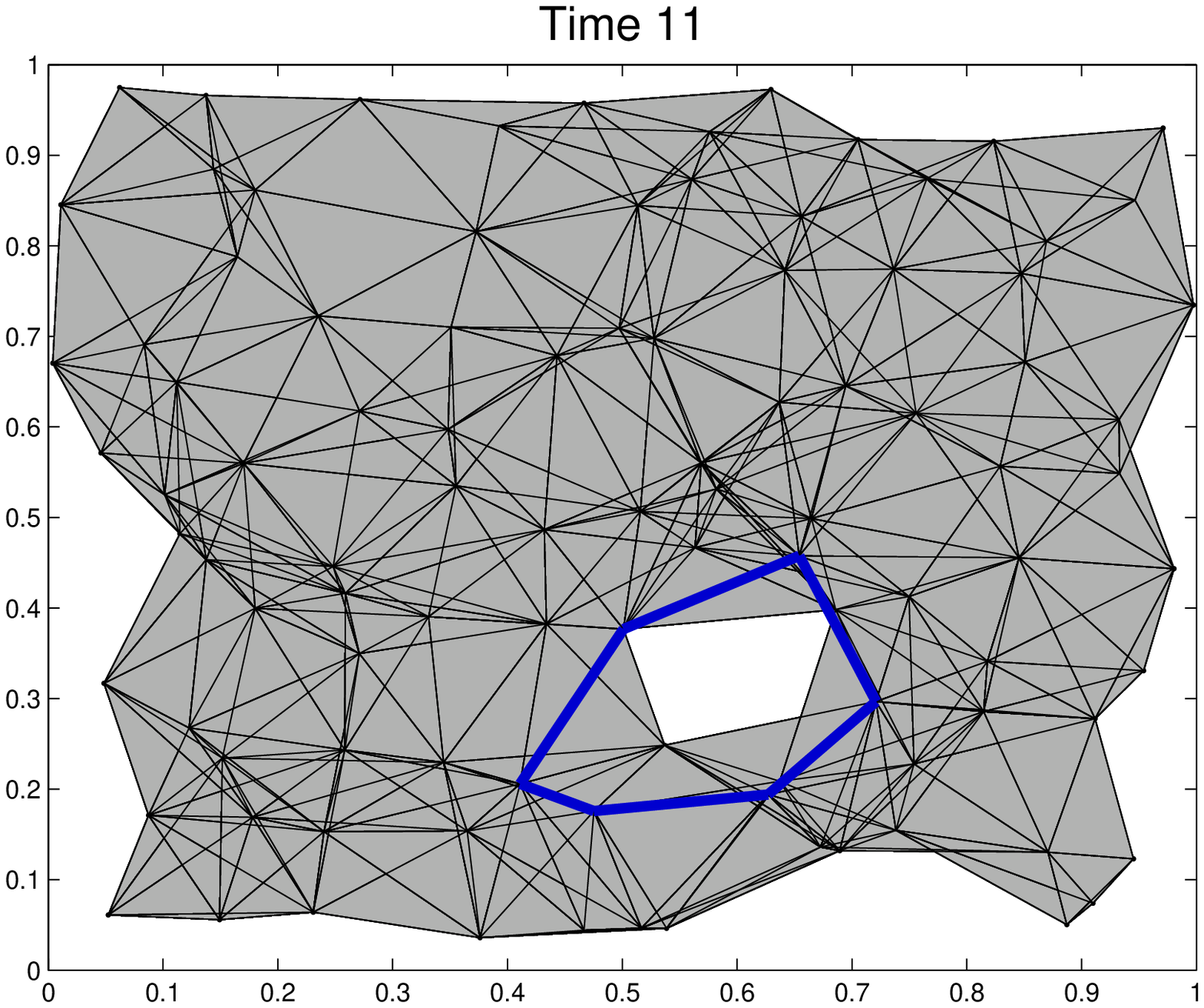}  \\
\includegraphics[scale=0.22]{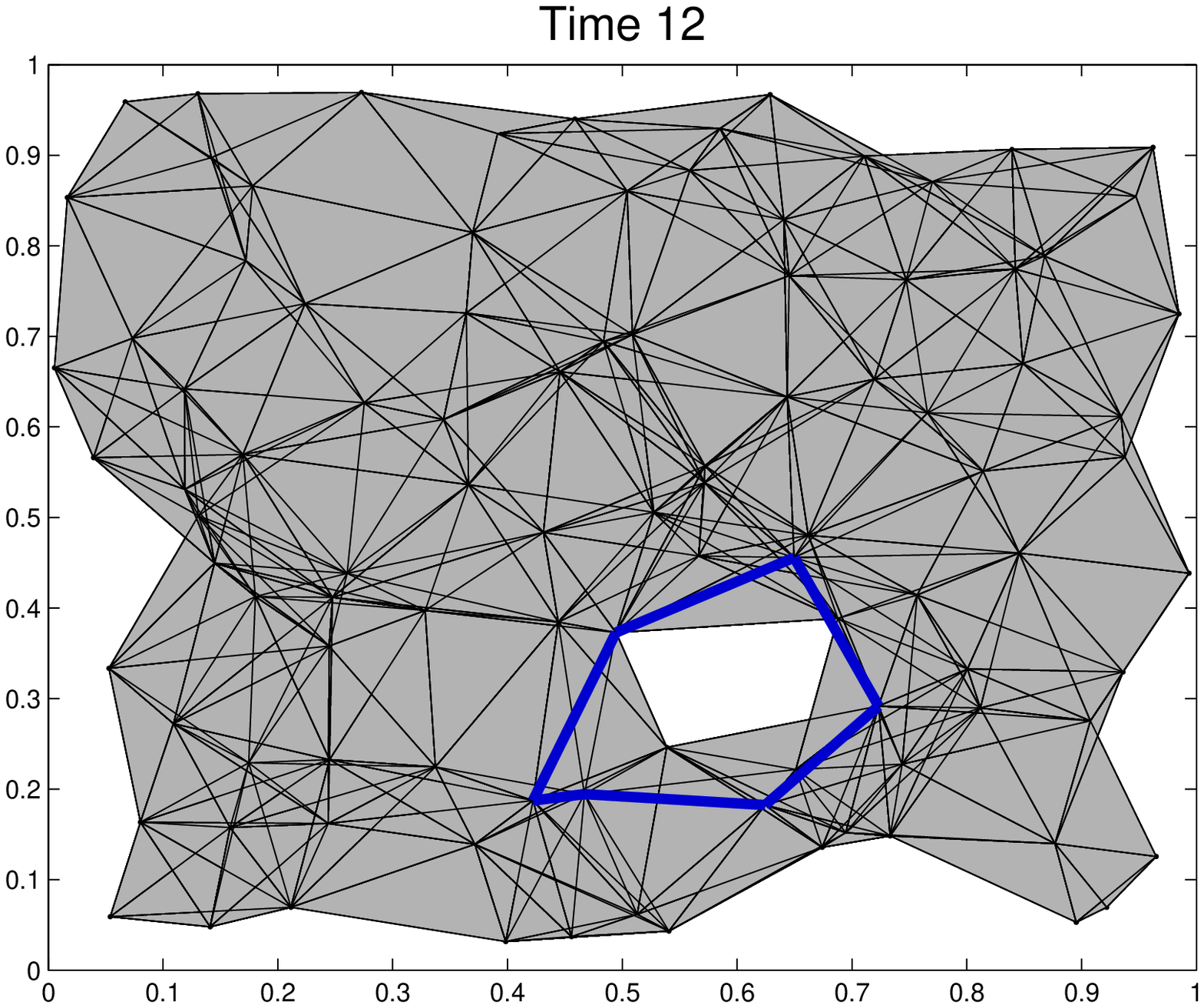} & \includegraphics[scale=0.22]{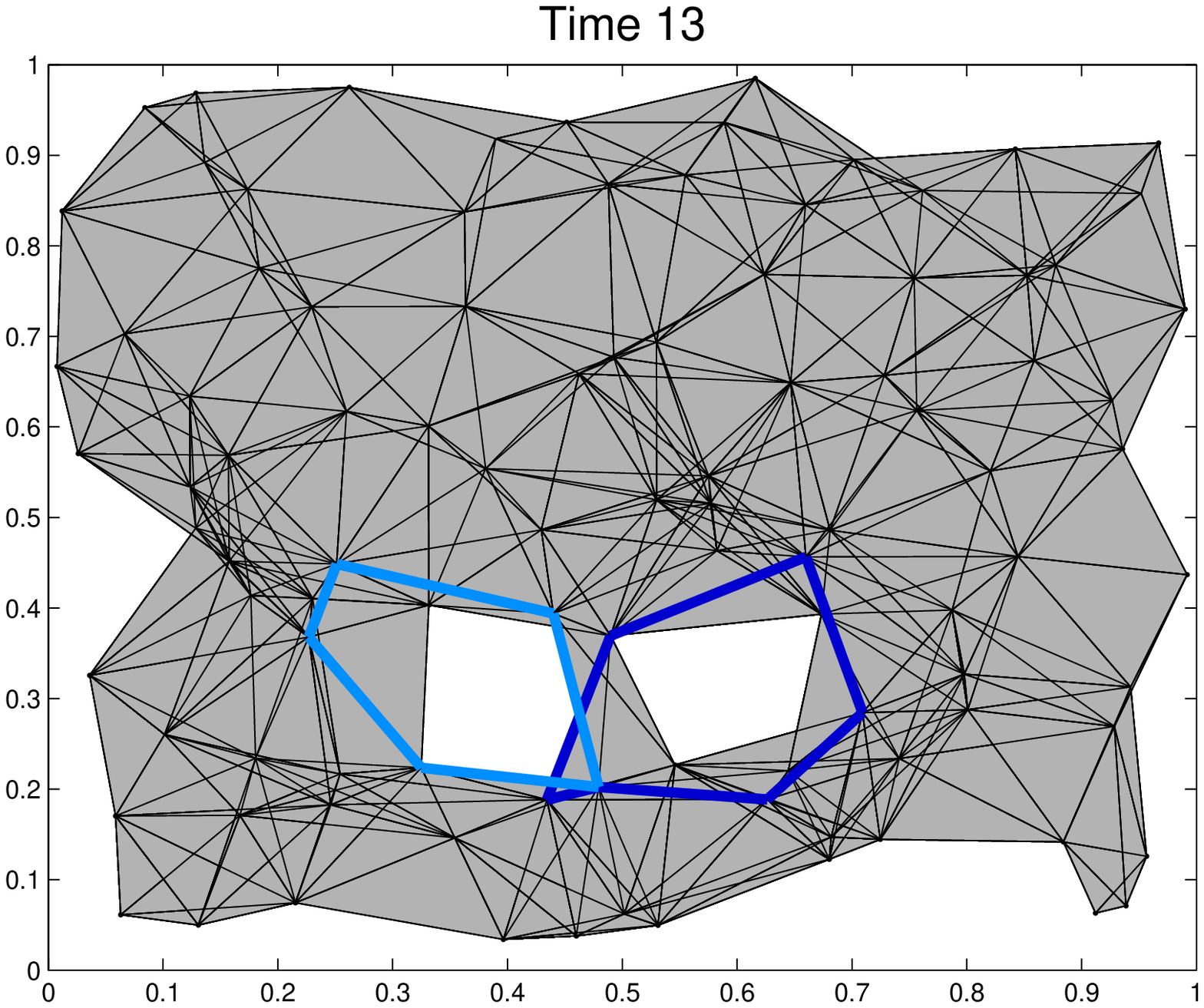} & \includegraphics[scale=0.22]{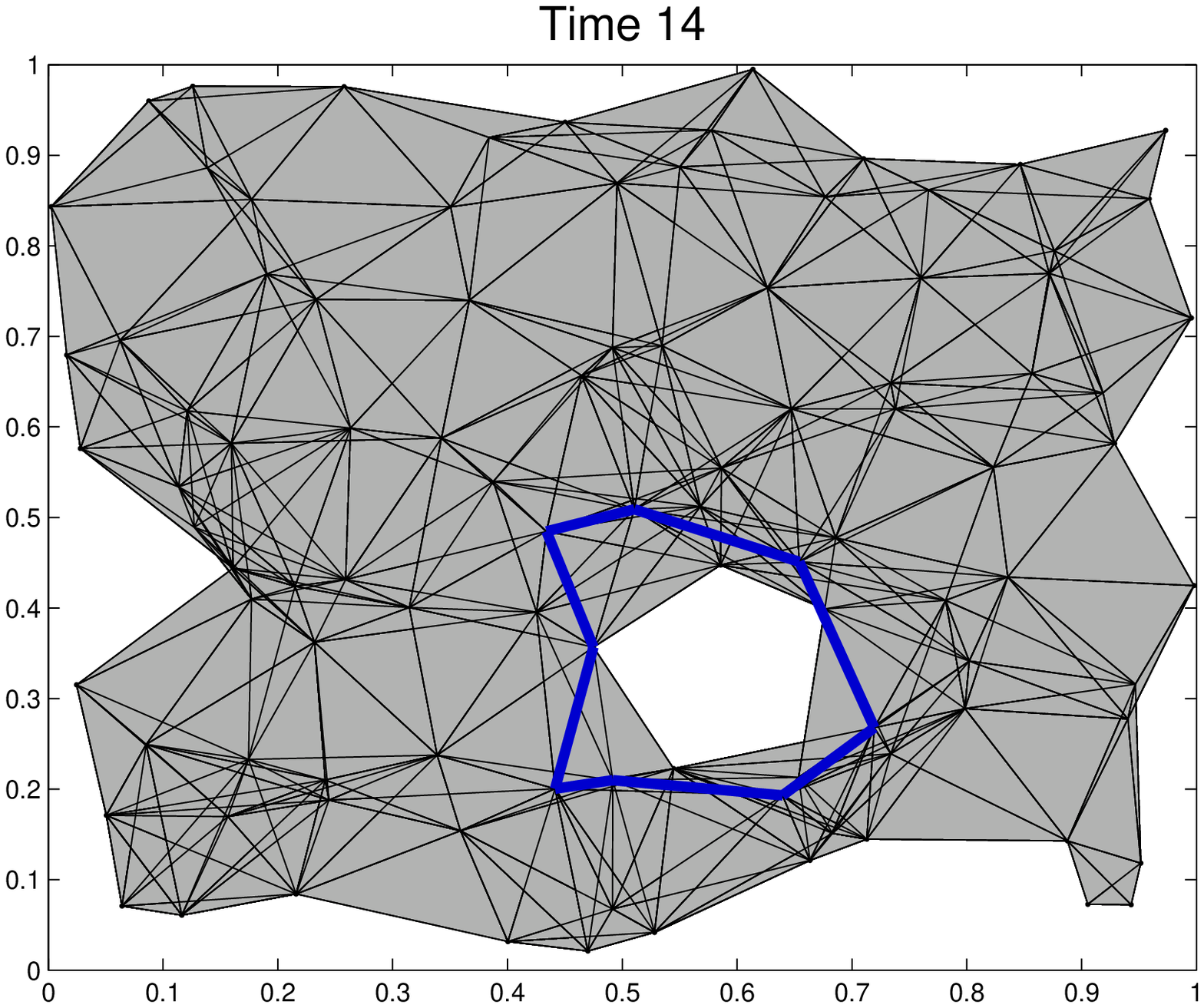} \\
\includegraphics[scale=0.22]{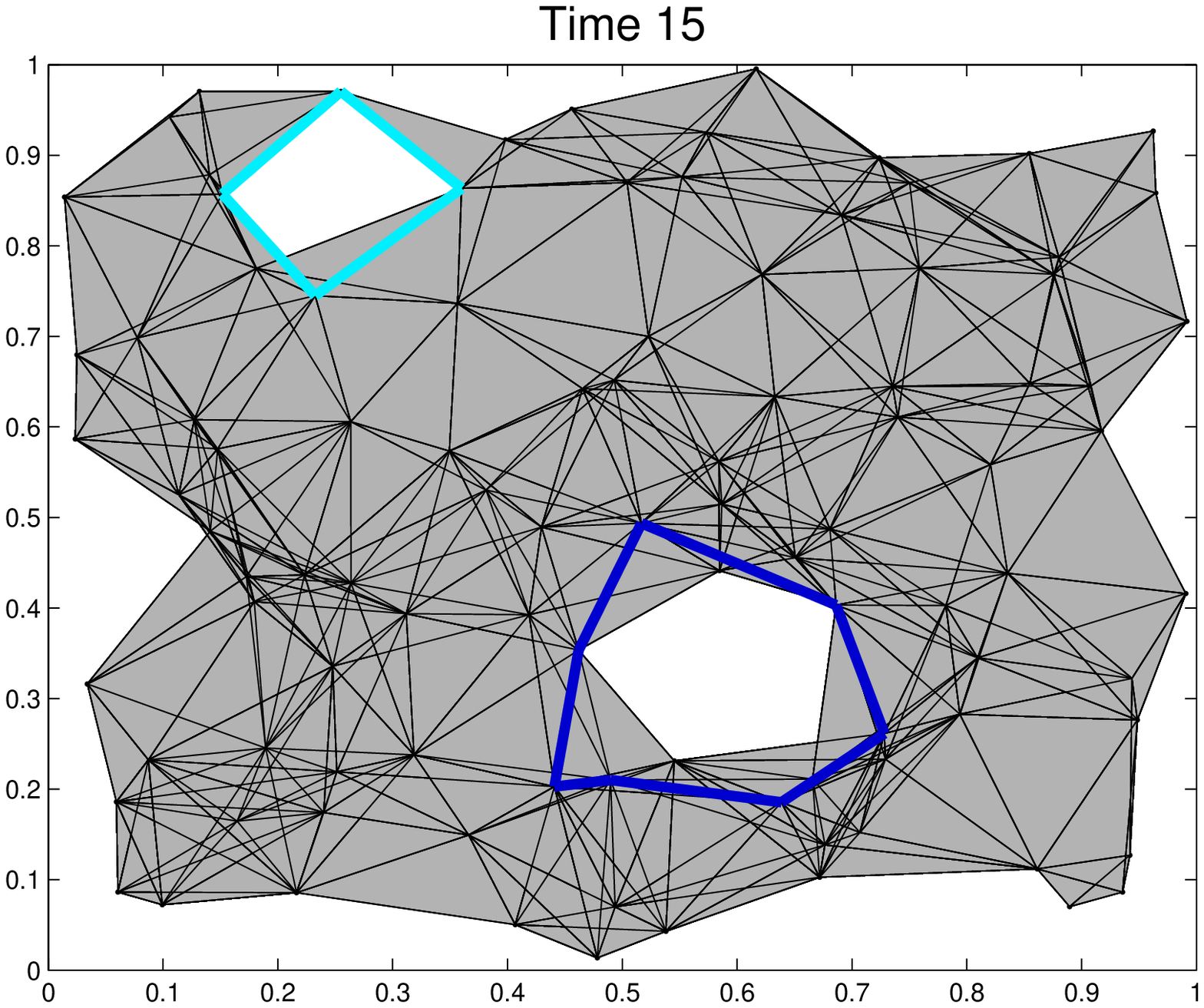} & \includegraphics[scale=0.22]{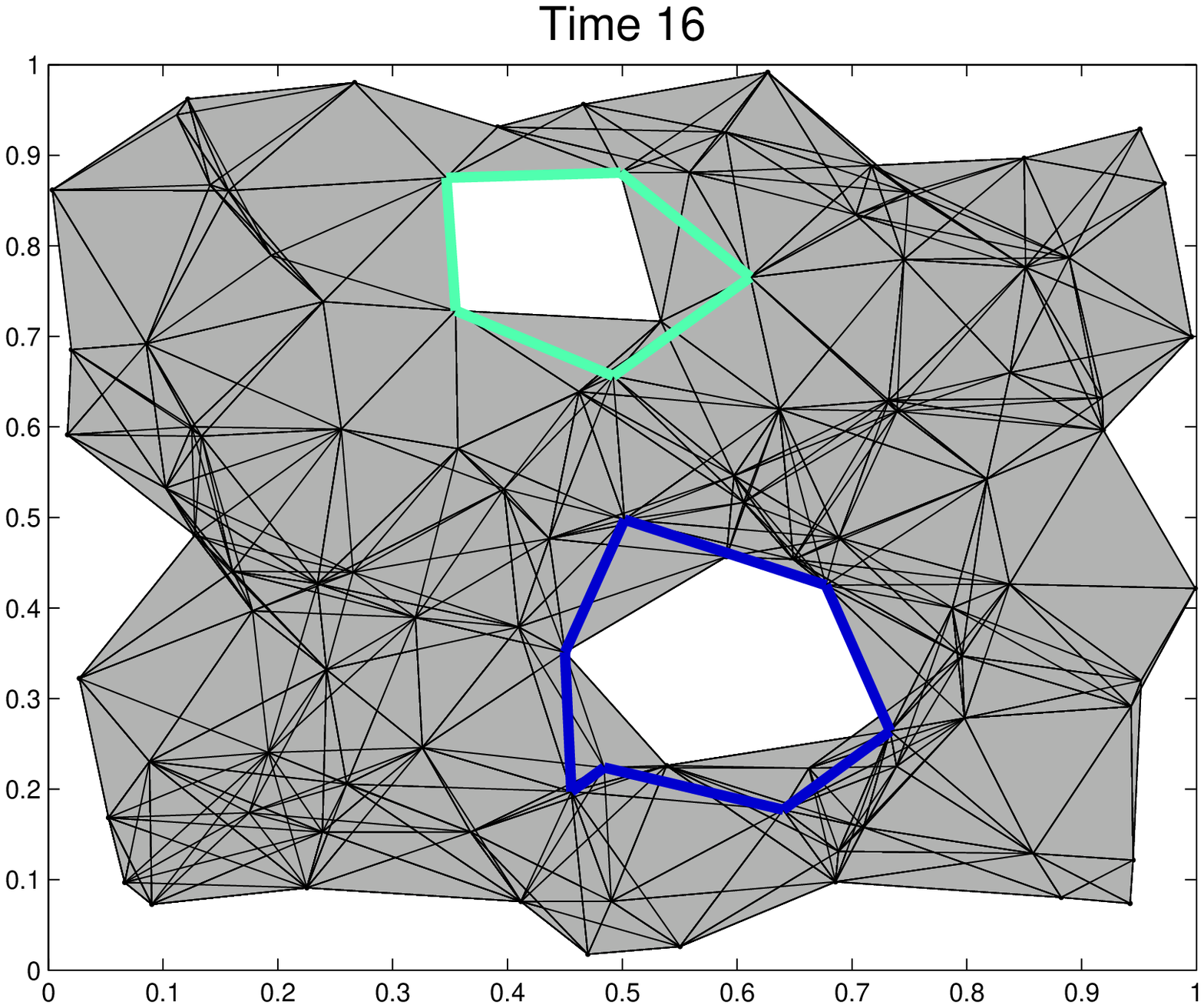} & \includegraphics[scale=0.22]{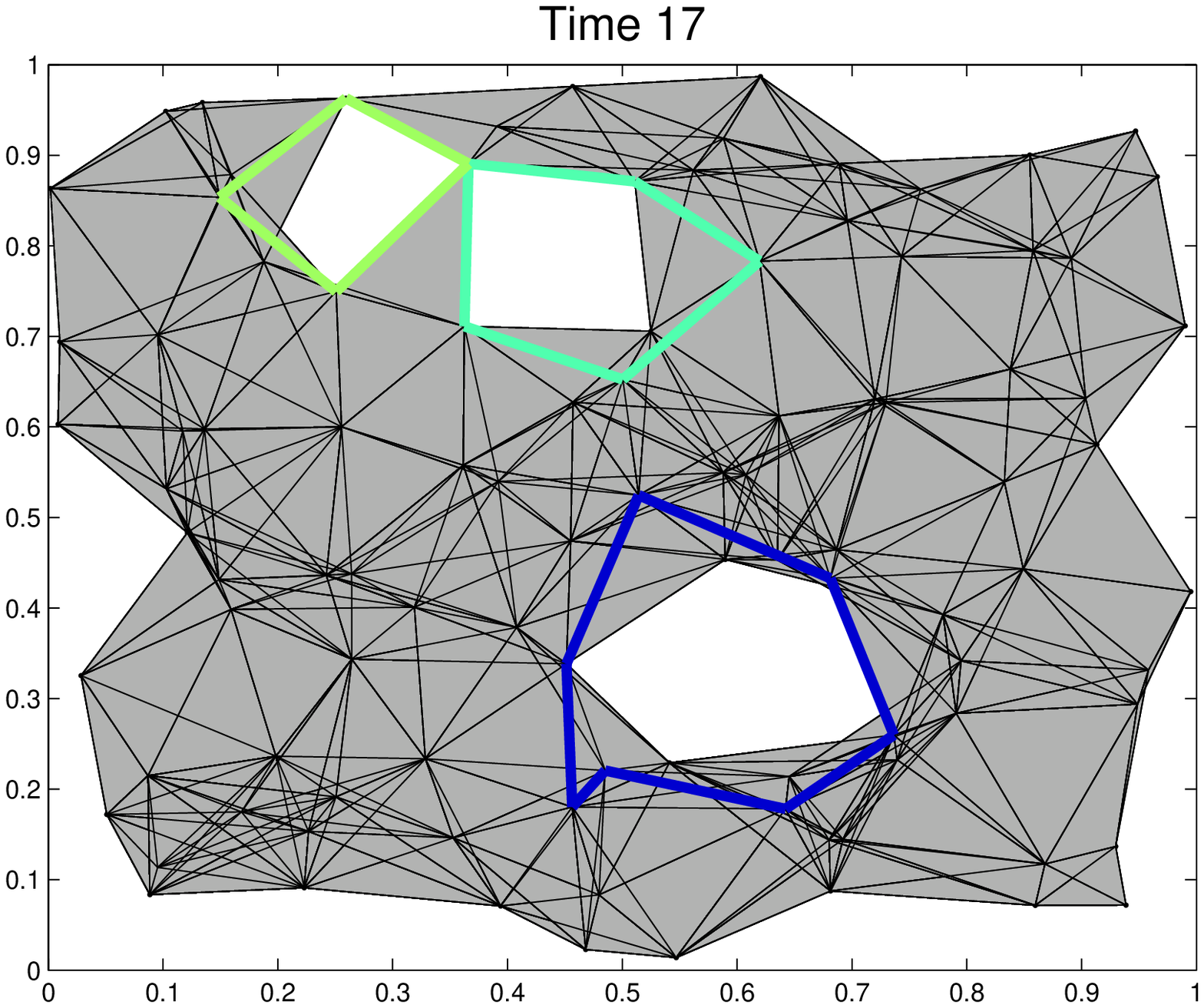} \\
\includegraphics[scale=0.22]{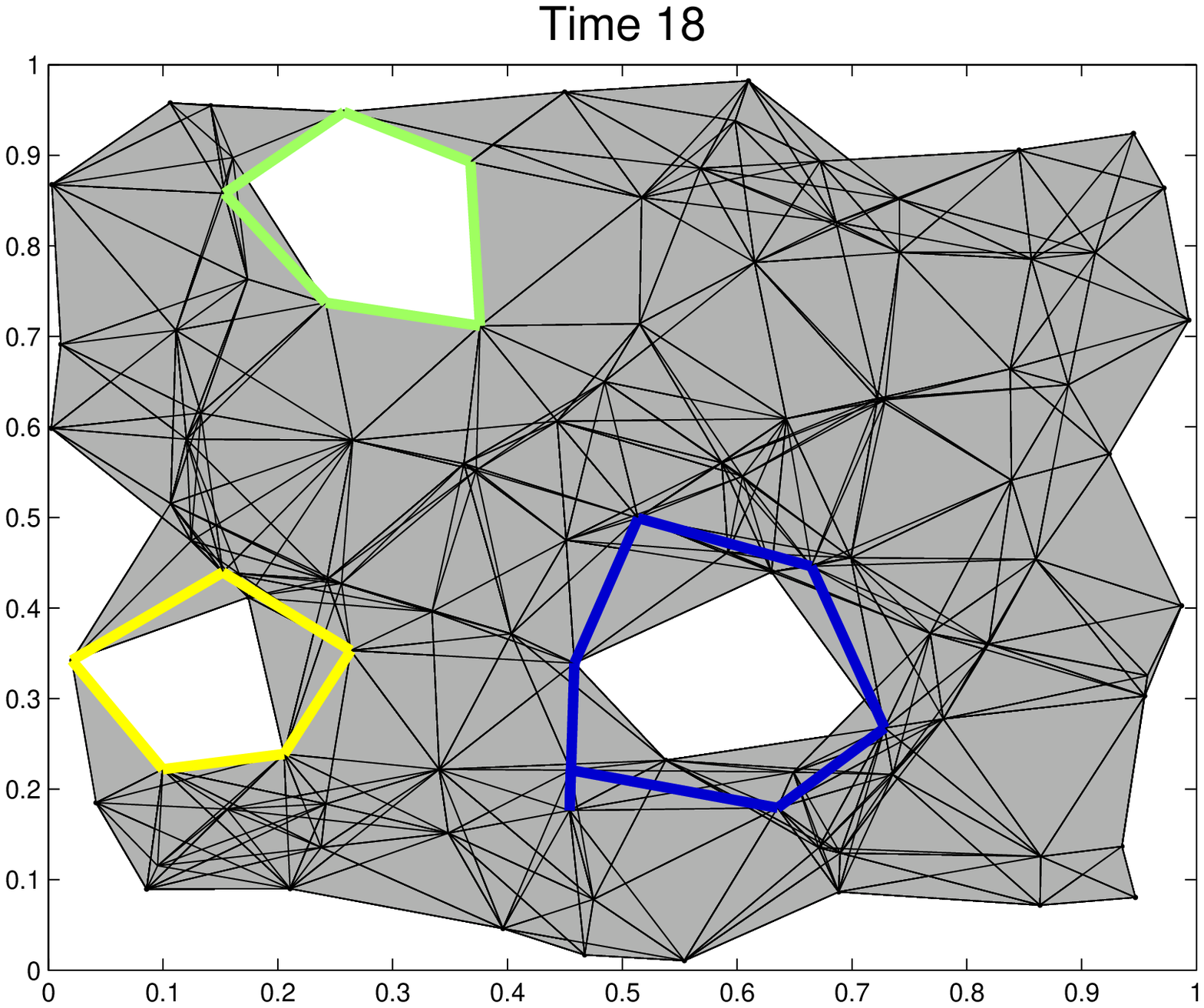} & \includegraphics[scale=0.22]{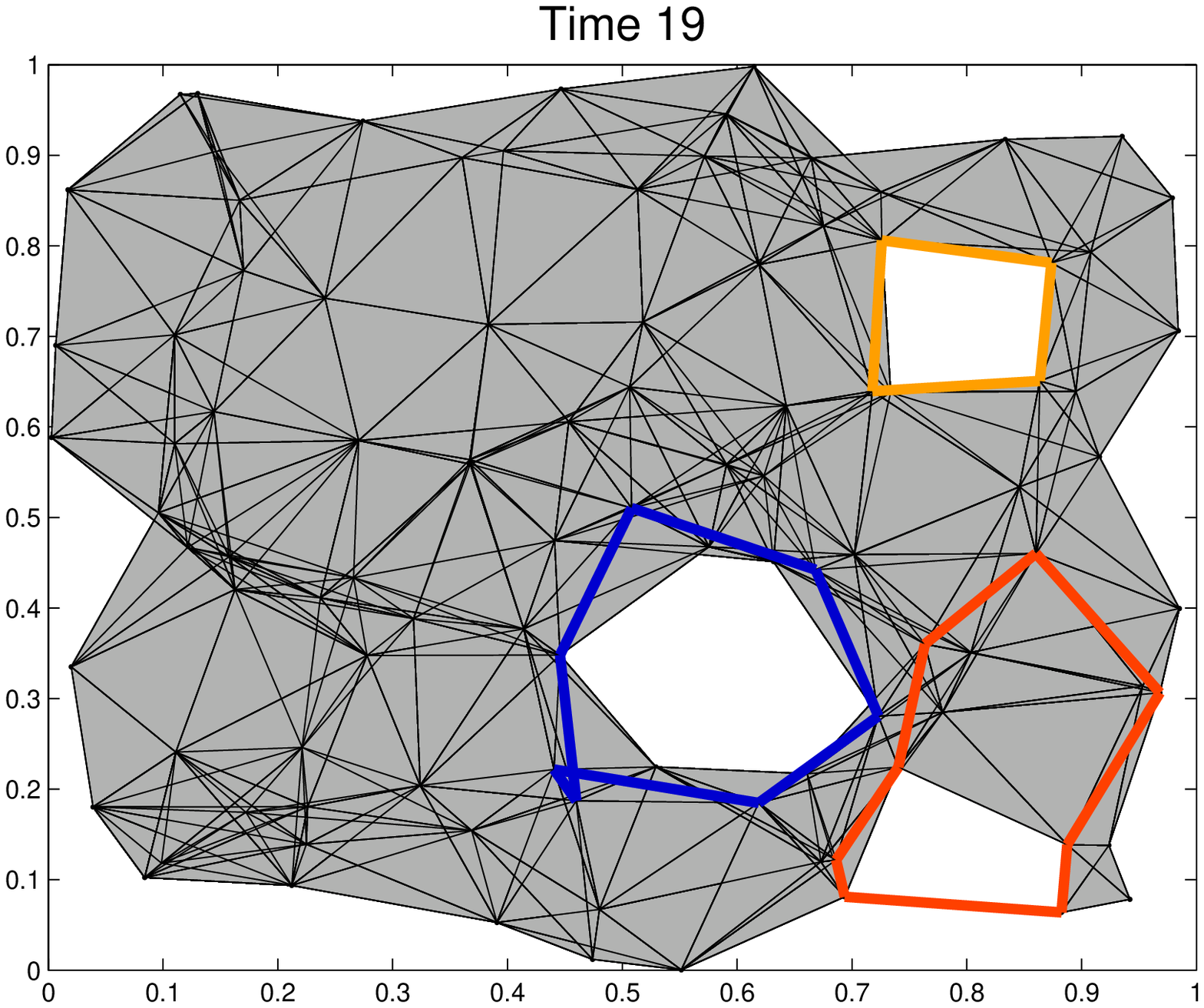} & \includegraphics[scale=0.22]{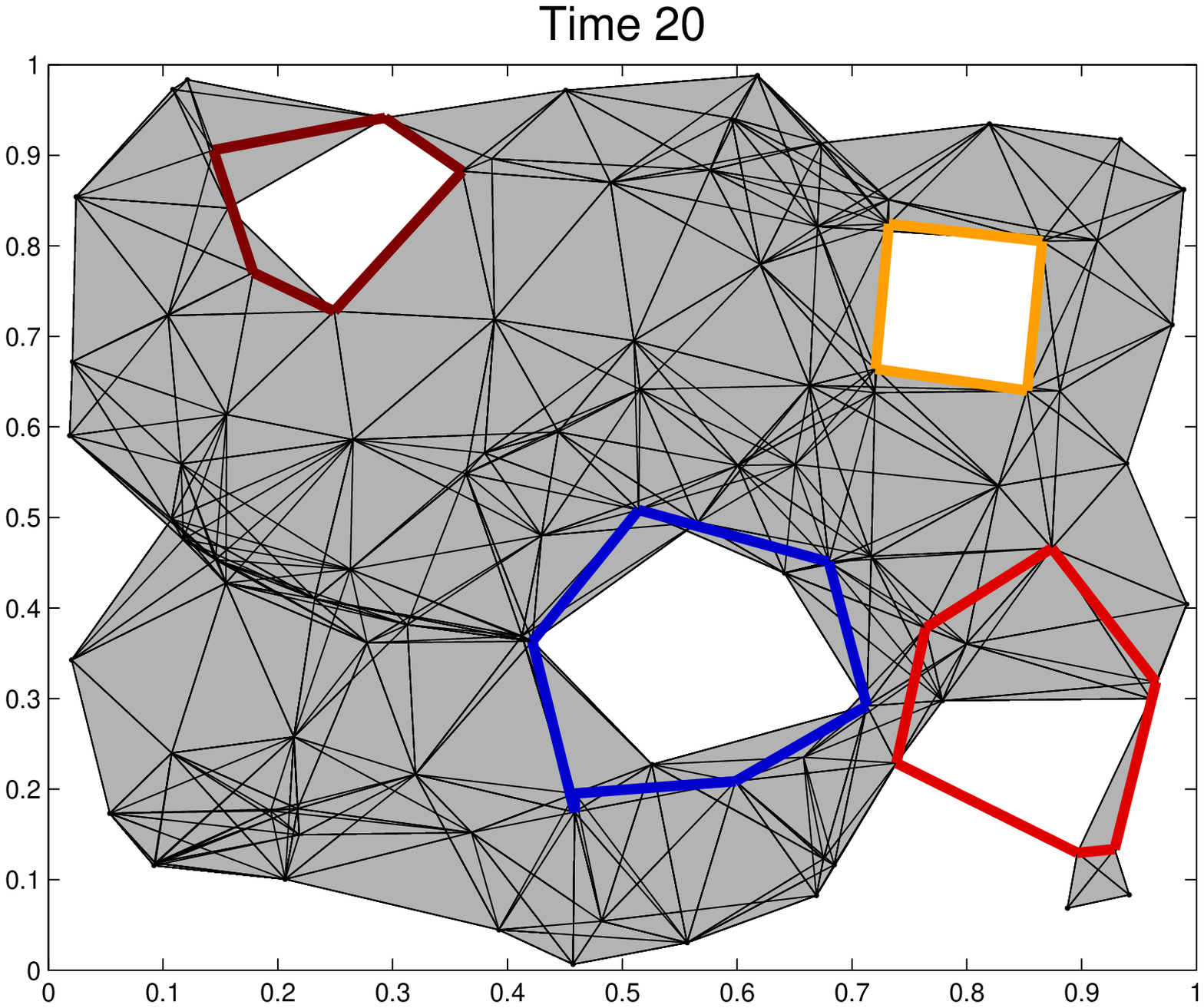} \\
\end{tabular}
\end{center}
\caption{Representative cycles, color-coded with their corresponding intervals (shown in the barcode - top left). \label{DenseNetwork}}
\end{figure}

Since canonical information is only available when a hole is formed due to the removal of a triangle, this method performs best when the network is relatively dense, and the holes are appearing and disappearing in relative isolation (without extensive merging and splitting of holes during their lifetimes). In such a setting the representative cycles typically surround each hole uniquely, and even encircle the holes relatively tightly. Figure \ref{DenseNetwork} illustrates this, with the color of each cycle indicating the bar that it corresponds to in the barcode.

\subsection{Enhancing barcode with estimated size information}

\begin{figure}[htp]
\begin{center}
\begin{tabular}{ccc}
\includegraphics[scale=0.35]{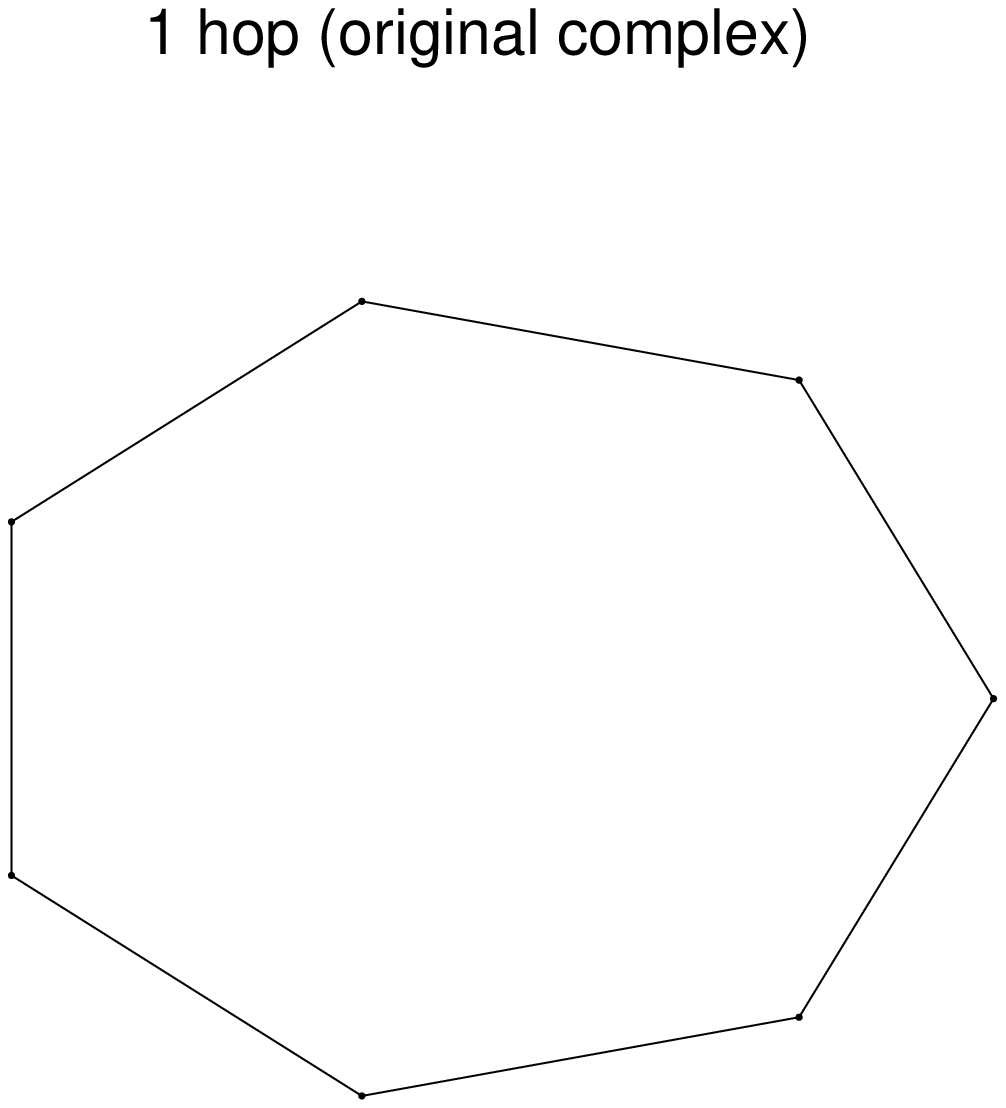} & \includegraphics[scale=0.35]{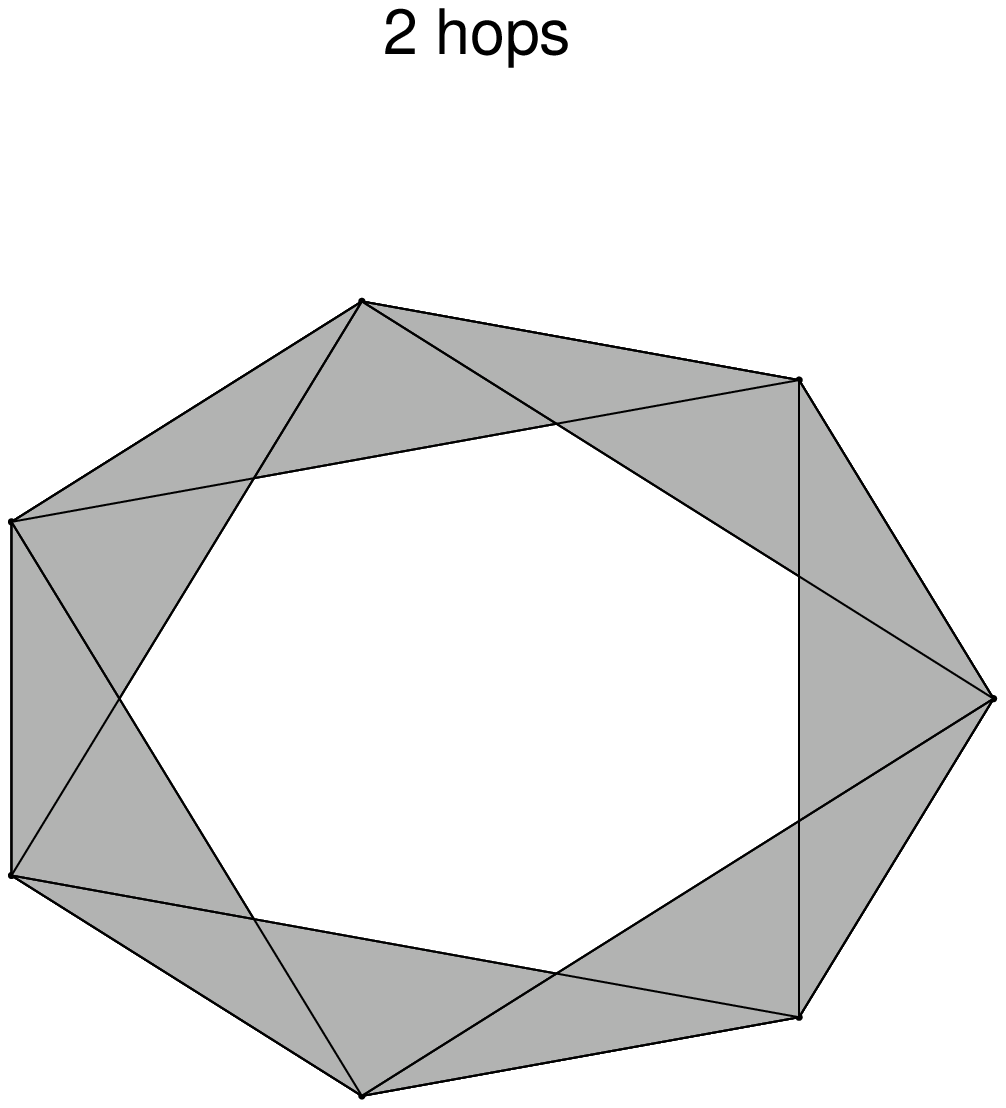} & \includegraphics[scale=0.35]{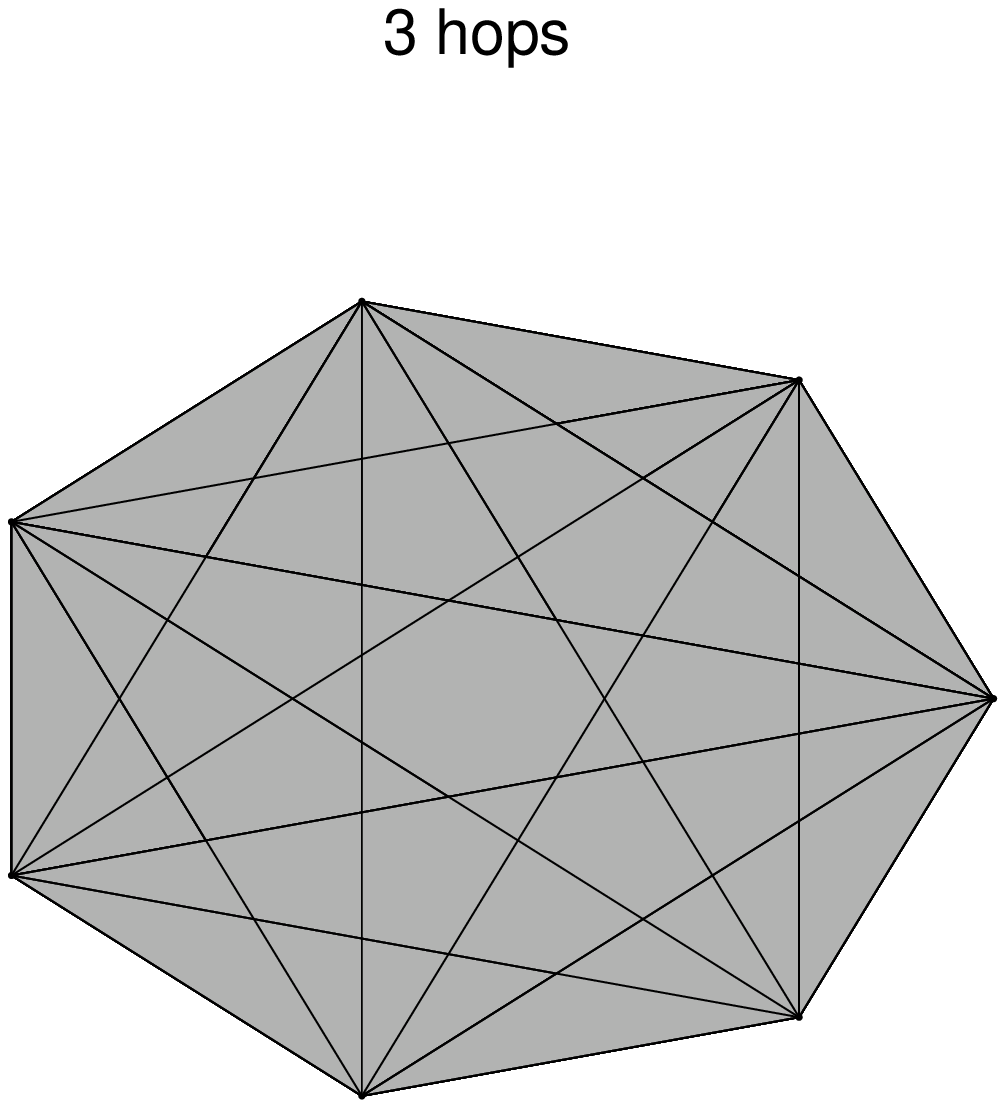} \\
\end{tabular}
\end{center}
\caption{Hop distance filtration at depth $k=1$ (original complex), left, $k=2$ (hole persists), middle, and $k=3$ (hole is filled in), right. \label{HopFilt}}
\end{figure}

In addition to tracking holes, choosing a specific representative cycle for each bar allows us to attach additional information onto the barcode, such as size estimates at each point. Given our coordinate-free setting, the only size estimate available for a hole is the length of the shortest cycle surrounding it. For a specific simplicial complex $K$, this may be obtained by performing a hop-distance based filtration. Such a filtration starts with the original simplicial complex $K^1 = K$, and then forms a nested sequence of complexes $K^2, K^3, \ldots$, where $K^h$ is formed by taking $K^{h-1}$ and adding an edge between nodes that are $h$ hops apart in $K$, and then filling in all higher-dimensional simplices. A hop-distance filtration on a simple simplicial complex consisting of a single loop is shown in Figure \ref{HopFilt}. Given a nontrivial cycle in $K$, the depth at which it becomes trivial in the hop-distance filtration is determined by the size (in shortest hop-length) of the largest hole that the cycle surrounds. Table \ref{HopSizes} gives this relationship.

Now, a hop-distance filtration is performed on the simplicial complex at each time point, and for each bar in the zigzag barcode, the persistence of its representative cycle in the hop distance filtration is computed. This additional size information may be added to the barcode visually by thickening the bar proportionally to the size value.

\begin{figure}[htp]
\begin{center}
\begin{tabular}{ccc}
\includegraphics[scale=0.22]{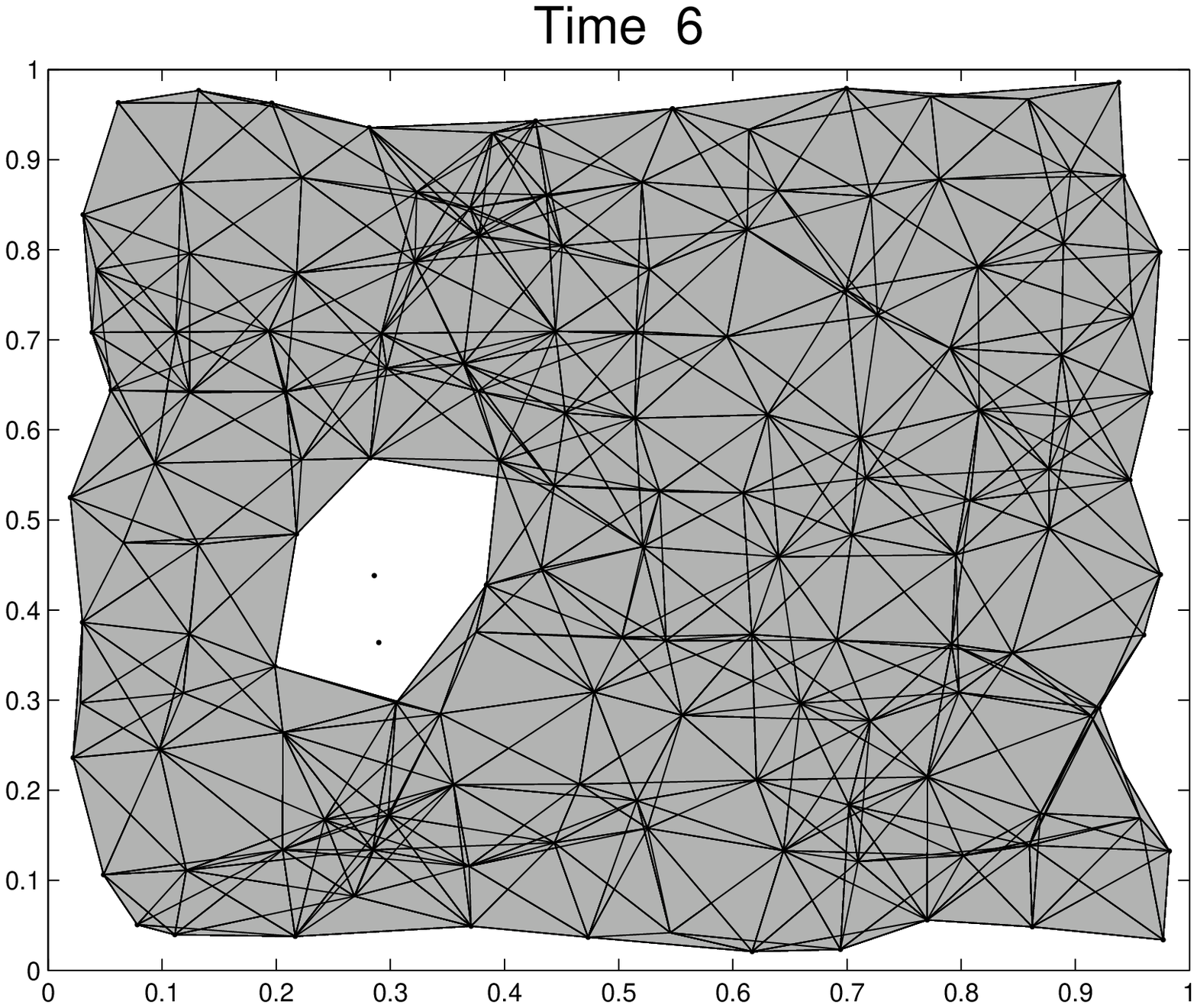} & \includegraphics[scale=0.22]{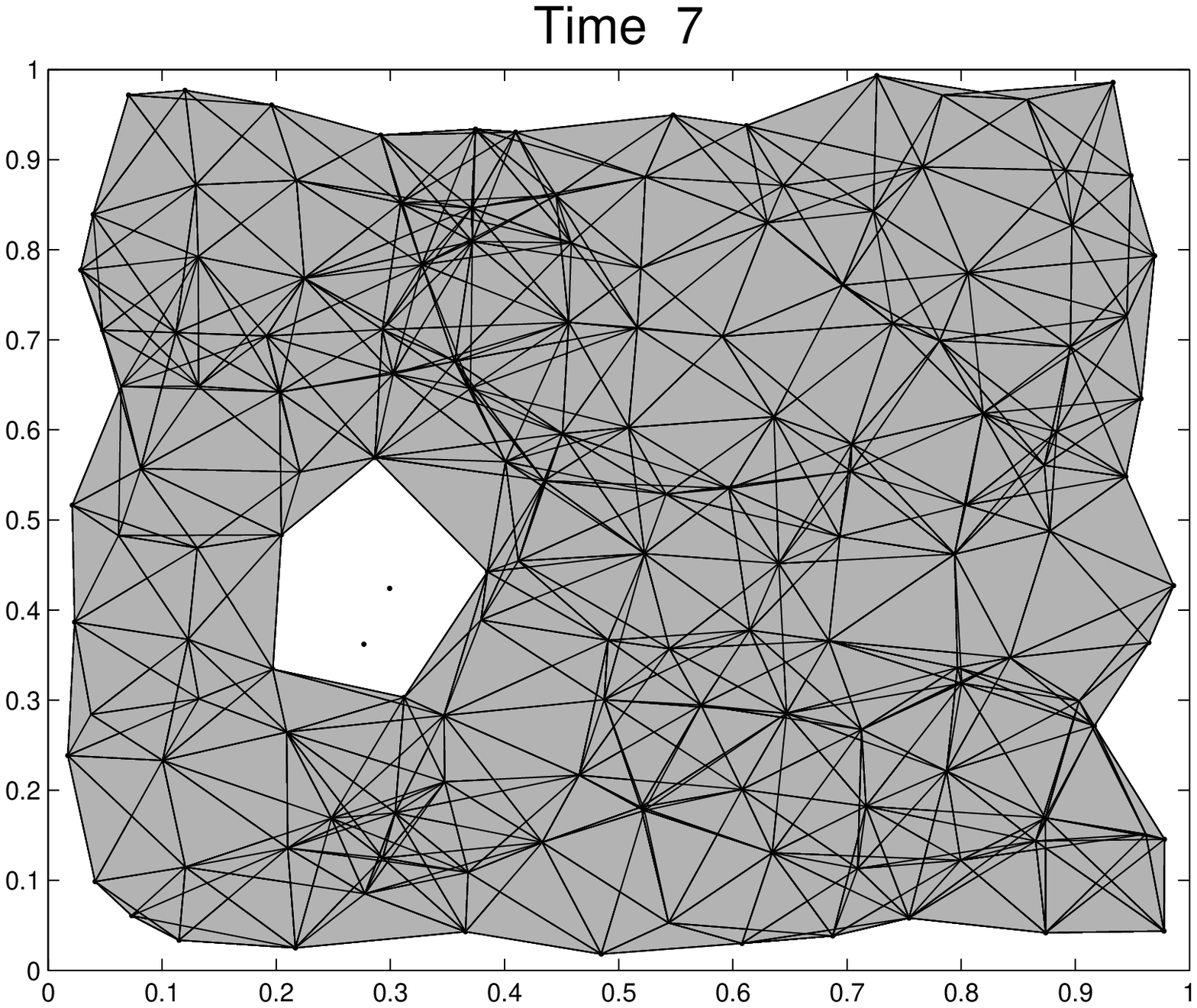} & \includegraphics[scale=0.22]{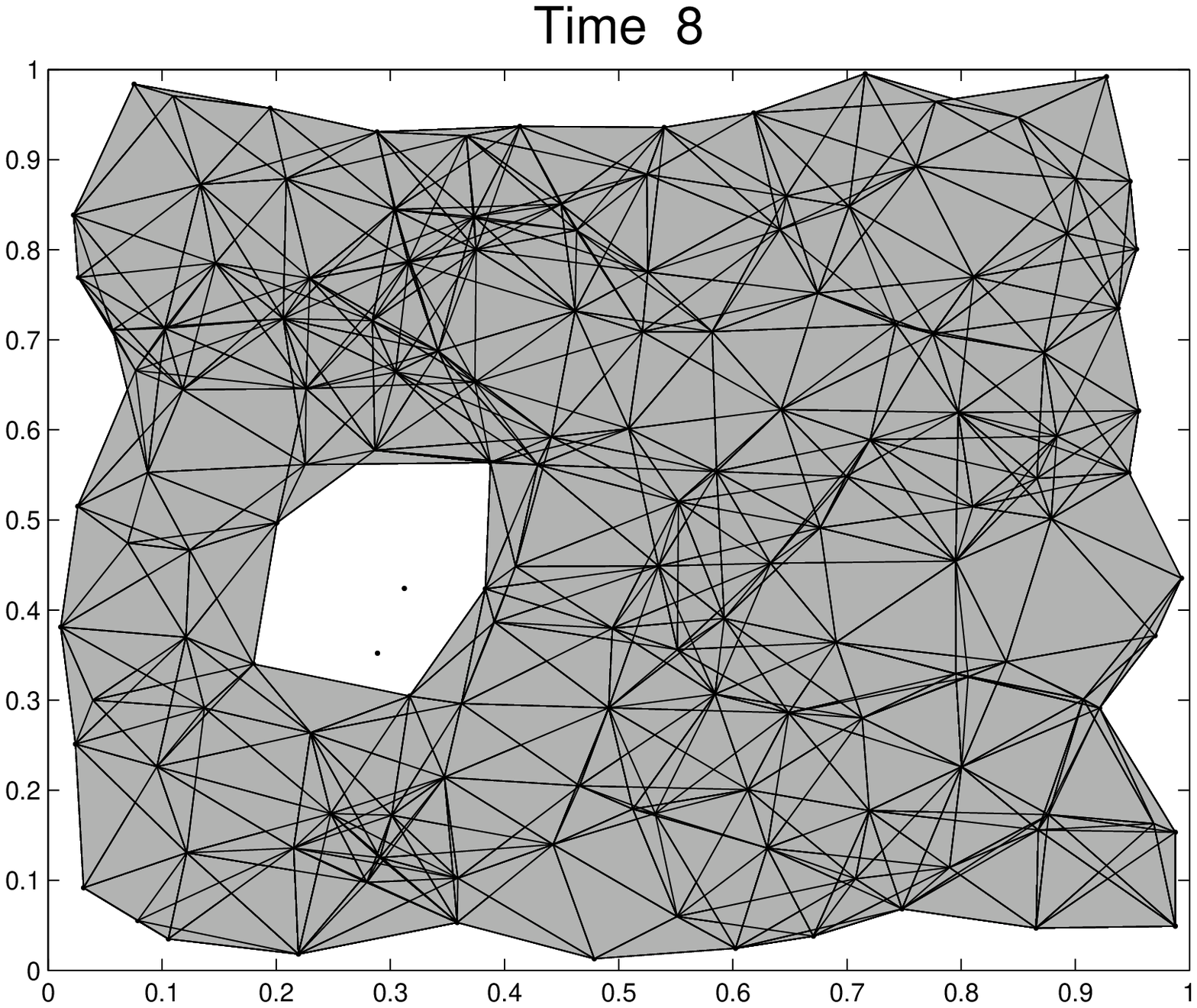} \\
\includegraphics[scale=0.22]{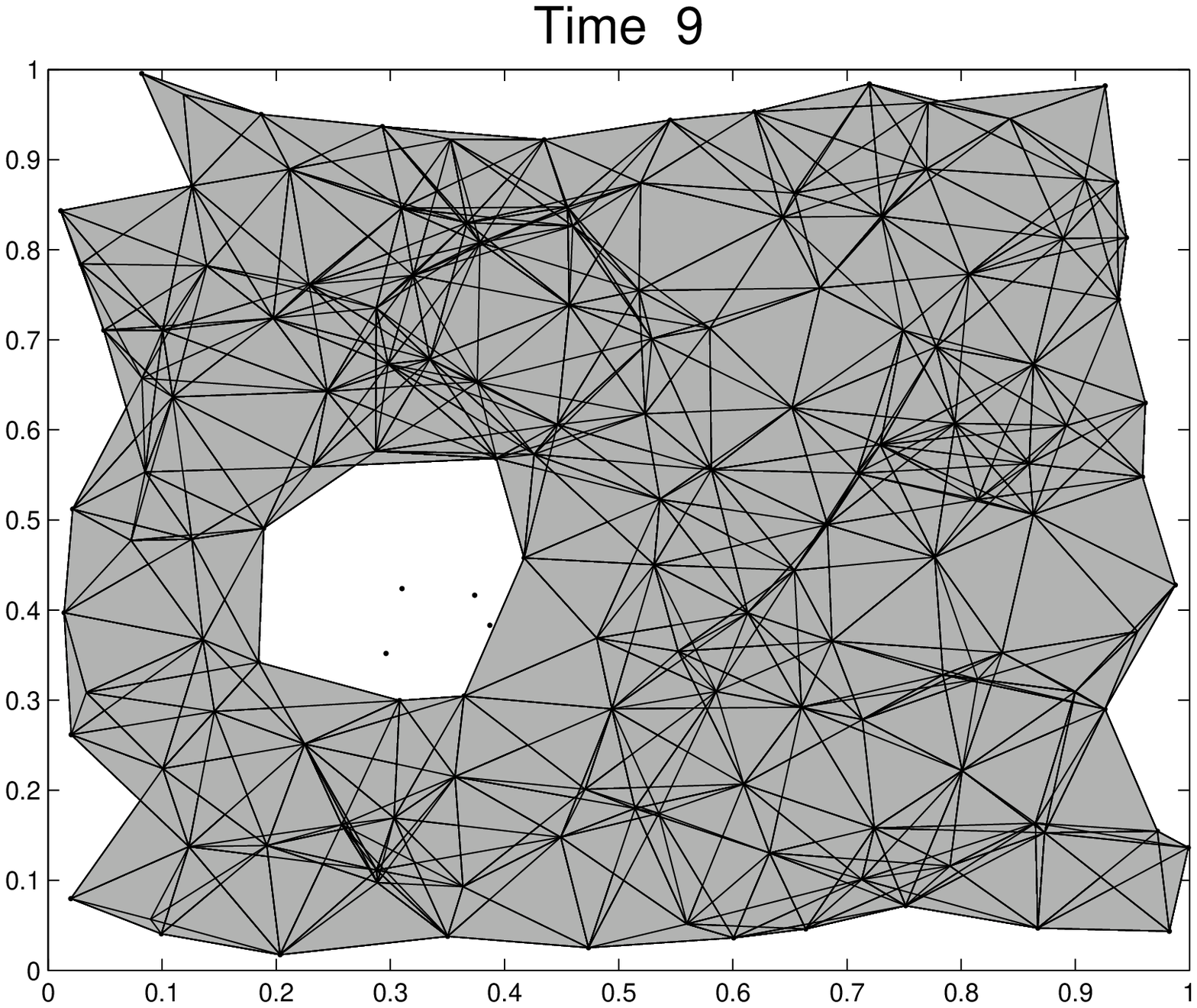} & \includegraphics[scale=0.22]{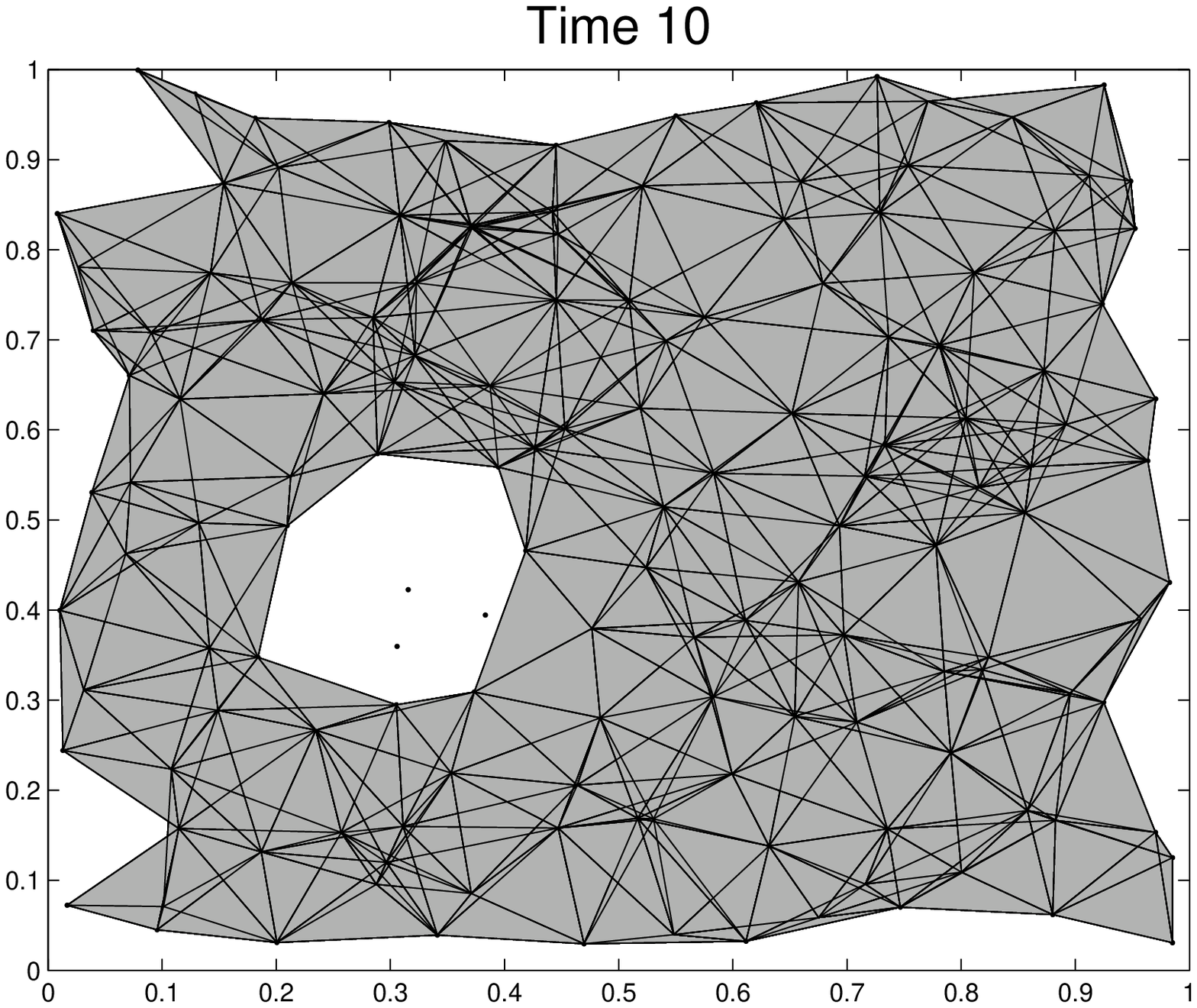} & \includegraphics[scale=0.22]{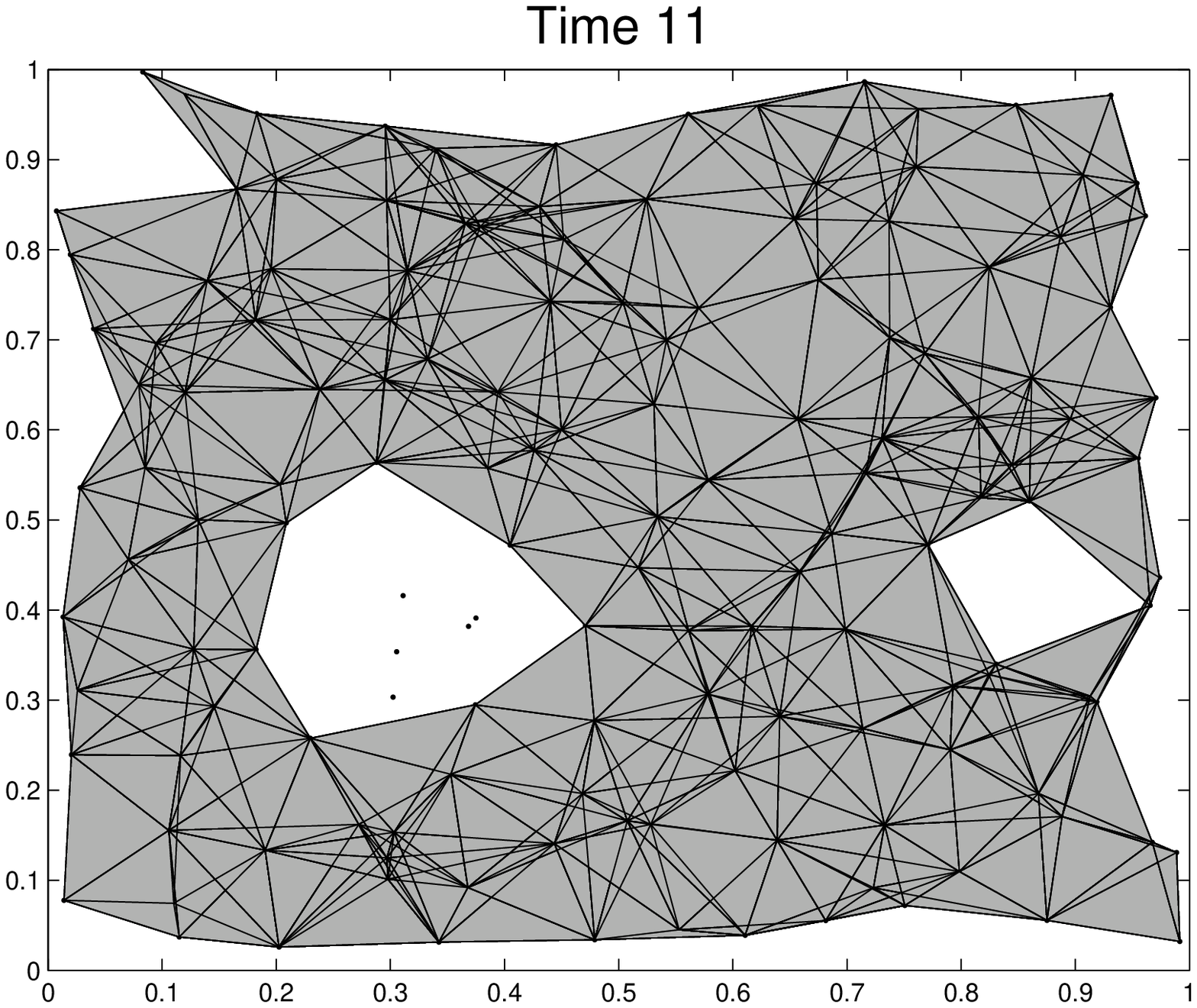}  \\
\includegraphics[scale=0.22]{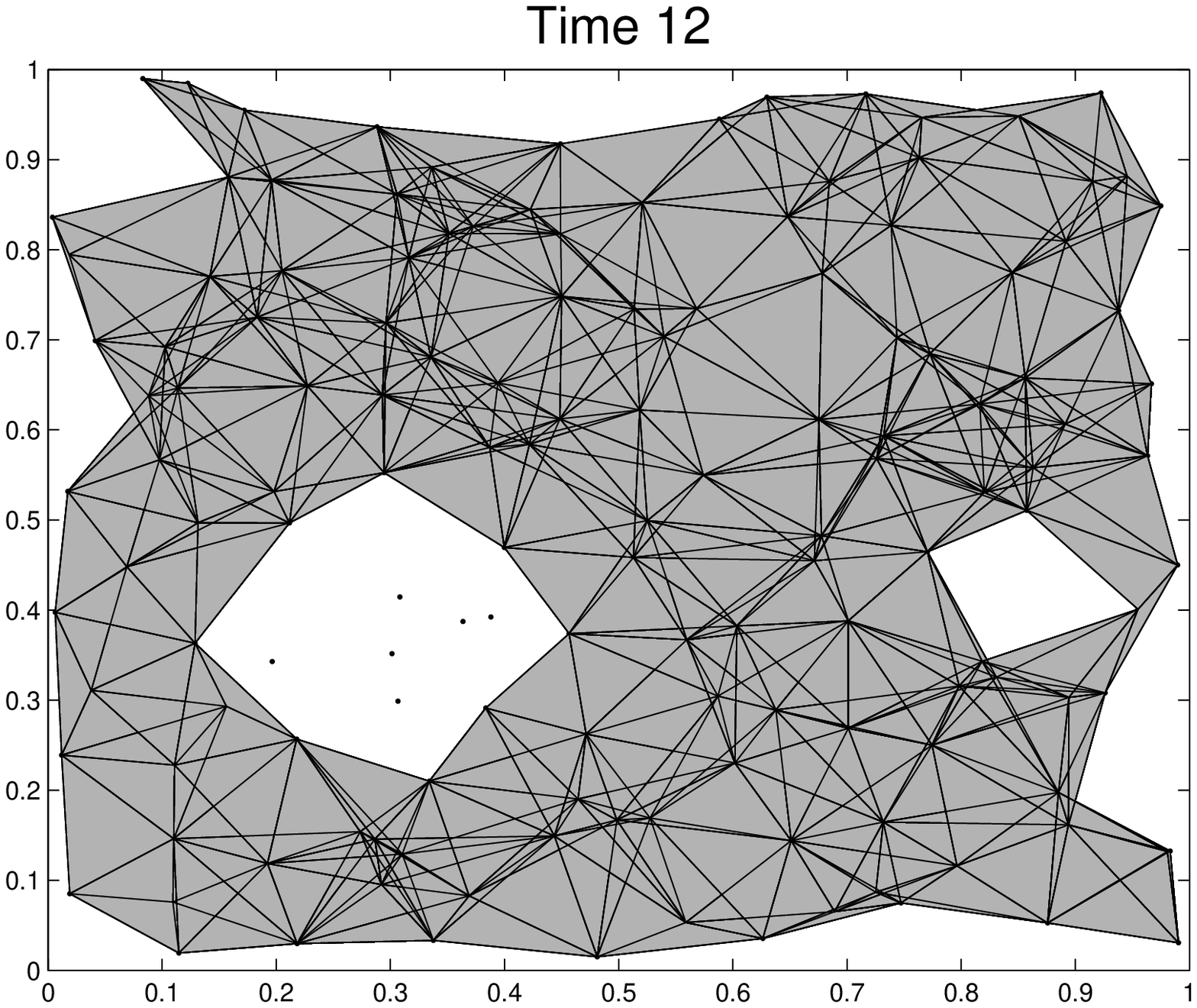} & \includegraphics[scale=0.22]{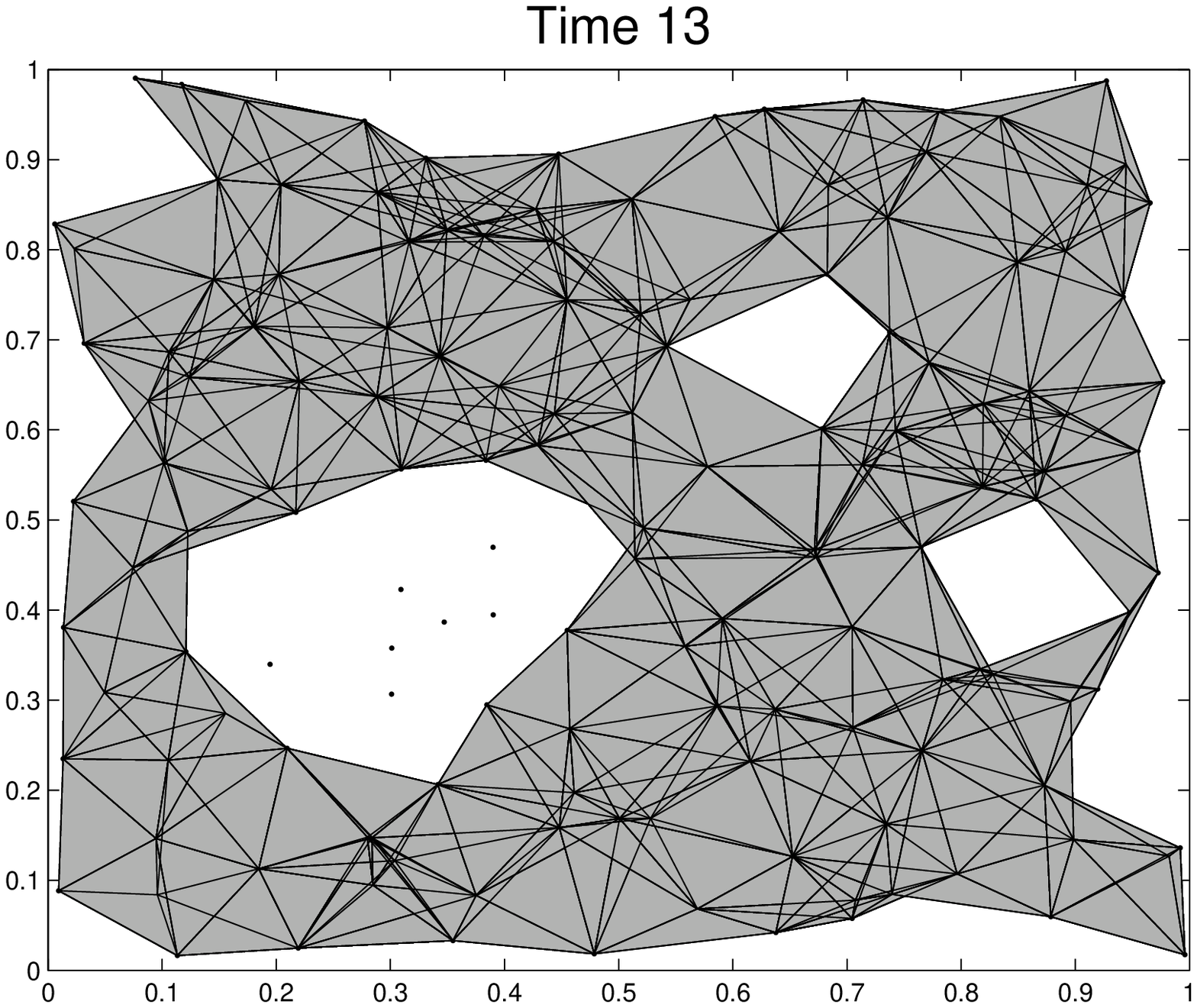} & \includegraphics[scale=0.22]{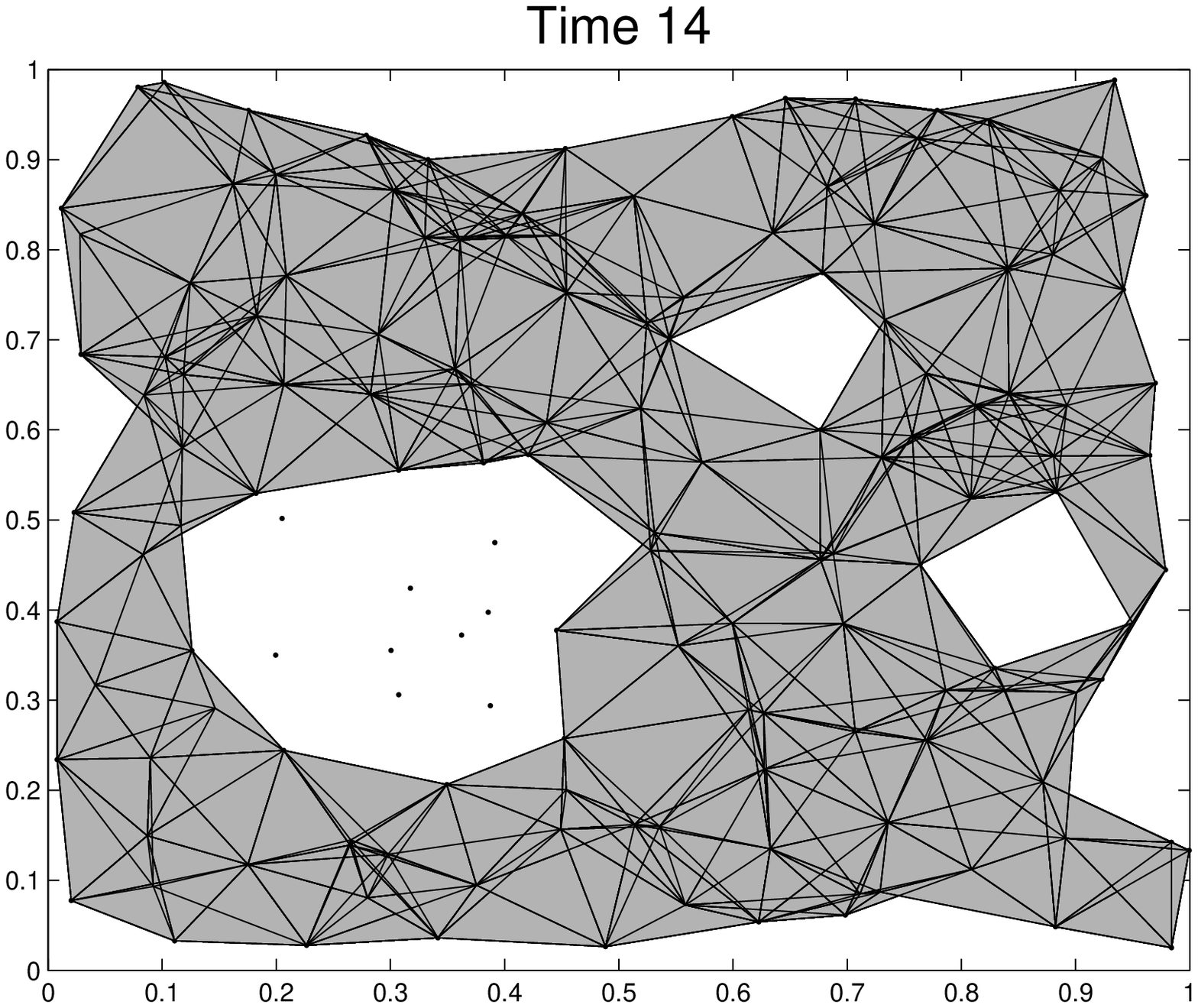} \\
\includegraphics[scale=0.22]{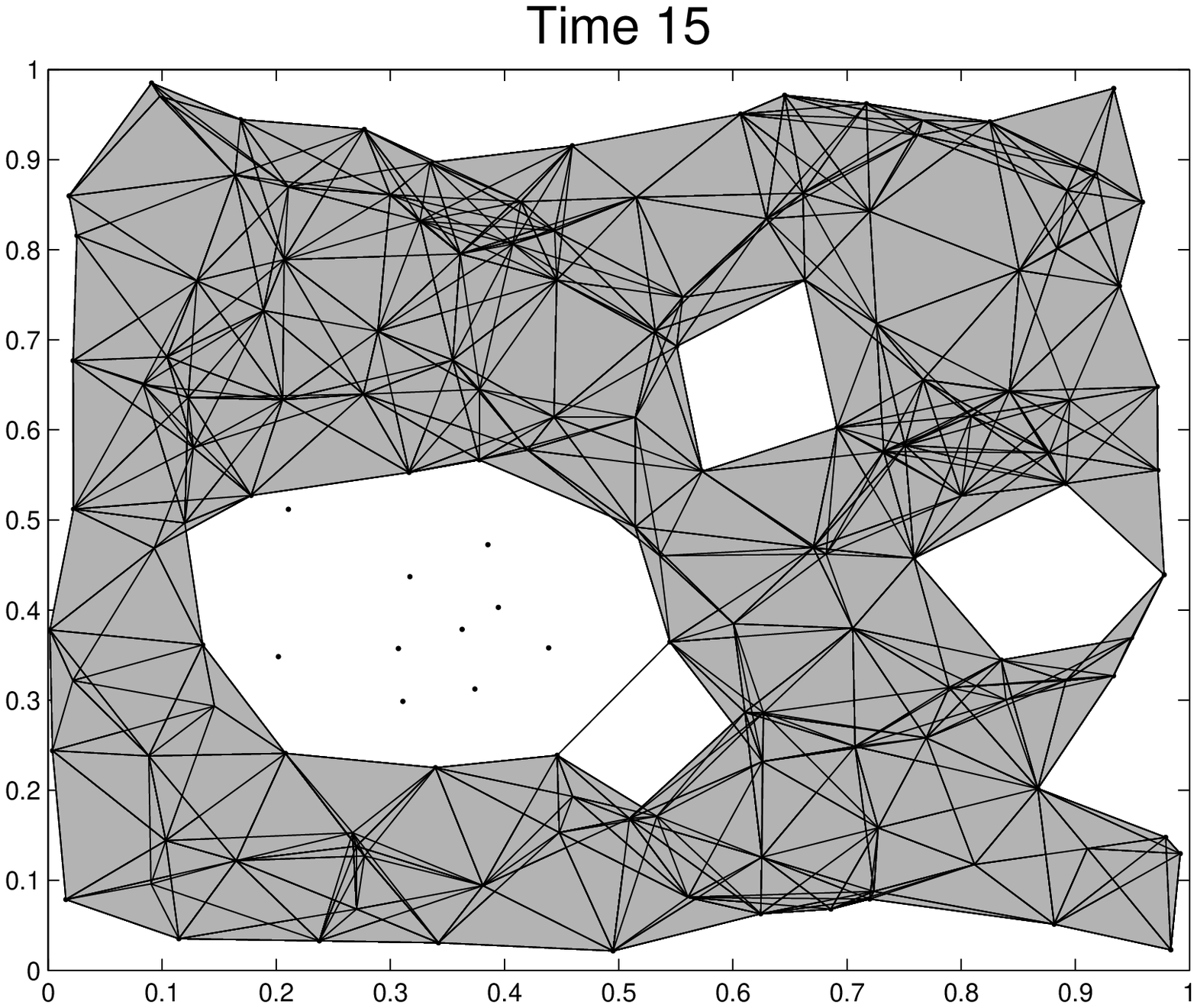} & \includegraphics[scale=0.22]{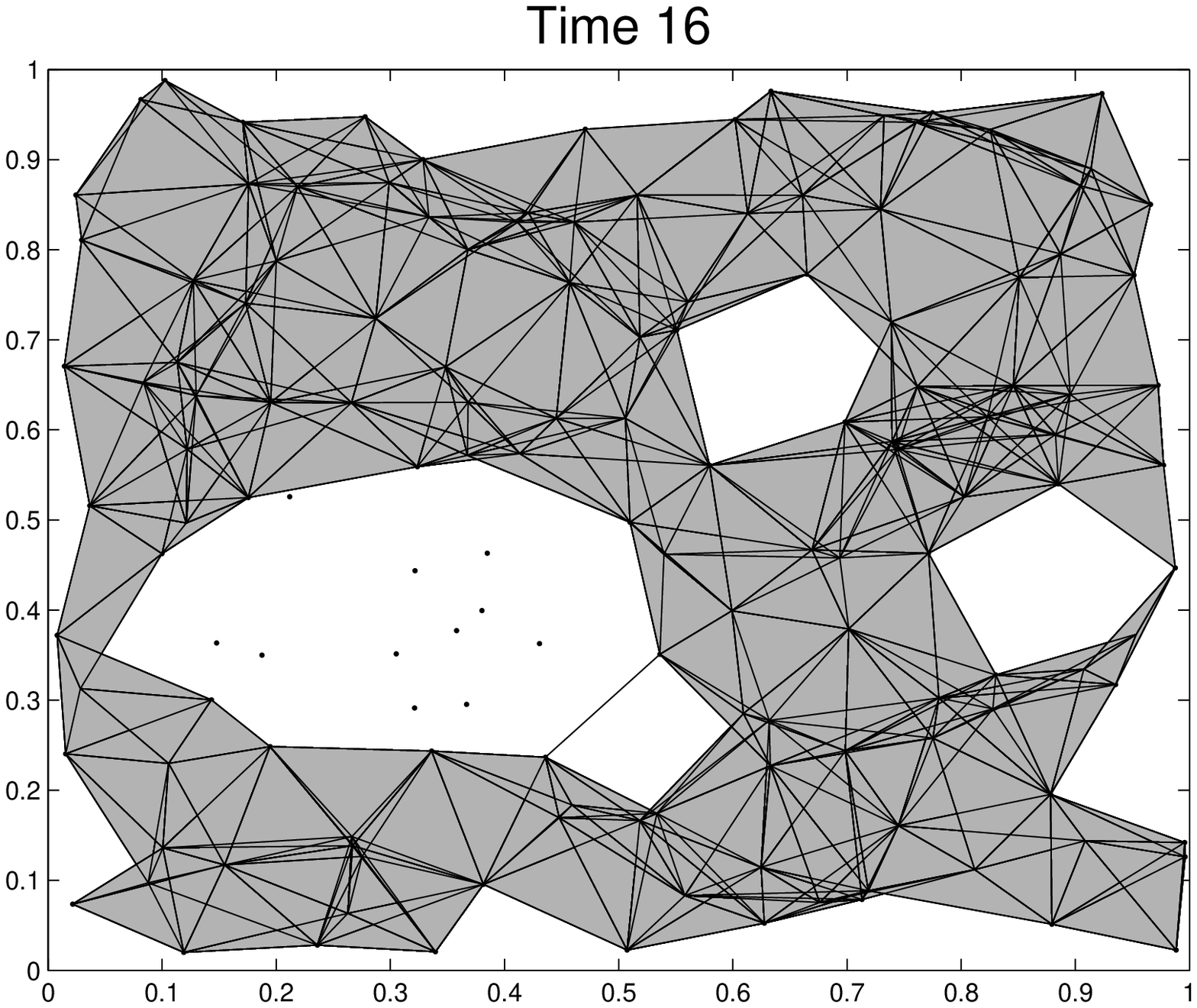} & \includegraphics[scale=0.22]{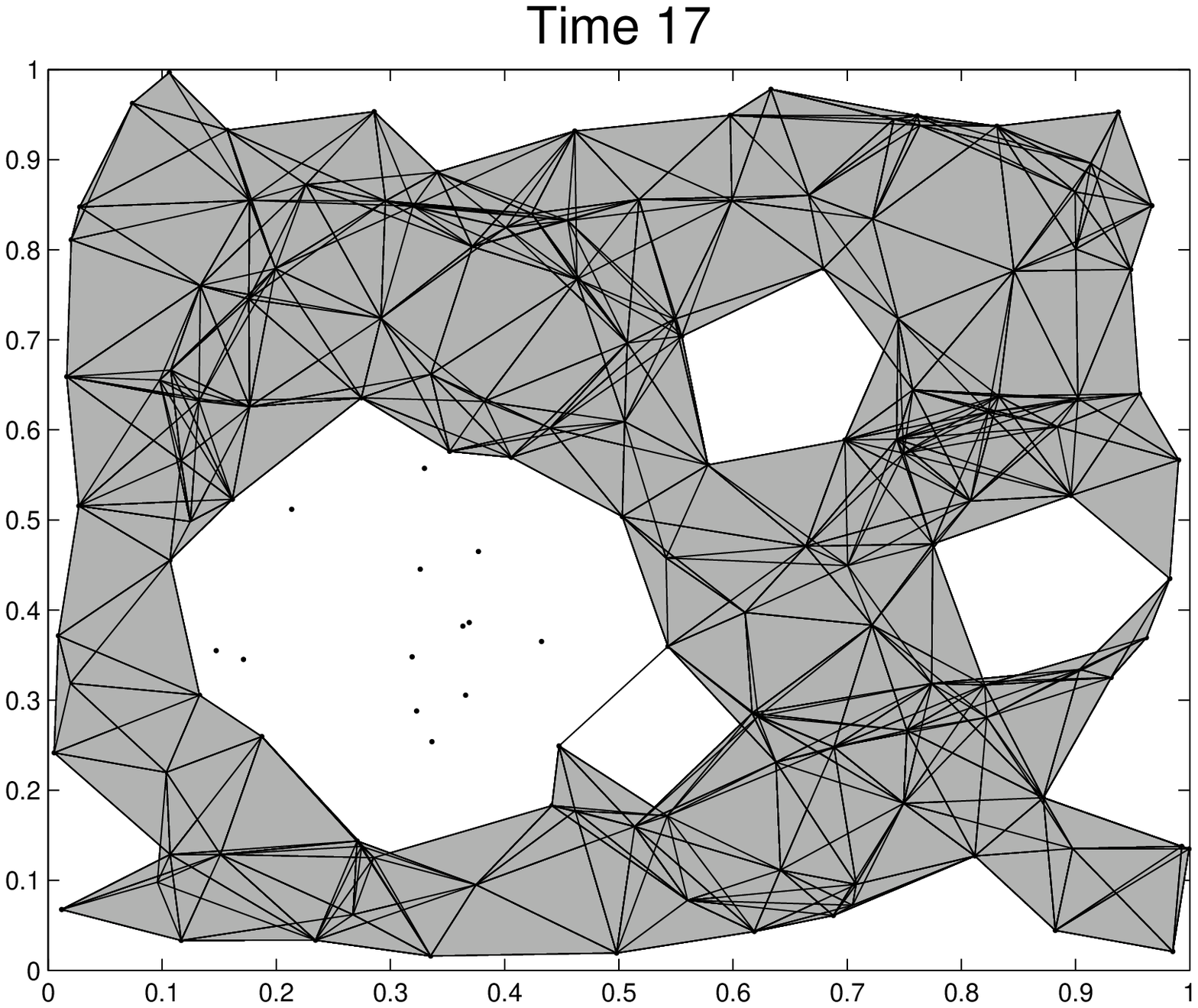} \\
\includegraphics[scale=0.22]{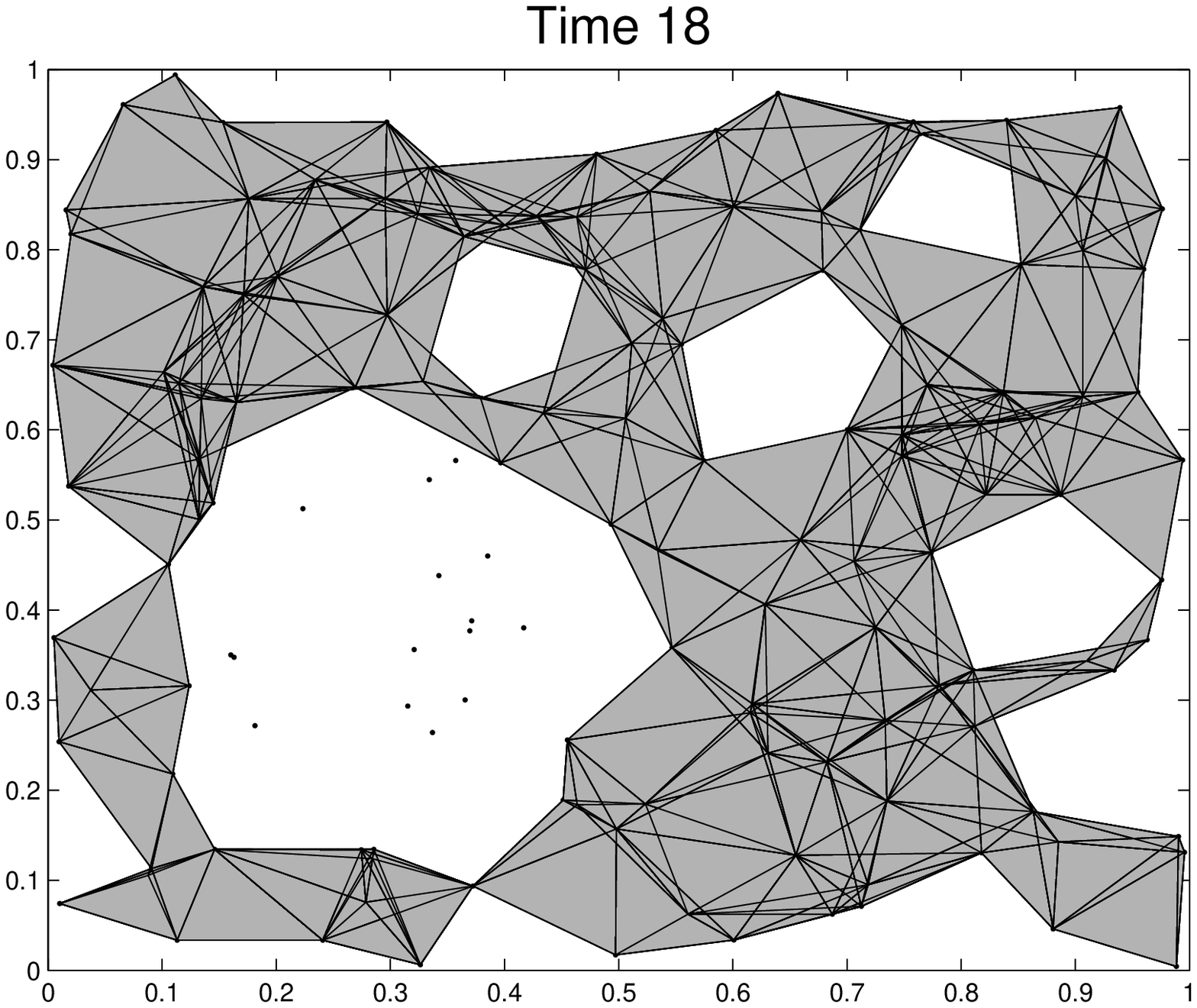} & \includegraphics[scale=0.22]{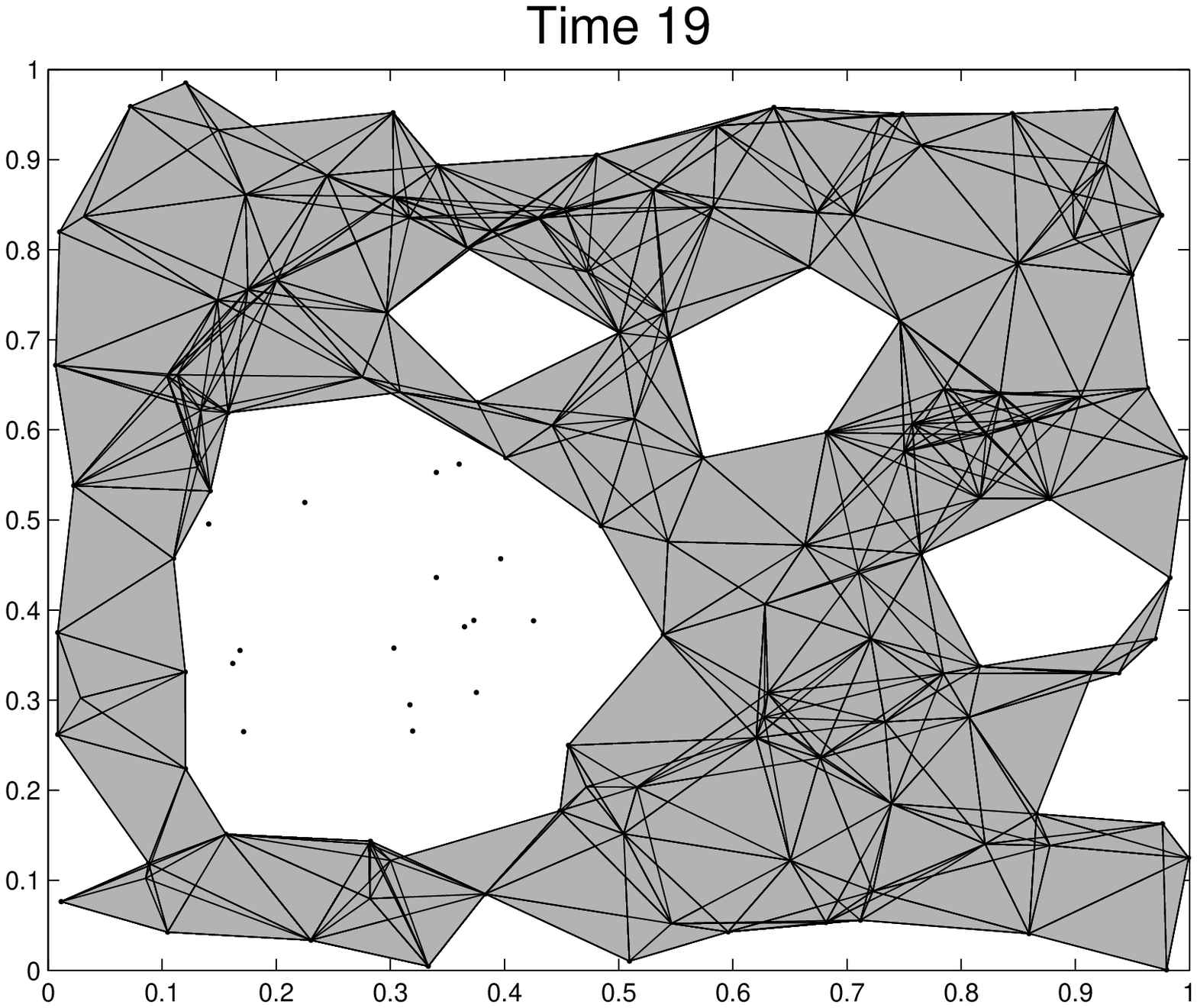} & \includegraphics[scale=0.22]{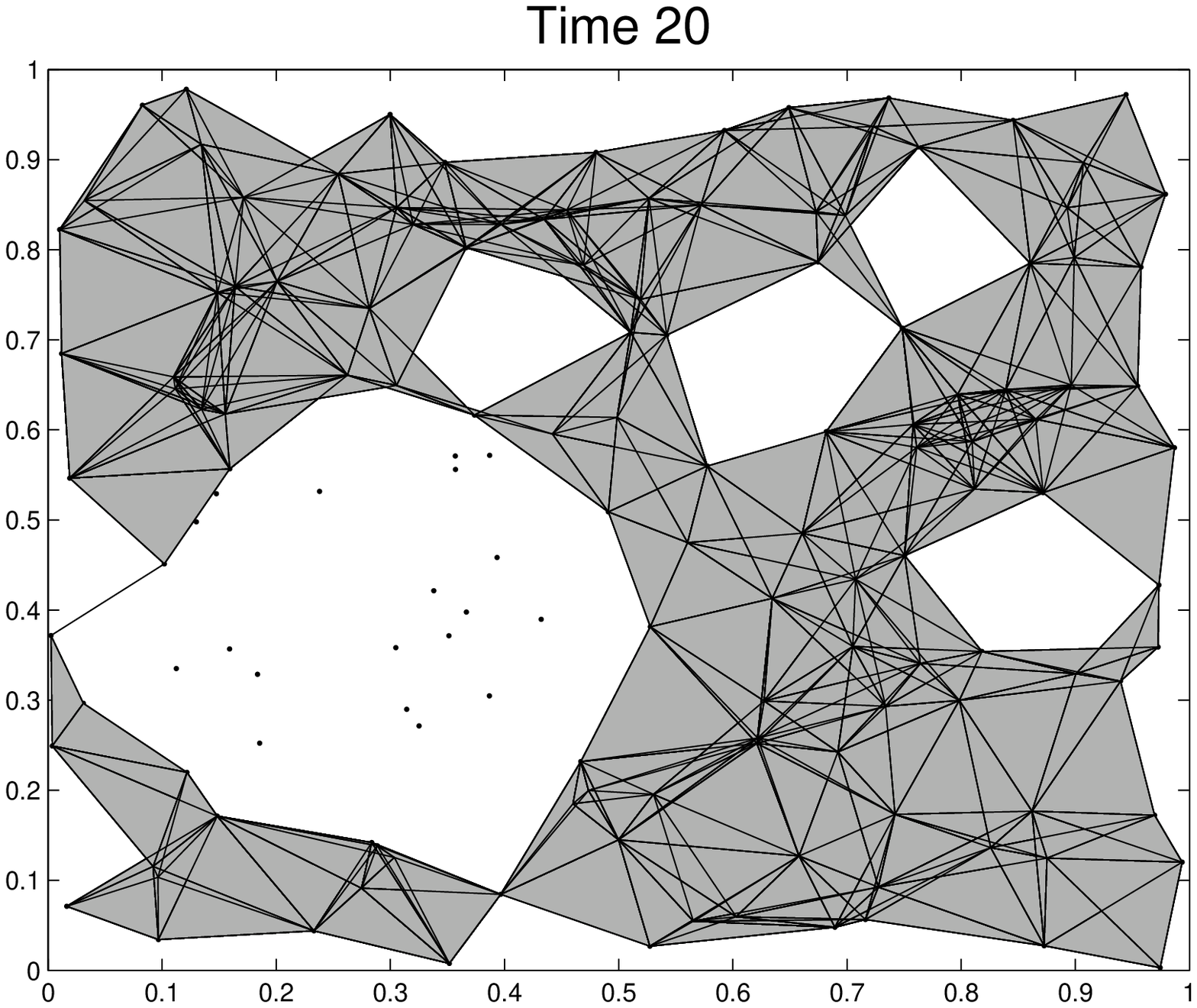} \\
\end{tabular}
\end{center}
\caption{A network with an expanding failure region, where nodes within the failure region can no longer sense or communicate.\label{ExpandingFailure}}
\end{figure}

\begin{figure}[htp]
\begin{center}
\begin{tabular}{l}
\includegraphics[scale=0.5]{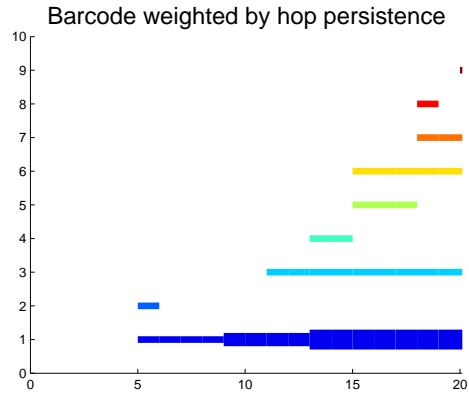} \\
\end{tabular}
\end{center}
\caption{Barcode from zigzag persistence on the network with the expanding failure region in Figure \ref{ExpandingFailure}. The thickness of each bar is proportional to the depth its adaptive representative cycle persists in the hop distance filtration at each time point. \label{WeightedBarcode}}
\end{figure}

As an example, Figure \ref{ExpandingFailure} shows a network with an expanding failure region. The barcode thickened by the estimated hole size is shown in Figure \ref{WeightedBarcode}, and it is clear that there is one hole that is becoming increasingly problematic.

\begin{table}
\begin{center}
\begin{tabular}{|c|c|}
\hline
Hop-length of shortest & Persistence of hole in \\
cycle surrounding hole & hop distance filtration \\
\hline
4, 5, 6 & 1 \\
7, 8, 9 & 2 \\
10, 11, 12 & 3 \\
\vdots & \vdots \\
$3k + 1$, $3k+2$, $3k+3$ & k \\
\hline
\end{tabular}
\end{center}
\caption{The relationship between persistence of a cycle in the hop-distance filtration, and size (in terms of hop-length) of the largest hole in surrounds. \label{HopSizes}}
\end{table}

\section{Conclusion}

Persistent homology and zigzag persistent homology represent the dynamic homology of a sequence of spaces by computing a set of intervals, describing the birth and death times of homological features in the sequence. We present here a method for assigning a representative cycle at each time point to each interval in this decomposition. The original choice and method for updating these representative cycles are geometrically motivated, so they are interpreted as `tracking' homological features. We describe the properties that the set of representative cycles must have to be compatible with the birth-death decomposition, and prove that the method presented maintains such properties.

Some applications of the method to track coverage holes in time-varying sensor networks are presented. For spaces in the plane, this method of tracking attempts to approximate the canonical basis for the first homology (where one homology class surrounds each hole), as best as possible, while still being compatible with the birth-death decomposition. Having chosen a specific representative cycle for each interval at each time point, additional features (such as estimates of hole size) can be attached onto the barcode, for a more comprehensive description of the dynamic coverage of the network.

\bibliographystyle{plain}
\bibliography{ZigzagCoverageBib}

\end{document}